\documentclass[10pt]{iopart}

\usepackage[breaklinks = true, colorlinks = true, linkcolor = blue, urlcolor = blue, citecolor = blue]{hyperref}
\usepackage{iopams}
\expandafter\let\csname equation*\endcsname\relax
\expandafter\let\csname endequation*\endcsname\relax
\usepackage{amsmath,amssymb}
\usepackage{url}
\usepackage[]{units}
\usepackage{graphicx}
\usepackage{epstopdf}
\usepackage{amsmath}
\usepackage{cite}
\begin{document}
\title[Strong lensing of explosive transients]{Strong gravitational lensing of explosive transients}

\author{Masamune Oguri$^{1,2,3}$}
\address{$^{1}$Research Center for the Early Universe, University of
  Tokyo, Tokyo 113-0033, Japan}
\address{$^{2}$Department of Physics, University of Tokyo, Tokyo
  113-0033, Japan}
\address{$^{3}$Kavli Institute for the Physics and Mathematics of the
  Universe (Kavli IPMU, WPI), University of Tokyo, Chiba 277-8582,
  Japan} 
\ead{masamune.oguri@ipmu.jp}

\begin{abstract}
Recent rapid progress in time domain surveys makes it possible to
detect various types of explosive transients in the Universe in large
numbers, some of which will be gravitationally lensed into multiple
images. Although a large number of strongly lensed distant galaxies
and quasars have already been discovered, strong lensing of explosive
transients opens up new applications, including improved measurements
of cosmological parameters, powerful probes of small scale structure of
the Universe, and new observational tests of dark matter scenarios, 
thanks to their rapidly evolving light curves as well as their compact
sizes. In particular, compact sizes of emitting regions of these
transient events indicate that wave optics effects play an
important role in some  cases, which can lead to totally new
applications of these lensing events. Recently we have witnessed first
discoveries of strongly lensed supernovae, and strong lensing events
of other types of explosive transients such as gamma-ray bursts, fast
radio bursts, and gravitational waves from compact binary mergers are
expected to be  observed soon. In this review article, we summarize
the current state of research on strong gravitational lensing of
explosive transients and discuss future prospects.
\end{abstract}

%
%
\noindent Keywords: cosmology, gravitational lensing, transients

\maketitle
\ioptwocol

\section{Introduction}
\label{sec:introduction}

Gravitational lensing is the deflection of light rays due to
intervening inhomogeneous matter distributions in the
Universe. The gravitational lensing effect is unambiguously predicted by
Einstein's General Relativity, and has actually been used to test the
validity of General Relativity as a gravitational theory. For
instance, the deflection angle at the surface of the Sun is predicted
to $1.7''$ in General  Relativity, which was confirmed by observations
during a solar eclipse in 1919 (see e.g., \cite{2001ASPC..252...21C}
for a historical review). 

When the defection angle is sufficiently large, it is possible that
multiple images of a distant source are observed. In order for such
{\it strong} gravitational lensing to be observed, a chance alignment
of a background source and a foreground object that acts as a lens
along the line-of-sight is needed. While the chance alignment of
multiple stars is quite rare \cite{1936Sci....84..506E}, strong
gravitational lensing (strong lensing) is observed to be more common
among galaxies and clusters of galaxies
\cite{1937PhRv...51..290Z,1937PhRv...51..679Z}.  
Galaxies and clusters of galaxies are massive enough to split multiple
images by more than an arcsecond on the sky, which can be resolved by
astronomical observations in various wavelengths. Observations of such
strong lensing events provide a unique opportunity to
accurately measure the mass of the foreground lensing object, as well
as to study the background object taking advantage of the
magnification due to the lensing effect, as noted by Zwicky
\cite{1937PhRv...51..290Z,1937PhRv...51..679Z}.  

Strong lensing was discovered for the first time in 1979 by Walsh {\it
  et al.} \cite{1979Natur.279..381W}. The background source is a
quasar, which is a very bright active galactic nucleus powered by a
supermassive black hole at the center of a galaxy. Quasars are bright
enough to be detected even at cosmological distances, and their
compact sizes suggest that their multiple images are well separated,
which make them as an ideal source for strong lensing. In the first
example, the quasar Q0957+561 at redshift $z=1.4$ is split into two
images separated by $6''$ due to the gravitational lensing effect of
a foreground group of galaxies. The lensing interpretation was
confirmed by the identical spectra of the two quasar images. 

Strong lensing of background galaxies has also been
discovered. Since galaxies are much larger in size than quasars,
lensed galaxies often form giant arcs, which are highly elongated
galaxy images due to strong lensing. Such giant arc was
discovered for the first time in the galaxy cluster A370 in the 1980s 
\cite{1986BAAS...18R1014L,1987A&A...172L..14S}.

Since the first discoveries, many strong lensing systems
have been discovered in various surveys. To date, more than 100
strongly lensed quasars have been discovered from radio and optical
surveys including Cosmic Lens All-Sky Survey
\cite{2003MNRAS.341....1M,2003MNRAS.341...13B} and Sloan Digital Sky
Survey Quasar Lens Search 
\cite{2006AJ....132..999O,2012AJ....143..119I,2016MNRAS.456.1595M}, 
from which evidence for the large value of cosmological constant has
been obtained \cite{2002PhRvL..89o1301C,2012AJ....143..120O}.
In addition, hundreds of gravitationally lensed galaxies have been
discovered in wide-field surveys including Sloan Digital Sky Survey 
\cite{2006ApJ...638..703B,2008ApJ...682..964B,2008AJ....135..664H,2009ApJ...696L..61K,2012ApJ...744...41B,2016ApJ...824...86S},
COSMOS \cite{2008ApJS..176...19F,2008MNRAS.389.1311J},
Canada-France-Hawaii Telescope Legacy Survey
\cite{2012ApJ...749...38M}, Herschel Astrophysical Terahertz Large
Area Survey \cite{2010Sci...330..800N,2013ApJ...762...59W}, and South
Pole  Telescope \cite{2013Natur.495..344V,2013ApJ...767..132H}.
These strongly lensed galaxies are used e.g., to constrain the dark
matter distribution in lensing galaxies as well as the initial mass
function of stars 
(e.g., \cite{2006ApJ...649..599K,2010ApJ...709.1195T}). 
A large number of gravitationally lensed distant galaxies have also
been discovered by deep imaging of central regions of massive clusters
of galaxies with {\it Hubble Space Telescope}
\cite{2012ApJS..199...25P,2017ApJ...837...97L,2019arXiv190302002C}.
Recently, more strong lens systems are being found in
various surveys such as Gaia
\cite{2018MNRAS.479.5060L,2019MNRAS.483.4242L,2019A&A...622A.165D},
Dark Energy Survey \cite{2017ApJS..232...15D,2018MNRAS.481.1041T,2018MNRAS.480.5017A,2019MNRAS.484.5330J},
Kilo-Degree Survey \cite{2019MNRAS.484.3879P}, Pan-STARRS1 
\cite{2019MNRAS.486.4987R}, and Subaru Hyper Suprime-Cam survey
\cite{2018PASJ...70S..29S,2018ApJ...867..107W}. 

One important application of strong lensing comes from
{\it time delays} between multiple images. The arrival time difference
between multiple images is naturally expected as they travel through
different paths, which serves as a very useful probe of the Universe. 
For instance, in 1964 Sjur Refsdal proposed to use 
measurements of time delays to constrain the Hubble constant $H_0$,
which is one of the most fundamental cosmological parameters 
\cite{1964MNRAS.128..307R}. This is possible because $H_0$ determines
the absolute length scale of the Universe, and therefore changes the
time delay between images by changing the difference of the light ray
paths. 

In order to measure time delays between multiple images, sources have
to be time-variable. Quasars are suited for this application, because
they are known to change their brightness, presumably due to the
variation of the gas inflow and accretion disk instabilities. Indeed
the time delay for the first gravitationally lensed quasar Q0957+561
is measured to be 417 days \cite{1997ApJ...482...75K}, and subsequently
time delays have been measured for more than 20 quasar lens systems
\cite{2007ApJ...660....1O,2010ApJ...712.1378P,2015LRR....18....2J}.
In combination with detailed modeling of mass distributions of lensing
galaxies, now quasar lens time delays constrain $H_0$ at better than
3\% precision 
\cite{2017MNRAS.468.2590S,2017MNRAS.465.4914B,2019MNRAS.484.4726B,2019arXiv190704869W}.

Recently measurements of $H_0$ attract a lot of attention given a
possible tension among them. One of the most traditional methods to
measure $H_0$ is the so-called distance ladder (e.g., 
\cite{2001ApJ...553...47F,2011ApJ...730..119R,2016ApJ...826...56R}),
with the most recent measurement yielding the best-fit value of
$H_0=74.03\pm 1.42$~km/s/Mpc including systematics
\cite{2019ApJ...876...85R}. On the other hand, $H_0$ can also be
inferred from observations of cosmic microwave background
anisotropies, yielding $H_0=67.4\pm 0.5$~km/s/Mpc assuming the
standard $\Lambda$-dominated cold dark matter model 
\cite{2018arXiv180706209P}. The discrepancy between these two
measurements might suggest new physics such as additional
relativistic particle species, or might be attributed to unknown
systematic errors in either or both of these two measurements.
While the latest measurement of $H_0$ from 6 quasar lens time delays
is $H_0=73.3^{+1.7}_{-1.8}$~km/s/Mpc \cite{2019arXiv190704869W} and is
consistent with the distance ladder result, more accurate and precise
measurements of $H_0$ from time delays as well as the exploration of a
possible dependence of the constraints on redshifts are important to 
understand the origin of the $H_0$ tension.  
 
In fact, the Refsdal's original proposal assumed to use strong
lensing of supernovae, rather than quasars, to measure
$H_0$ from time delays. Because of the relatively small number of
distant supernovae observed so far, strong lensing of
supernovae has not been discovered until recently, which is the
reason why strongly lensed supernovae have not been used to obtain
competitive constraints on $H_0$. However, strong lensing of
supernovae has several advantages over strong lensing of quasars, as
will be discussed below. These advantages make strongly lensed
supernovae an alternative powerful probe of the Universe. 

In addition to supernovae, there are other types of explosive
transients known, including gamma-ray bursts, fast radio bursts, and
gravitational waves from compact binary mergers. Here we refer
``explosive'' transients to as astronomical events with relatively
short time scales, which we adopt in order to distinguish them from
long term variable objects such as quasars. These transients are
observed at cosmological distances, and therefore are subject to
strong lensing applications. A notable difference of these transients
from supernovae is that their typical time scales of light curves,
seconds or milliseconds, are much shorter than the time scale of
supernova light curves, a month to several months. The shorter time
scales indicate that the measurement precision of time delays is much
better and that they can in principle probe much smaller mass
scale of the lensing object for which a typical time delay is much
shorter. 

In most applications of strong lensing, we can assume
geometric optics, which is a good approximation when the wavelength is
sufficiently small compared with the scale of the structure of
interest. However, there are cases where we have to take account of
wave optics effects, which is more fundamental than the geometric
optics (e.g., \cite{1992grle.book.....S}). Wave optics effects
produce interesting observable features such as the interference
pattern, which may provide additional useful information on the
lensing object. In order for this effect to be observed, the source
must be sufficiently compact compared to the lens, as the finite
source size smears the interference pattern. Since these explosive
transients have compact sizes as compared with quasars and galaxies,
strong lensing of explosive transients may open up the possibility of
using wave optics effects as additional applications. 

In this review article, we focus on strong lensing of 
explosive transients, which will be discovered in large numbers in the
future. We discuss how these events can be used to address several
outstanding questions in modern cosmology, such as the nature of dark
matter and dark energy. In addition, strong lensing can
be used to understand these explosive transients better, with the help
of gravitational lensing magnification. We also discuss the prospect
for detecting these events in the future. We note that this review
article focuses on a limited aspect of strong lensing, and
in fact there are many reviews and textbooks 
\cite{1992grle.book.....S,1992ARA&A..30..311B,2001stgl.book.....P,2006glsw.conf...91K,2010CQGra..27w3001B,2010ARA&A..48...87T,2011A&ARv..19...47K,2013SSRv..177...31M,2016A&ARv..24...11T,2017grle.book.....D,2018pgl..book.....C,2018PhR...778....1B}
that are useful to cover the broader aspects of strong lensing.  

The rest of this review article is organized as follows. In 
Section~\ref{sec:basics}, we briefly review basic theory of strong
lensing. In Section~\ref{sec:transients}, we summarize
explosive transients that we discuss in this review article. In
Section~\ref{sec:applications}, we discuss possible applications of
strong lensing of these explosive transients. In
Section~\ref{sec:observations}, we summarize observations so far and
also present future prospects. We give a brief summary in
Section~\ref{sec:conclusions}. Unless otherwise stated, we assume a
flat cosmological model with matter density $\Omega_M=0.3156$,
cosmological constant $\Omega_\Lambda=0.6844$, and the dimensionless
Hubble constant $h=0.6727$ \cite{2016A&A...594A..13P}. 

\section{The basics of strong gravitational lensing}
\label{sec:basics}

\subsection{Multiple images and the Einstein radius}
\label{sec:basics_ein}

We begin with a brief overview of the formulation of strong
lensing. While the path of light rays in arbitrary
matter distributions in the Universe is calculated by the geodesic
equation in General Relativity, which is a differential equation that
connects the path of light ray (i.e., null geodesics) to the space
time metric, in most astronomical situations where
deflection angles are small we can apply weak-field approximation to
linearize the geodesic equation and obtain the so-called lens
equation. Although the weak-field limit does not apply to the source
regions for e.g., binary black hole mergers, in the far-field
gravitational lensing regime considered in this article the 
weak-field limit is still valid. The lens equation can be regarded as
mapping between positions of the source (that would be observed in
absence of the gravitational lensing effect) and the image (that is
actually observed) on the sky. In the analysis of strong lensing, it
is also common to assume that the deflection is dominated by a single object
along the line-of-sight whose size is thin as compared with
cosmological distances, although in some cases line-of-sight mass
structures are not negligible. Under these approximations, the
gravitational lensing effect is fully described by the following lens
equation  
\begin{equation}
\boldsymbol{\beta}=\boldsymbol{\theta}-\boldsymbol{\alpha}(\boldsymbol{\theta}),
\label{eq:lenseq}
\end{equation}
where two-dimensional vectors $\boldsymbol{\beta}$ and
$\boldsymbol{\theta}$ denote positions of the source and the image on the
sky, respectively, and $\boldsymbol{\alpha}$ is the deflection angle 
\begin{equation}
\boldsymbol{\alpha}(\boldsymbol{\theta})=\frac{1}{\pi}\int
d\boldsymbol{\theta}'\frac{\boldsymbol{\theta}-\boldsymbol{\theta}'}{\left|\boldsymbol{\theta}-\boldsymbol{\theta}'\right|^2}\kappa(\boldsymbol{\theta}'),
\label{eq:defangle}
\end{equation}
where $\kappa$, which is sometimes referred to as convergence, is
essentially the projected surface mass density distribution of the
lensing object $\Sigma(\boldsymbol{\theta})$ normalized by the
critical surface density $\Sigma_{\rm cr}$
\begin{equation}
\kappa(\boldsymbol{\theta})=\frac{\Sigma(\boldsymbol{\theta})}{\Sigma_{\rm cr}}=\frac{1}{\Sigma_{\rm cr}}\int_{-\infty}^{\infty} dz\,
\rho(\boldsymbol{\theta}, z),
\end{equation}
\begin{equation}
\Sigma_{\rm cr}=\frac{c^2}{4\pi G}\frac{D_{\rm os}}{D_{\rm ol}D_{\rm ls}},
\end{equation}
where $\rho$ is the three-dimensional density profile of the lensing
object, $z$ denotes the line-of-sight direction, $c$ is the speed of
light, $G$ is the gravitational constant, and $D_{\rm os}$, $D_{\rm
  ol}$, and $D_{\rm ls}$ are angular diameter distances (see e.g.,
\cite{2008cosm.book.....W} for the definition) from the
observer to the source, from the observer to the lens, and from the
lens to the source, respectively.
We note that the integrand in Equation~(\ref{eq:defangle}) is the
deflection angle from a differential mass element such that the
integral sums over deflection angles from different mass elements.
Figure~\ref{eq:lenseq} shows a
schematic illustration of a gravitational lens system, including
definitions of the angles involved in the lens equation. 

\begin{figure}
\begin{center}
\includegraphics[width=8.0cm]{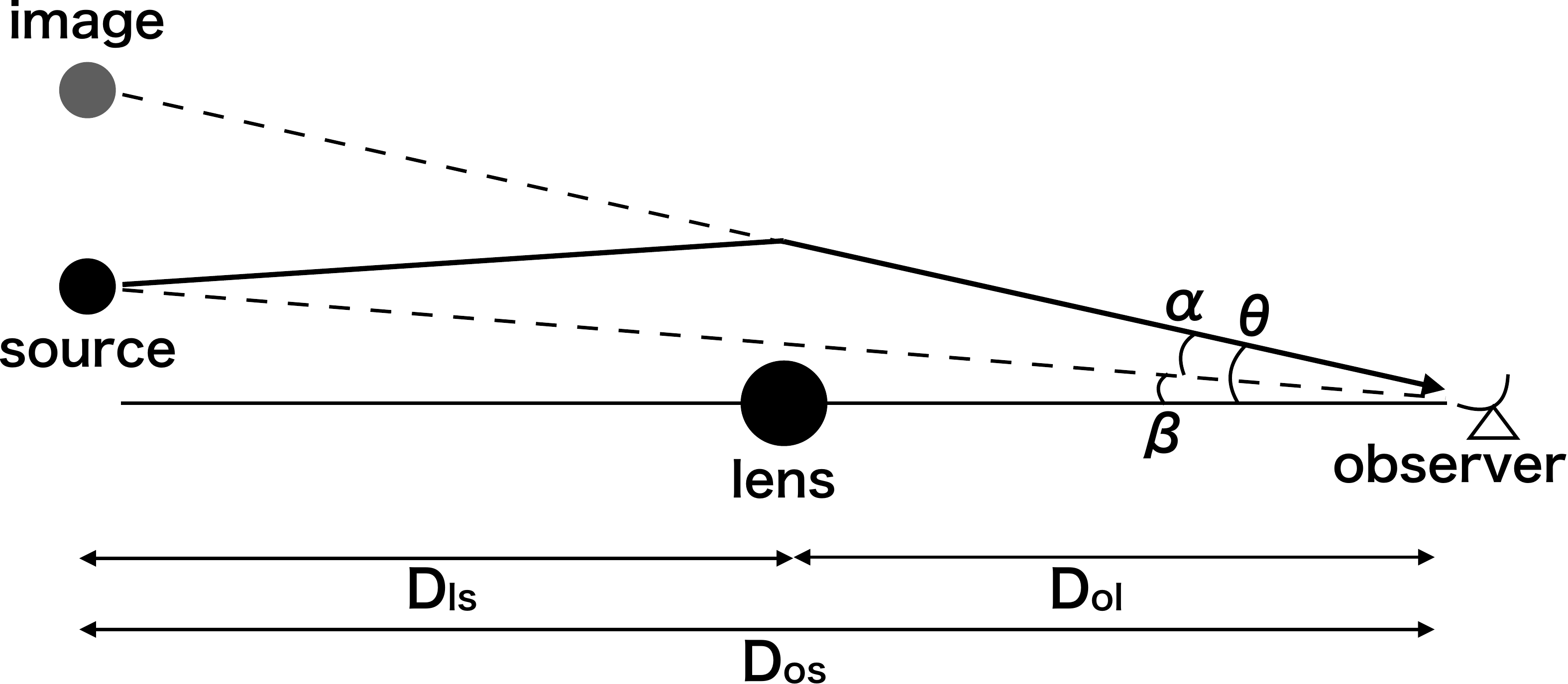}
\caption{Schematic illustration of a gravitational lens system.}
\label{fig:gl_config}
\end{center}
\end{figure}

Given the mass distribution of the lens, the lens
equation~(\ref{eq:lenseq}) predicts the image position 
$\boldsymbol{\theta}$ for the source position $\boldsymbol{\beta}$.  
Importantly, the lens equation is in general nonlinear in
$\boldsymbol{\theta}$, which suggests that multiple
$\boldsymbol{\theta}$ can satisfy the lens equation for a given
$\boldsymbol{\beta}$. These multiple solutions of the lens equation
correspond to multiple images. Such multiple images can be produced
where the deflection angle $\boldsymbol{\alpha}$ is sufficiently
large, i.e., in high density regions such as centers of galaxies and
clusters of galaxies.

Gravitational lensing not only changes the observed position on the
sky, but also changes the observed brightness of the source. The
change of the brightness is determined by the Jacobi matrix from the
lens equation
\begin{equation}
A(\boldsymbol{\theta})=\frac{\partial\boldsymbol{\beta}}{\partial\boldsymbol{\theta}},
\label{eq:jacobi}
\end{equation}
from which the magnification $\mu$ of each image is computed as
\begin{equation}
\mu(\boldsymbol{\theta})=\frac{1}{{\rm det}A(\boldsymbol{\theta})}.
\label{eq:mag_def}
\end{equation}
This means that the image at $\boldsymbol{\theta}$ is magnified by a
factor of $|\mu(\boldsymbol{\theta})|$. The sign of $\mu$ corresponds
to the parity of the image such that the parity of the image is
flipped when $\mu$ is negative.

\begin{figure}
\begin{center}
\includegraphics[width=7.0cm]{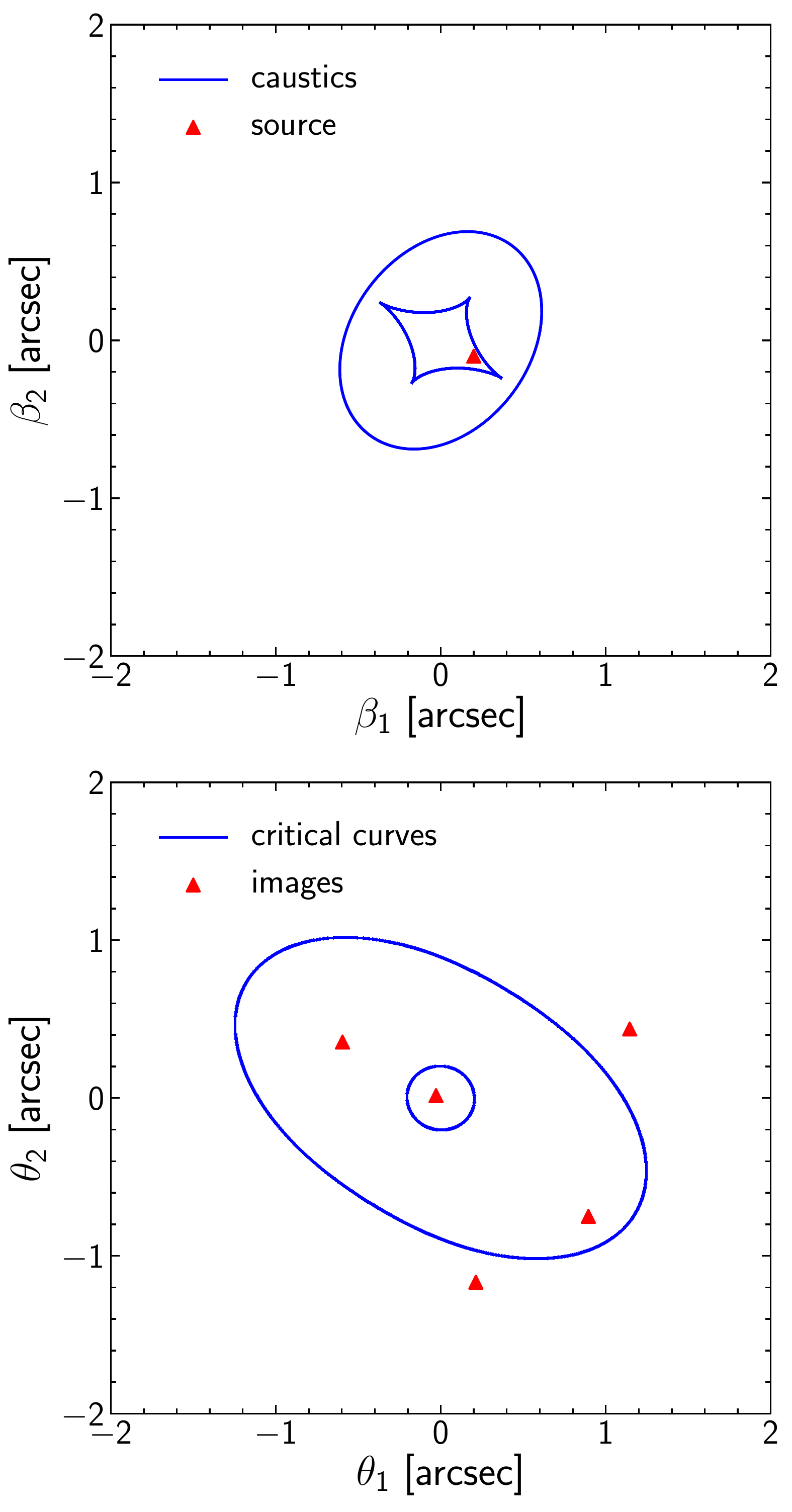}
\caption{An example of the configuration of multiple images produced
  by strong lensing. The upper panel show the location of
  the source and caustics in the source plane,
  $\boldsymbol{\beta}=(\beta_1,\,\beta_2)$. The lower panel show
  multiple images and critical curves in the image plane
  $\boldsymbol{\theta}=(\theta_1,\,\theta_2)$. In this example, 5
  images are produced. The lens equation is solved using {\tt glafic}
  \cite{2010PASJ...62.1017O}.} 
\label{fig:crit_example}
\end{center}
\end{figure}

Equation~(\ref{eq:mag_def}) indicates that magnification factors
formally diverge at points satisfying ${\rm det}A(\boldsymbol{\theta})=0$.
These points in the image plane form closed curves, which are called
critical curves. Corresponding curves in the source plane obtained via
the lens equation~(\ref{eq:lenseq}) are called caustics. These curves
are important in strong lensing studies because they are closely
related to the image multiplicity. When a source is located far from
caustics, there is only one image. Once a source crosses a caustic,
the number of images increases or decreases by 2. Therefore, we can
infer the number of images and image configuration by checking
the position of a source with respect to caustics. We show an example
in Figure~\ref{fig:crit_example}, in which we can see 5 images as the
source crosses caustics twice (see also \cite{1981ApJ...244L...1B}).

When the mass distribution of the lensing object is spherically
symmetric (i.e., $\kappa(\boldsymbol{\theta})=\kappa(\theta)$), 
the deflection angle~(\ref{eq:defangle}) reduces to 
\begin{equation}
\boldsymbol{\alpha}(\boldsymbol{\theta})=\frac{\boldsymbol{\theta}}{\theta}\alpha(\theta)=\frac{2\boldsymbol{\theta}}{\theta^2}\int_0^\theta d\theta'\,\theta'\kappa(\theta'),
\end{equation}
and therefore the lens equation~(\ref{eq:lenseq}) reduces to the
one-dimensional equation
\begin{equation}
\beta=\theta-\alpha(\theta).
\end{equation}
This indicates that in the limit of $\beta\rightarrow 0$ the lensed
image forms a ring with the radius $\theta_{\rm Ein}$ that satisfies
\begin{equation}
\theta_{\rm Ein}-\alpha(\theta_{\rm Ein})=0.
\label{eq:tein_def}
\end{equation}
This radius $\theta_{\rm Ein}$ is called the Einstein radius. We can
rewrite equation~(\ref{eq:tein_def}) to obtain
\begin{equation}
M(<\theta_{\rm Ein})=D_{\rm ol}^2\int_0^{\theta_{\rm Ein}}
d\theta'\,2\pi\theta'\Sigma(\theta')=\pi D_{\rm ol}^2\theta_{\rm Ein}^2\Sigma_{\rm
  cr},
\label{eq:mein}
\end{equation}
which indicates that the Einstein radius probes the total projected
mass of the lensing object within the Einstein radius, as long as lens
and source redshifts are known.

In fact, equation~(\ref{eq:mein}) has important implications. It has
been known that image separations between multiple images are
typically twice the Einstein radius, which is approximately true even
when the lens mass distribution is not spherically symmetric. This
means that we can estimate the Einstein radius from
observations of multiple images (see also Figure~\ref{fig:crit_example}). 
We can then use equation~(\ref{eq:mein}) to translate
the observed Einstein radius into the total projected mass within the
Einstein radius $M(<\theta_{\rm Ein})$. Remarkably, the relation given
by equation~(\ref{eq:mein}) does not depend on the radial density
profile of the lens object. Therefore, $M(<\theta_{\rm Ein})$ is one
of the most robust quantities we can extract from observations of
strong lensing, and hence plays a central role in the strong lens
analysis. 

It is useful to present Einstein radii for some lens mass distributions. 
For instance, the simplest mass model is a point mass lens with mass
$M$, $\rho(\boldsymbol{r})=M\delta(\boldsymbol{r})$. From
equation~(\ref{eq:mein}), the Einstein radius is found to
\begin{eqnarray}
\theta_{\rm Ein}&=&\frac{1}{D_{\rm ol}}\sqrt{\frac{M}{\pi\Sigma_{\rm
      cr}}}\nonumber\\ 
&\sim& 1.63''\times 10^{-6}\left(\frac{M}{M_\odot}\right)^{1/2}
\left(\frac{D_{\rm ol}D_{\rm os}/D_{\rm ls}}{3.06\,{\rm Gpc}}\right)^{-1/2},
\label{eq:tein_pm}
\end{eqnarray}
where distances (Gpc stands for gigaparsec) are normalized to values
at the lens redshift $z_{\rm l}=0.5$ and the source redshift $z_{\rm
  s}=1.0$. Another lens model that is commonly used is a singular
isothermal sphere (SIS) model whose three-dimensional radial density
profile is given by $\rho(r)=\sigma^2/2\pi G r^2$, where $\sigma$ is
the velocity dispersion. The Einstein radius of the SIS model is
computed as 
\begin{eqnarray}
\theta_{\rm Ein}&=&\frac{4\pi \sigma^2}{c^2}\frac{D_{\rm ls}}{D_{\rm os}}\nonumber\\ 
&\sim& 0.492''\left(\frac{\sigma}{200\,{\rm km\,s^{-1}}}\right)^2
\left(\frac{D_{\rm ls}/D_{\rm os}}{0.426}\right),
\end{eqnarray}
where the distances are again normalized to values at the lens
redshift $z_{\rm l}=0.5$ and the source redshift $z_{\rm s}=1.0$.

\begin{figure}
\begin{center}
\includegraphics[width=8.0cm]{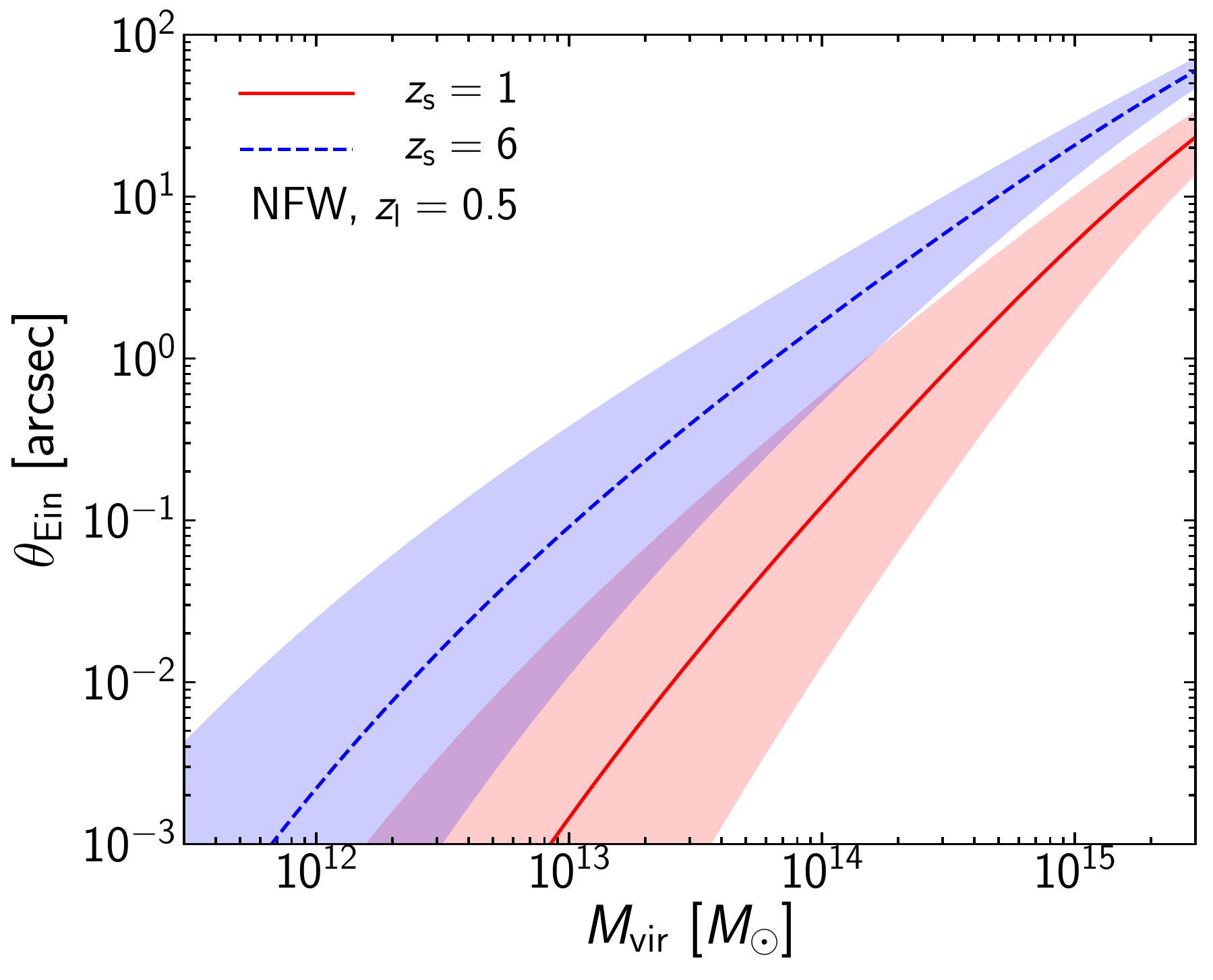}
\caption{The Einstein radius $\theta_{\rm Ein}$ as a function of the
  halo mass $M_{\rm vir}$ for an NFW profile. The lens redshift is
  fixed to $z_{\rm l}=0.5$, whereas source redshifts of $z_{\rm s}=1$
  ({\it solid red}) and $6$ ({\it dashed blue}) are considered. Here the
  mass-concentration relation presented in \cite{2015ApJ...799..108D}
  is adopted. The shaded regions represent the uncertainty of
  $\theta_{\rm Ein}$ originating from 1$\sigma$ scatter ($\sigma_{\ln
    c}=0.3$) of the concentration parameter. }
\label{fig:nfw_ein}
\end{center}
\end{figure}

The so-called Navarro-Frenk-White (NFW) profile
\cite{1996ApJ...462..563N,1997ApJ...490..493N} is yet another mass
profile that is commonly used in the analysis of strong lensing. The
NFW profile is used to model the density profile of dark 
matter halos, with its three-dimensional radial density profile given
by $\rho(r)\propto r^{-1}(r+r_{\rm s})^{-2}$, where $r_{\rm s}$ is the
scale radius. While the deflection angle of the spherical NFW profile 
can be computed analytically (e.g., \cite{1996A&A...313..697B}), no
simple analytical expression for the Einstein radius is known.
Figure~\ref{fig:nfw_ein} shows the relation between the halo mass and
the Einstein radius for the NFW profile. It is found that the Einstein
radius is sensitive to not only the halo mass but also the
concentration parameter $c_{\rm vir}=r_{\rm vir}/r_{\rm s}$, where
$r_{\rm vir}$ is the virial radius. Since its inner density profile is
not very steep, the Einstein radius for the NFW profile encloses
relatively little of the total halo mass. 

\subsection{Time delays}
\label{sec:timedelay}

The light comprising the multiple images produced by strong lensing
travels different paths, and hence take different amounts of time to
propagate to us. The arrival time difference between multiple images
can be observed if the source is time-variable such as quasars and
explosive transients. The time delay for each image is computed as
\begin{equation}
\Delta t = \frac{1+z_{\rm l}}{c}\frac{D_{\rm ol}D_{\rm os}}{D_{\rm ls}}\left[\frac{(\boldsymbol{\theta}-\boldsymbol{\beta})^2}{2}-\phi(\boldsymbol{\theta})\right],
\label{eq:tdelay}
\end{equation}
where $\phi(\boldsymbol{\theta})$ is the lens potential that is
related to the deflection angle as 
\begin{equation}
\boldsymbol{\alpha}=\boldsymbol{\nabla}_{\boldsymbol{\theta}}\phi.
\end{equation}
The time delay involves contributions both from a geometric delay
originating from different path lengths and a gravitational time delay
originating from the gravitational potential of the lens.  
Note that we can observe only the time delay between different images
e.g., we can observe $\Delta t_{ij}=\Delta t_i-\Delta t_j$ for
image $i$ and $j$. 

Again, it is useful to present time delays for some lens models such
as point mass lens and SIS. Using equation~(\ref{eq:tdelay}), we can
rewrite $\Delta t_{ij}$ as 
\begin{equation}
\Delta t_{ij} = \Delta t_{\rm fid}\Phi(\boldsymbol{\theta}_i, \boldsymbol{\theta}_j),
\label{eq:deltat}
\end{equation}
\begin{equation}
\Delta t_{\rm fid}=\frac{1+z_{\rm l}}{c}\frac{D_{\rm ol}D_{\rm os}}{D_{\rm
    ls}}\theta_{\rm Ein}^2,
\label{eq:deltat_fid}
\end{equation}
\begin{equation}
\Phi(\boldsymbol{\theta}_i,
\boldsymbol{\theta}_j)=\frac{(\boldsymbol{\theta}_i-\boldsymbol{\beta})^2}{2\theta_{\rm
    Ein}^2}-\frac{(\boldsymbol{\theta}_j-\boldsymbol{\beta})^2}{2\theta_{\rm
    Ein}^2}-\frac{\phi(\boldsymbol{\theta}_i)}{\theta_{\rm
    Ein}^2}+\frac{\phi(\boldsymbol{\theta}_j)}{\theta_{\rm Ein}^2}.
\label{eq:deltat_phi}
\end{equation}
The factor $\Delta t_{\rm fid}$ represents a typical size of the time
delay for the lens, and $\Phi(\boldsymbol{\theta}_i,
\boldsymbol{\theta}_j)$ is a $\mathcal{O}(1)$ function that represents
the dependence of the time delay on the image configuration. For
instance, if the multiple image configuration is {\it symmetric}, we have
$|\boldsymbol{\theta}_i|\sim |\boldsymbol{\theta}_j|$ and
$|\boldsymbol{\beta}|\sim 0$, leading to $\Phi(\boldsymbol{\theta}_i,
\boldsymbol{\theta}_j)\sim 0$. Put another way, time delays are larger
when the image configuration is more asymmetric, and are smaller when
the image configuration is more symmetric.

We can also reinterpret the gravitational lensing effect from this
expression of the time delay 
\cite{1985A&A...143..413S,1986ApJ...310..568B}. 
Fermat's principle in geometric optics states that a light ray takes a
path with a stationary path length. This immediately suggests that
observed images should satisfy the following condition
\begin{equation}
\boldsymbol{\nabla}_{\boldsymbol{\theta}}\Delta t=0.
\label{eq:fermat}
\end{equation}
By inserting equation~(\ref{eq:tdelay}) to equation~(\ref{eq:fermat}) 
it is easily found that we can recover the lens
equation~(\ref{eq:lenseq}) from this condition.

The typical time delay value is encapsulated by $\Delta t_{\rm fid}$.
Using equation~(\ref{eq:mein}), $\Delta t_{\rm fid}$ is rewritten as 
\begin{equation}
  \Delta t_{\rm fid}=(1+z_{\rm l})\frac{4G M(<\theta_{\rm Ein})}{c^3},
\label{eq:tfid_mein}
\end{equation}
which suggests that $\Delta t_{\rm fid}$ is on the order of the light
crossing time of the ``gravitational radius'' for the mass defined by
$M(<\theta_{\rm Ein})$. 
Again, by normalizing distances to those at the lens redshift $z_{\rm
  l}=0.5$ and the source redshift $z_{\rm s}=1.0$, 
$\Delta t_{\rm fid}$ is estimated as
\begin{equation}
\Delta t_{\rm fid} \sim 128\,{\rm day}\times\left(\frac{\theta_{\rm
    Ein}}{1''}\right)^2\left[\frac{(1+z_{\rm l})D_{\rm ol}D_{\rm
      os}/D_{\rm ls}}{4.59\,{\rm Gpc}}\right].
\label{eq:tfid_value}
\end{equation}
Therefore, for typical galaxy-scale strong lens systems with
$\theta_{\rm Ein}\sim 0.5''-1''$, we expect time delays on the order
of a month to a few months. For reference, by inserting
the typical Einstein radius of a solar mass compact object (see
equation~\ref{eq:tein_pm}), $\theta_{\rm Ein}/1''=10^{-6}$, to
equation~(\ref{eq:tfid_value}), we obtain $\Delta t_{\rm fid}\sim
1.1\times 10^{-5}$~sec for the fiducial distances. More generally,
from equation~(\ref{eq:tfid_mein}) we can estimate
$\Delta t_{\rm fid}$ for a point mass lens with mass $M$ as 
\begin{equation}
\Delta t_{\rm fid} \sim 1.97\times 10^{-5}~{\rm sec}\times (1+z_{\rm l})\left(\frac{M}{M_\odot}\right),
\label{eq:tfid_value_point}
\end{equation}
which can also be applicable to other lens models if we replace $M$ to $M(<\theta_{\rm Ein})$.

On the other hand, the function
$\Phi(\boldsymbol{\theta}_i, \boldsymbol{\theta}_j)$ depends on
the assumed mass model. For a point mass lens, 
\begin{equation}
\Phi(\boldsymbol{\theta}_i, \boldsymbol{\theta}_j)=\ln\left(\frac{\theta_j}{\theta_i}\right)+\frac{\theta_j^2-\theta_i^2}{2\theta_i\theta_j},
\label{eq:phi_point}
\end{equation}
and for an SIS lens
\begin{equation}
\Phi(\boldsymbol{\theta}_i,
\boldsymbol{\theta}_j)=\frac{2(\theta_j^2-\theta_i^2)}{(\theta_j+\theta_i)^2},
\label{eq:phi_sis}
\end{equation}
 where $\theta_i=|\boldsymbol{\theta}_i|$ and $\theta_j=|\boldsymbol{\theta}_j|$. 
Again, no simple analytic expression of $\Phi$ for the NFW profile is
known. 

As mentioned in Section~\ref{sec:introduction}, time delays provide a
powerful means of measuring the Hubble constant $H_0$, which is
sometimes referred to as time delay cosmography. Given that
$\theta_{\rm Ein}$ is well constrained from the data, observations of
time delays between multiple images put direct constraints on the
distance ratio $D_{\rm ol}D_{\rm os}/D_{\rm ls}$, which is inversely
proportional to $H_0$, but only if $\Phi$ is accurately known. The
examples above already indicate that values of $\Phi$ depend on the
underlying lens mass model, which implies that accurate determinations
of lens mass distributions are a key for successful time delay
cosmography.  

When multiple images are observed, we can constrain the lens mass
distribution from positions and flux ratios of multiple images.
However, in most cases these constraints are insufficient to robustly
constrain the lens mass distribution, and we need additional
constraints. For instance, host galaxies of quasars or any explosive
transients are also expected to be lensed into extended arcs, which
may provide useful additional constraints (e.g., 
\cite{2001ApJ...547...50K}). Furthermore, the velocity dispersion of
the lensing galaxy, which can be observed by deep spectroscopy of the
lensing galaxy, is sometimes used as additional constraints on the
lens mass distribution (e.g., \cite{2002MNRAS.337L...6T}).

However, there is a fundamental difficulty in the strong lensing
analysis, which originates from various degeneracies inherent to the
lens equation. One such example is the mass-sheet degeneracy 
\cite{1985ApJ...289L...1F}, in which the following transform is
considered
\begin{equation}
\phi(\boldsymbol{\theta})\rightarrow (1-\kappa_{\rm
  ext})\phi(\boldsymbol{\theta})+\kappa_{\rm ext}\frac{\theta^2}{2},
\end{equation}
\begin{equation}
\boldsymbol{\beta}\rightarrow (1-\kappa_{\rm ext})\boldsymbol{\beta},
\end{equation}
where $\kappa_{\rm ext}$ is constant. It is straightforward to see
that this transform keeps the lens equation~(\ref{eq:lenseq})
unchanged.  This transform corresponds to an operation that rescales
the mass of the lensing object and instead inserts a constant mass
sheet $\kappa_{\rm ext}$. Importantly, this transform also changes 
time delays (\ref{eq:tdelay}) between any multiple image pairs as
\begin{equation}
\Delta t_{ij}\rightarrow (1-\kappa_{\rm ext})\Delta t_{ij},
\end{equation}
which indicates that $H_0$ estimated from {\it observed} time delay
should scales as $H_0\rightarrow (1-\kappa_{\rm ext})H_0$. Therefore,
$H_0$ measured from time delays is subject to the uncertainty of
$\kappa_{\rm ext}$ that cannot be constrained from strong lensing
observations. As we will discuss later, one way to break the degeneracy
is to observe the magnification factor $\mu$, because the transform
changes $\mu$ as
\begin{equation}
\mu\rightarrow (1-\kappa_{\rm ext})^{-2}\mu.
\label{eq:mass_sheet_mu}
\end{equation}
Note that this transform does not change the ratio of magnification
factors between multiple images. 

This mass-sheet degeneracy implies other approximate degeneracies. For
instance, for a power-law mass model with $\phi\propto r^\beta$ ($\beta=1$
corresponds to an SIS profile), the change of $\beta$ around $\beta=1$
can be approximated by the mass-sheet transform with 
$1-\kappa_{\rm ext}=2-\beta$ (e.g., \cite{2002MNRAS.332..951W}), 
which implies that the Hubble
constant from time delays is sensitive to the radial slope of the
density profile of the lensing object, which is difficult to be
constrained from strong lensing observations. Furthermore, the
mass-sheet transform is generalized to the source-position transform
\cite{2014A&A...564A.103S,2017A&A...601A..77U}, which is essentially a 
global mapping of the source plane that keeps observed image
positions unchanged. In order to measure $H_0$ robustly from time
delays, it is essential to explore these degeneracies carefully, and
to make use of additional observational constraints that can break
these degeneracies.  

\subsection{Microlensing and Substructure lensing}
\label{sec:microlens}

While calculations of multiple image positions and time delays so far
assumed that the lens mass distribution is smooth, this assumption is
in fact not valid. Small scale structures in lensing objects often
affect strong lensing observables significantly, and therefore should
be taken into account.

One such example is microlensing by stars in lensing galaxies (see
e.g.,
\cite{2002ApJ...580..685S,2004ApJ...605...58K,2006glsw.conf..453W}).
The contribution to the lens potential includes both dark and
baryonic matter in the lensing object, and some fraction of the
baryonic matter consists of stars. Lensing by these individual stars
can significantly affect magnifications of individual multiple images
if the source size is comparable or smaller than the Einstein radii
of the stars (see e.g., \cite{1979Natur.282..561C,2005ApJ...628..594M}).

\begin{figure}
\begin{center}
\includegraphics[width=8.0cm]{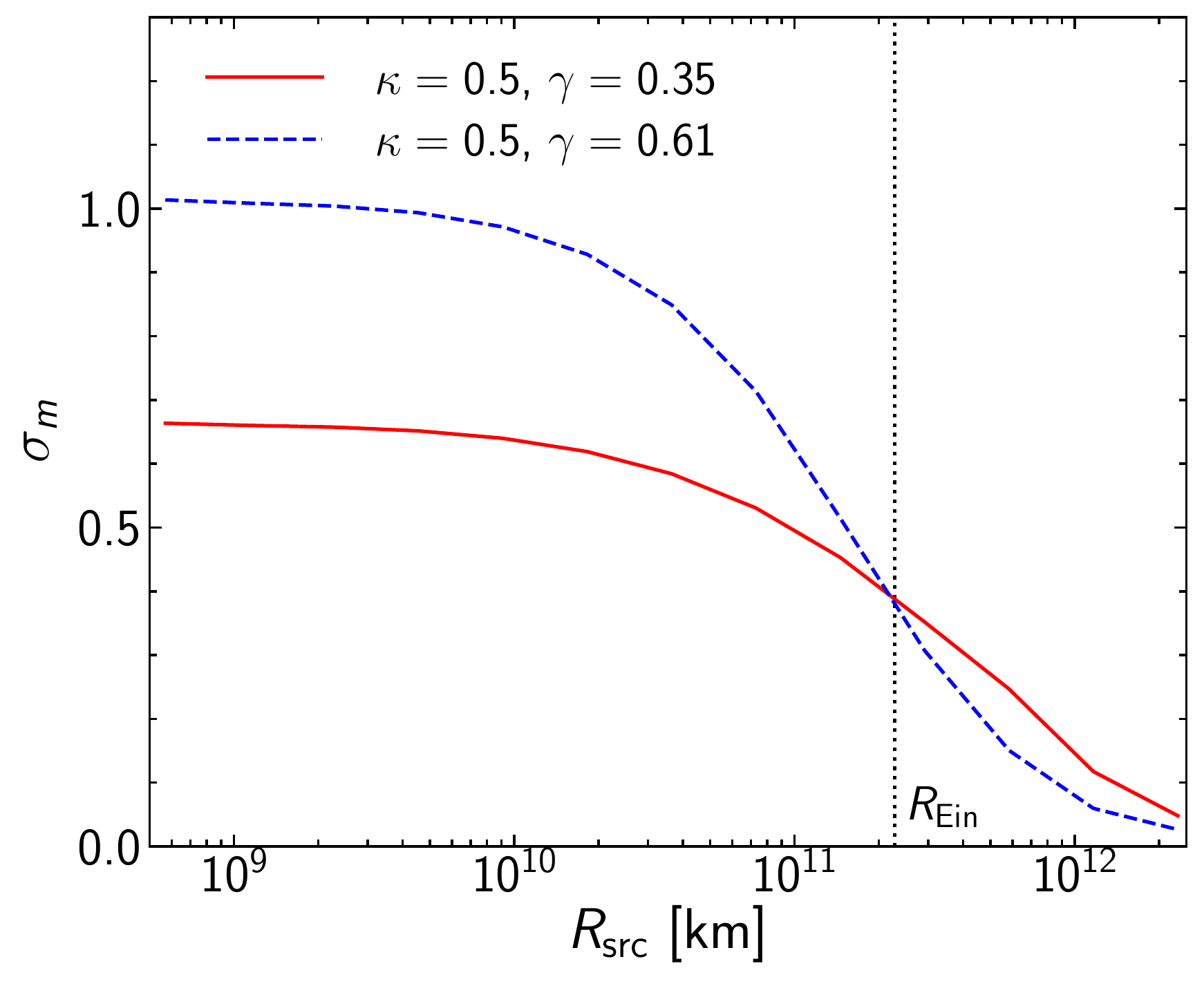}
\caption{Flux variabilities $\sigma_m$ (the standard deviation in
  magnitude) due to microlensing as a function of the source size,
  which are computed using the GERLUMPH microlensing magnification
  maps \cite{2014ApJS..211...16V,2015ApJS..217...23V} with the smooth
  matter fraction of $0.5$. Note that the distributions of flux
  variabilities are not Gaussian particularly for small source
  sizes. We show results both for positive parity (convergence
  $\kappa=0.5$ and shear $\gamma=0.35$) and negative parity
  ($\kappa=0.5$ and $\gamma=0.61$) cases. The source is assumed to
  have a top-hat profile with the radius $R_{\rm src}$. We assume
  $z_{\rm l}=0.5$, $z_{\rm s}=1$, and the microlens mass
  $M=0.3\,M_\odot$ to convert the simulation results into physical
  units. For comparison, the Einstein radius in the source plane is
  indicated by a vertical dotted line.} 
\label{fig:size_dm}
\end{center}
\end{figure}

To illustrate the sensitivity of microlensing variabilities on the
source size, in Figure~\ref{fig:size_dm} we show examples of flux
variabilities as a function of the source size $R_{\rm src}$, which is
computed using the GPU-Enabled, High Resolution cosmological
MicroLensing parameter survey (GERLUMPH) microlensing magnification maps
\cite{2014ApJS..211...16V,2015ApJS..217...23V}. It is clear that
microlensing variabilities are suppressed at $R_{\rm src}\gtrsim R_{\rm
  Ein}$ due to the large source size, where $R_{\rm Ein}=D_{\rm
  os}\theta_{\rm Ein}$ is the Einstein radius of the microlens in the
source plane.

Strong lensing allows us to probe the small-scale structure of the
dark matter distribution as well. In the standard cold dark matter
(CDM) model, the dark matter distribution in galaxies and clusters is
predicted to be lumpy rather than smooth. In the CDM model, the mass
function of such substructures extends to very small masses in which
no star is formed (e.g., \cite{2005Natur.433..389D}). The detection of
very small mass substructures in observations therefore serves as a
critical test of the CDM model (e.g., \cite{2017ARA&A..55..343B}). 

Substructures can be detected by strong lensing via flux ratios between
multiple images \cite{1998MNRAS.295..587M}. This is because
substructures can affect the magnification of one of multiple images to
produce anomalous flux ratios that cannot be reproduced by mass models
assuming smooth mass distributions. However, in order to study the
abundance of substructures from flux ratio anomalies, lensing by
substructures should be distinguishable from microlensing by stars in
lensing galaxies, because microlensing also changes flux ratios
between multiple images as discussed above. One way to overcome this
issue is to use sources whose sizes are sufficiently large so that
they are insensitive to microlensing (see
Figure~\ref{fig:size_dm}). 

In addition to flux ratios, substructures affect time delays between
multiple images as well, particularly for time delays between multiple
images with small angular separations, such as merging image pairs near
the critical curve \cite{2007ApJ...660....1O,2009ApJ...699.1720K}.
This is because the smaller time delays between such merging pairs
make them more sensitive to perturbations.

\subsection{Lensing rates}
\label{sec:lensrate}

Strong lensing is a rare event that occurs only when the
light ray from a distant object passes through (or near) high density
regions such as centers of galaxies and clusters. The chance
probability of strong lensing can be calculated as long as the mass
distribution and abundance of putative lensing objects are known. From the
density profile one can derive the lensing cross section, i.e., the
area on the sky within which strongly lensed multiple images are
produced, which is on the order of $\theta_{\rm Ein}^2$.

Historically, strong lensing probabilities are calculated assuming
lensing by galaxies. For instance, the detailed calculation in
Turner {\it et al.} \cite{1984ApJ...284....1T} indicates that strong
lensing events are dominated by those due to field elliptical galaxies. They
also show that strong lensing probabilities are a steeply increasing
function of the source redshift. Calculations of strong lensing
probabilities have been improved partly due to improved measurements
of velocity dispersion functions of galaxies in observations
\cite{1992ApJ...393....3F,1996ApJ...466..638K,1999ApJ...510...42C,2002PhRvL..89o1301C,2003MNRAS.346..746C,2005ApJ...622...81M,2005ApJ...624...34H,2005MNRAS.364.1451O,2010MNRAS.405.2579O,2012AJ....143..120O,2015ApJ...811...20C}.

It has been known that clusters of galaxies also produce strong
lensing. While individual clusters have larger lensing
cross sections than galaxies, clusters are much less abundant than
galaxies. Narayan and White \cite{1988MNRAS.231P..97N} discussed the
image separation distribution in the standard CDM
cosmology to argue that the contribution of clusters to the total
strong lensing probability is small but non-negligible. This
calculation has been updated following the improved knowledge of
the density profiles and the abundance of clusters  
\cite{1995Sci...268..274W,1995ApJ...453..545K,2001ApJ...555..504W,2001ApJ...549L..25K,2001ApJ...563..489T,2004ApJ...605...78O,2004ApJ...610..663O,2007MNRAS.382..121H,2007MNRAS.378..469L,2008MNRAS.391..653M,2008MNRAS.390.1647B,2009MNRAS.392..930O,2012MNRAS.423.2308Z,2012A&A...547A..66R,2014A&A...565A..28W}. 

As briefly mentioned in Section~\ref{sec:basics_ein}, $N$-body
simulations of the structure formation in the CDM model have
revealed that the density profile of dark matter halos is universal
and is well approximated by the NFW profile
\cite{1996ApJ...462..563N,1997ApJ...490..493N}. As shown in
Figure~\ref{fig:nfw_ein}, the Einstein radius of the NFW profile is a
steep function of the halo mass such that it becomes too small
for galaxy-scale dark matter halos, $M_{\rm vir}\lesssim 10^{13}M_\odot$,
which appears to contradict observations in which there are many
strong lens systems with $\theta_{\rm Ein}\sim 1''$ due to isolated
galaxies.

This issue is resolved by taking proper account of the baryonic
component. Dissipative cooling of gas makes the spatial distribution
of stars much more compact that that of dark matter. At the galaxy
scale this effect is more efficient such that the total density
profile of dark matter and the baryonic component resembles an SIS
profile that was also mentioned in Section~\ref{sec:basics_ein}. 
Indeed calculations based on this idea successfully reproduce the
observed image separation distribution of strong lenses for a wide
mass range from the galaxy to cluster scales
\cite{1998PhDT.........6K,2000ApJ...532..679P,2001ApJ...559..531K,2002ApJ...566..652L,2002ApJ...580....2O,2003ApJ...584L...1M,2004ApJ...600L...7H,2004ApJ...601..104K,2005ApJ...621..559K,2006MNRAS.367.1241O,2012ApJ...749...38M,2018MNRAS.475.4939A},
which suggest that the contribution of clusters to the total strong
lensing probability is $\sim 1-10$\% and that strong lensing events
are dominated by those due to single galaxies.

In practice, we need to take account of selection effects when we
compare expected strong lensing probabilities with observations.
The best-known example is the magnification bias
\cite{1980ApJ...242L.135T}, which originates from the fact that
in any survey objects are detected only above some flux
threshold. Because of  gravitational lensing magnifications, faint
objects that fall below the threshold in absence of gravitational
lensing can in fact be observed thanks to the magnification.  
This effect increases the observed strong lensing rates and hence
should be taken into account. For sources with the flux $f$ and
differential number counts $N(f)=dN/df$, the magnification bias factor
$B$ is computed as 
\begin{equation}
B=\frac{1}{N(f)}\int_{\mu_{\rm min}}^\infty
\frac{d\mu}{\mu}\frac{dP}{d\mu}N(f/\mu),
\label{eq:mag_bias}
\end{equation}
where $dP/d\mu$ denotes the magnification probability distribution.
As a simple example, assuming power-law number counts $N(f)\propto
f^{-\alpha}$ and an SIS lens for which $dP/d\mu=8/\mu^3$
($\mu>\mu_{\rm min}=2$),
we can compute $B$ as
\begin{equation}
  B=\frac{2^\alpha}{3-\alpha}.
  \label{eq:mag_bias_pl}
\end{equation}
From this expression it is found that steeper number counts (larger
$\alpha$) lead to the larger magnification bias factor. 

In addition to the magnification bias, there are other possible
selection effects. Multiple images with large differences in their
fluxes are difficult to be identified in observations, and any cut on
the flux ratio of multiple images reduces the strong lensing
probability. For lens systems with small Einstein radii, image
separations of multiple images can be too small to be resolved in
observations, depending on the spatial resolutions of observations. Also
when the size of the source is comparable or larger than the Einstein
radius, the gravitational lensing effect on the source is quite
inefficient. These effects remove strong lenses with small image
separations from the sample, leading to the smaller strong lensing
probability. Therefore, depending on the threshold on the image
separation, they can significantly change the relative contribution of
cluster lenses to the whole strong lens sample. In order to make fair
comparisons with observations, any theoretical calculations of strong
lensing probabilities should take proper account of these selection
effects. 

Here we present some examples of calculations of strong lensing
probabilities, following recent calculations presented in
Oguri \cite{2018MNRAS.480.3842O}. In short, we compute strong lensing  
probabilities due to single galaxies, because galaxies dominate 
the total strong lensing probability as discussed above. 
The strong lensing probability $P_{\rm sl}(z_{\rm s})$ 
for a source at redshift $z_{\rm s}$ is computed as
\begin{equation}
P_{\rm sl}(z_{\rm s})=\int_0^{z_{\rm s}} dz_{\rm l} \frac{d^2V}{dz_{\rm
    l}d\Omega}\int_0^\infty d\sigma \frac{dn}{d\sigma} B\sigma_{\rm
  sl}(\sigma),
\label{eq:p_sl}
\end{equation}
where $d^2V/dz_{\rm l}d\Omega$ is the comoving volume element per
redshift and steradian, $dn/d\sigma$ is the velocity dispersion
function of galaxies, $\sigma_{\rm sl}(\sigma)$ is the strong lensing 
cross section in units of steradian for galaxies at redshift 
$z_{\rm l}$ with the velocity dispersion $\sigma$, and $B$
encapsulates various selection effects such as the magnification
bias. To compute the strong lensing cross section, we assume
that the mass distribution of lensing galaxies follow a Singular
Isothermal Ellipsoid, which is an extension of an SIS to include the
ellipticity in the projected mass distribution. We also add external
shear perturbation. The velocity dispersion function $dn/d\sigma$ is
taken from the one measured in the Sloan Digital Sky Survey
\cite{2010MNRAS.404.2087B} with the redshift evolution predicted by
the Illustris cosmological hydrodynamical simulation
\cite{2015MNRAS.454.2770T}. The strong lensing probability is derived
in the Monte-Carlo approach in which many lenses and sources are
randomly generated and the lens equation is solved numerically using
{\tt glafic} \cite{2010PASJ...62.1017O}. Interested readers are
referred to \cite{2018MNRAS.480.3842O} for more details.

\begin{figure}
\begin{center}
\includegraphics[width=8.0cm]{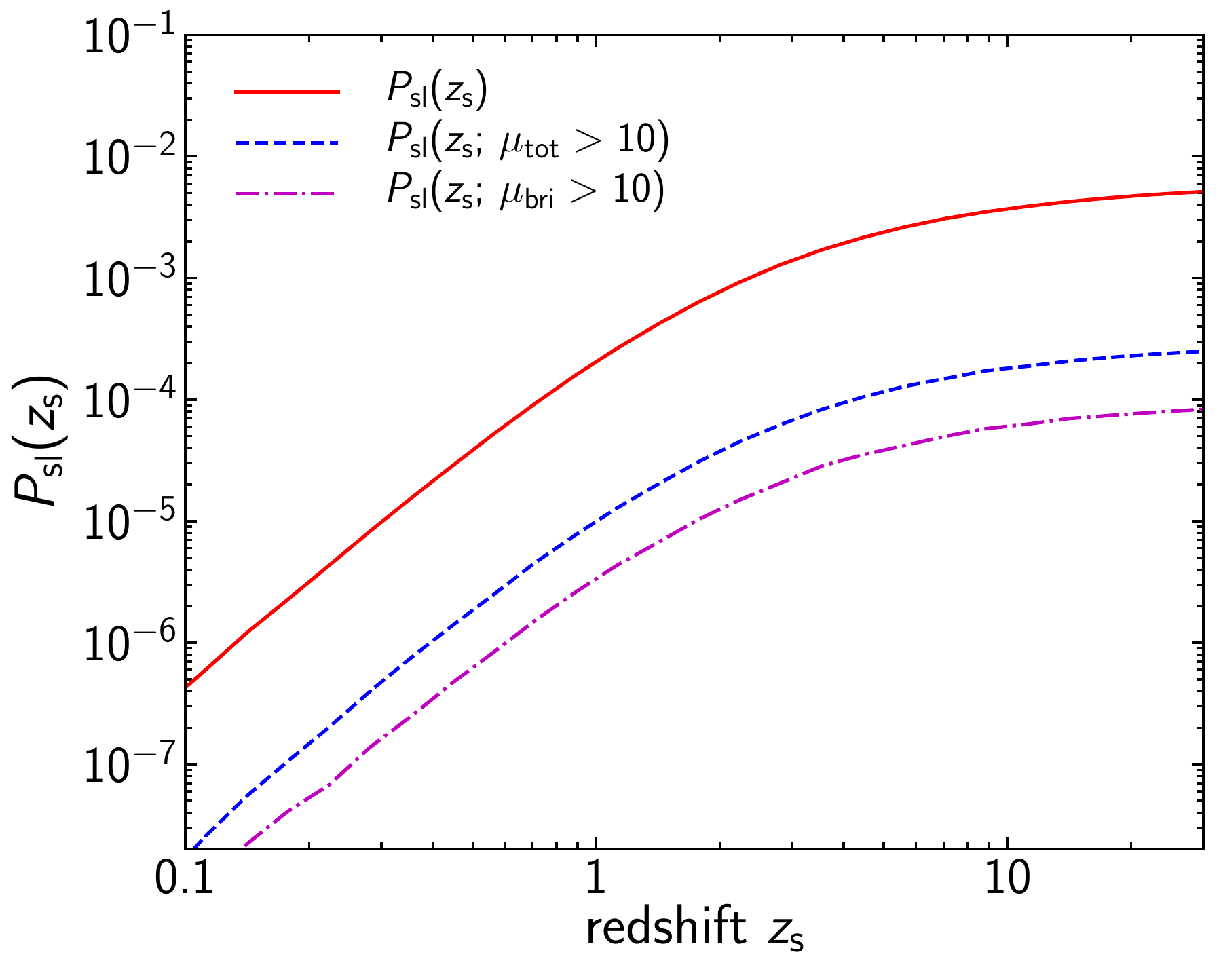}
\caption{Strong lensing probabilities $P_{\rm sl}(z_{\rm s})$ defined
  by equation~(\ref{eq:p_sl}) as a function of the source redshift
  $z_{\rm s}$. See the text for the setup of the calculations. The
  solid line shows strong lensing probabilities without any selection
  bias (i.e., $B=1$), whereas dashed and dash-dotted lines show
  probabilities of strong lensing with the total magnification
  $\mu_{\rm tot}>10$ and the magnification of the brightest image
  $\mu_{\rm bri}>10$, respectively.}
\label{fig:slprob}
\end{center}
\end{figure}

Figure~\ref{fig:slprob} shows the strong lensing probability
$P_{\rm sl}(z_{\rm s})$ computed with the setup described above,
without any selection bias i.e., $B=1$. As noted in e.g.,
\cite{1984ApJ...284....1T}, the strong lensing probability is a
steeply increasing function of the source redshift at low redshifts,
$z\lesssim 1$. The dependence on the redshift becomes somewhat weaker
at higher redshifts. The redshift dependence mainly comes from the
total volume, $\int_0^{z_{\rm s}} dz_{\rm l} d^2V/dz_{\rm l}d\Omega$ in
equation~(\ref{eq:p_sl}), which suggests that the redshift dependence
is $\propto z_{\rm s}^3$ at low redshifts. We find that the strong
lensing probability shown in Figure~\ref{fig:slprob} is 
crudely approximated by the following functional form
\begin{equation}
P_{\rm sl}(z_{\rm s}; B=1)\approx \frac{(5\times 10^{-4})z_{\rm
    s}^3}{(1+0.41z_{\rm s}^{1.1})^{2.7}},
\label{eq:p_sl_approx}
\end{equation}
which may be useful for quick estimates of the occurrence of strong
lensing events in various situations.

As discussed above, in most cases we have to take account of the
magnification bias, which can significantly enhance the strong lensing
probability. For instance, the radio source sample used in Cosmic Lens
All-Sky Survey \cite{2003MNRAS.341....1M,2003MNRAS.341...13B}
approximately has power low number counts with $\alpha=2.1$, which
leads to, from equation~(\ref{eq:mag_bias_pl}), the magnification bias
factor of $B\sim 4.8$ for an SIS lens.

The magnification bias factor is larger when number counts are
steeper. An extreme example is found in the bright ends of number
counts or luminosity functions. For many types of sources, there are
exponential cutoffs in their number counts or luminosity functions,
and beyond those exponential cutoffs the magnification bias is
infinitely large as without gravitational lensing magnifications we
would not expect any sources observed far beyond the cutoffs. 
Therefore at these luminosity or flux ranges, almost all the observed
sources are strong lensing events, suggesting that the
strong lens search among such brightest sources is highly efficient. A
good example of this is brightest galaxies in the submm wavelength,
which indeed have been found to be dominated by strong lensing (e.g.,
\cite{2010Sci...330..800N}). 

In Figure~\ref{fig:slprob}, we also show probabilities of strong
lensing events with magnifications $\mu>10$. Since magnification
probabilities are approximately $dP/d\mu\propto \mu^{-3}$ for most
situations, we naively expect $P_{\rm sl}(z_{\rm s})\propto \mu_{\rm
  min}^{-2}$. In our fiducial case without the selection effect we
have $\mu_{\rm min}\approx 2$, suggesting that the larger
magnification threshold of $\mu_{\rm min}=10$ leads to $\approx 1/25$
smaller strong lensing probabilities (i.e., $B\approx 1/25$), which
appears to hold approximately in Figure~\ref{fig:slprob}.

However, there are some subtleties in computing the magnification
bias. The magnification factor used in the calculation of the
magnification bias can be either the total magnification of all the
multiple images, $\mu_{\rm tot}$, or the magnification of one of the
multiple images. To illustrate this point, in Figure~\ref{fig:slprob}
we consider two cases, one is $\mu=\mu_{\rm tot}$ and the other is the
magnification of the brightest image, $\mu=\mu_{\rm bri}$, which
clearly make a quantitative difference. The choice should be made
depending on the strong lens search strategy in the observations. If
multiple images are unresolved when searching for strong lensing
the total magnification should be used, whereas multiple images are
well resolved, either the magnification of the brighter or the fainter
image should be used.

In the case of strong lensing of explosive transients,
there may be another selection bias associated with time delays. A
transient survey is conducted during some period, and we may miss some
of the multiple images that fall outside the survey period. This
effect is more significant for multiple images with longer time
delays. This time delay bias \cite{2003ApJ...583..584O} may also be
important in future statistical analysis of lensed explosive
transients. 

\subsection{Wave optics effects}
\label{sec:waveoptics}

In this review article, thus far we implicitly assumed geometric
optics in all 
the calculations of gravitational lensing. Indeed, geometric optics
serve as an excellent approximation in most astronomical situations of
interest. However, there are some exceptional cases where wave optics
effects play a crucial role (e.g., 
\cite{1992ApJ...386L...5G,1998PhRvL..80.1138N,2003ApJ...595.1039T,2013MNRAS.431.1264M,2019NatAs...3..524N,2019arXiv190506066S,2019arXiv190605950M}),
especially for strong lensing of explosive transients
that is the topic of this review article. Here we briefly review the
wave aspect of gravitational lensing theory. For more details, see e.g.,
\cite{1992grle.book.....S,1999PThPS.133..137N}.

We consider the propagation of monochromatic waves
$\psi(\boldsymbol{x},t)=\tilde{\psi}(\boldsymbol{x})e^{-2\pi if t}$
with frequency $f$. In the presence of a weak gravitational field
that is characterized by the gravitational potential
$U(\boldsymbol{x})$ ($|U|\ll 1$), the propagation equation is 
\begin{equation}
(\boldsymbol{\nabla}^2+\omega^2)\tilde{\psi}=4\omega^2U\tilde{\psi},
\end{equation}
with $\omega=2\pi f$. Defining the amplification factor as
\begin{equation}
F=\frac{\tilde{\psi}^{\rm L}}{\tilde{\psi}},
\end{equation}
where $\tilde{\psi}^{\rm L}$ and $\tilde{\psi}$ denote wave amplitudes
with and without gravitational lensing, respectively, we obtain the
diffraction integral formula for the amplification factor of
gravitationally lensed waves in the expanding Universe as a function
of frequency $f$ and the source position $\boldsymbol{\beta}$
\begin{equation}
  F(f, \boldsymbol{\beta})
  =\frac{1+z_{\rm l}}{c}\frac{D_{\rm ol}D_{\rm os}}{D_{\rm ls}}\frac{f}{i}
  \int d^2\boldsymbol{\theta}\exp\left[2\pi i f \Delta
    t(\boldsymbol{\theta}, \boldsymbol{\beta})\right],
  \label{eq:wave_f}
\end{equation}
where $\Delta t$ is the arrival time defined by equation~(\ref{eq:tdelay}).
This formula allows us to compute both amplitude and phase shifts of
the wave due to gravitational lensing. 
Note that the wave intensity is amplified by $|F|^2$. We can simplify
this equation by defining the dimensionless parameter $w$ using
$\Delta t_{\rm fid}$ defined in equation~(\ref{eq:deltat_fid})
\begin{equation}
  w=2\pi f \Delta t_{\rm fid}=2\pi f
  \frac{1+z_{\rm l}}{c}\frac{D_{\rm ol}D_{\rm os}}{D_{\rm
      ls}}\theta_{\rm Ein}^2.
\label{eq:wave_w}
\end{equation}
From equation~(\ref{eq:tfid_mein}), it is found that $w$ is also expressed as 
\begin{equation}
  w=2\pi f (1+z_{\rm l})\frac{4G M(<\theta_{\rm Ein})}{c^3}.
\label{eq:wave_w_mein}
\end{equation}
By defining $\hat{\boldsymbol{\theta}}=\boldsymbol{\theta}/\theta_{\rm
  Ein}$ and $\hat{\boldsymbol{\beta}}=\boldsymbol{\beta}/\theta_{\rm
  Ein}$, we can rewrite equation~(\ref{eq:wave_f}) as
\begin{equation}
  F(f, \boldsymbol{\beta})
  =\frac{w}{2\pi i}
  \int d^2\hat{\boldsymbol{\theta}}\exp\left[i w T(\hat{\boldsymbol{\theta}}, \hat{\boldsymbol{\beta}})\right],
  \label{eq:wave_f_dl}
\end{equation}
where $T$ is similar to $\Phi$ defined in
equation~(\ref{eq:deltat_phi}) and is described as
\begin{equation}
 T(\hat{\boldsymbol{\theta}}, \hat{\boldsymbol{\beta}})=\frac{(\hat{\boldsymbol{\theta}}-\hat{\boldsymbol{\beta}})^2}{2}-\frac{\phi(\hat{\boldsymbol{\theta}})}{\theta_{\rm
    Ein}^2},
\end{equation}
which is also dimensionless.

The geometric optics limit corresponds to $f\rightarrow\infty$. In
this limit, we can evaluate equation~(\ref{eq:wave_f_dl}) using the
stationary phase approximation, where only critical points satisfying
\begin{equation}
\boldsymbol{\nabla}_{\hat{\boldsymbol{\theta}}}T(\hat{\boldsymbol{\theta}}, \hat{\boldsymbol{\beta}})=0,
\end{equation}
contribute to the integral in equation~(\ref{eq:wave_f_dl}). This is same as
equation~(\ref{eq:fermat}) and hence the lens equation, indicating
that the contributions comes from only multiple image positions 
$\hat{\boldsymbol{\theta}}_j$. In this limit we can approximate
equation~(\ref{eq:wave_f_dl})  as
\begin{equation}
  F(f, \boldsymbol{\beta}) \approx \sum_j \frac{1}{i}\left|{\rm det}
  A(\boldsymbol{\theta}_j)\right|^{-1/2}\exp\left[i w
    T(\hat{\boldsymbol{\theta}}_j,
    \hat{\boldsymbol{\beta}})+\frac{i\pi\sigma_{\rm s}}{4}\right],
\end{equation}
where $j$ runs over multiple images, $A(\boldsymbol{\theta})$ is the
Jacobi matrix defined in equation~(\ref{eq:jacobi}), and $\sigma_{\rm s}$ is
the signature of $A$ i.e., the number of positive eigenvalues minus
the number of negative eigenvalues. Given the definition of the
magnification factor~(\ref{eq:mag_def}), we can simplify this further
as 
\begin{equation}
  F(f, \boldsymbol{\beta}) \approx \sum_j
  \left|\mu(\boldsymbol{\theta}_j)\right|^{1/2}\exp\left[i w
    T(\hat{\boldsymbol{\theta}}_j, \hat{\boldsymbol{\beta}})-i\pi n_j\right],
\end{equation}
where $\mu(\boldsymbol{\theta})$ is the (signed) magnification factor
and $n_j=0$, $1/2$, and $1$ correspond to the cases where
$\boldsymbol{\theta}_j$ is a minimum, saddle, and maximum point of
$T(\boldsymbol{\theta})$, respectively. From this expression, 
we can derive the amplification of the wave intensity as
\begin{eqnarray}
  \left|F(f, \boldsymbol{\beta})\right|^2 &\approx& \sum_j
  \left|\mu(\boldsymbol{\theta}_j)\right|\nonumber\\
&&
\hspace*{-20mm}
+2\sum_{j<k}\left|\mu(\boldsymbol{\theta}_j)\mu(\boldsymbol{\theta}_k)\right|^{1/2}\cos\left[w
    \Phi(\boldsymbol{\theta}_j, \boldsymbol{\theta}_k)-\pi\Delta n_{jk}\right],
\label{eq:amp_wave}
\end{eqnarray}
where $\Phi(\boldsymbol{\theta}_j, \boldsymbol{\theta}_k)$ 
is defined in equation~(\ref{eq:deltat_phi}) and $\Delta n_{jk}=n_j-n_k$.
The first term in the right hand side of equation~(\ref{eq:amp_wave})
agrees with magnifications in the geometric optics, whereas the second
term represents wave optics effects and is the interference between
multiple images. In the limit $f\rightarrow\infty$, however, this term
rapidly oscillates such that e.g., averaging over a small finite source
size easily eliminates this term.

We caution that equation~(\ref{eq:amp_wave}) is valid only
approximately, and in order to take full account of wave optics
effects we should evaluate equation~(\ref{eq:wave_f_dl}) directly. 
For instance, in the case of a point mass lens, the amplification 
can be computed analytically \cite{1986ApJ...307...30D,1986PhRvD..34.1708D}
\begin{equation}
  \left|F(f, \boldsymbol{\beta})\right|^2=\frac{\pi w}{1-e^{-\pi w}}
\left|{}_1F_1\left(\frac{i}{2}w, 1; \frac{i}{2}w\hat{\beta}^2\right)\right|^2,
\label{eq:f_amp_point}
\end{equation}
where ${}_1F_1$ is the confluent hypergeometric function. In this
case, the dimensionless parameter $w$ reduces to
\begin{eqnarray}
  w&=&2\pi f \frac{4GM(1+z_{\rm l})}{c^3}\nonumber\\
 &\approx&
1.24 \times 10^{-4}(1+z_{\rm l})\left(\frac{M}{M_\odot}\right)\left(\frac{f}{\rm Hz}\right).
\label{eq:wave_w_point}
\end{eqnarray}
Equation~(\ref{eq:f_amp_point}) indicates that the maximum
amplification at $\hat{\beta}=0$ is 
\begin{equation}
  \left|F(f, \beta=0)\right|^2=\frac{\pi w}{1-e^{-\pi w}},
\end{equation}
which becomes $\left|F(f, \beta=0)\right|^2\rightarrow 1$ for
$w\rightarrow 0$. This is essentially diffraction of waves i.e.,
any obstacle whose size is much smaller than the wavelength does not
affect the propagation of waves. From this expression it is found
that the gravitational lensing magnification becomes quite inefficient
for $w\lesssim 1$ due to wave optics effects. We note that the similar
analytic expression of the amplification factor for an SIS lens is
also available \cite{2006JCAP...01..023M}
\begin{equation}
   \left|F(f, \boldsymbol{\beta})\right|^2=
\left|\sum_{n=0}^\infty \frac{\Gamma(1+n/2)}{n!}g(w, \hat{\beta})\right|^2,
\end{equation}
\begin{equation}
  g(w, \hat{\beta}) = \left(2 w e^{(3\pi/2)i}\right)^{n/2}{}_1F_1\left(-\frac{n}{2}, 1; \frac{i}{2}w\hat{\beta}^2\right),
\end{equation}
where the dimensionless parameter $w$ for an SIS reduces to
\begin{equation}
w=2\pi f (1+z_{\rm
  l})\frac{1}{c}\left(\frac{4\pi\sigma^2}{c^2}\right)^2\frac{D_{\rm
    ol}D_{\rm ls}}{D_{\rm os}}.
\end{equation}

\begin{figure}
\begin{center}
\includegraphics[width=8.0cm]{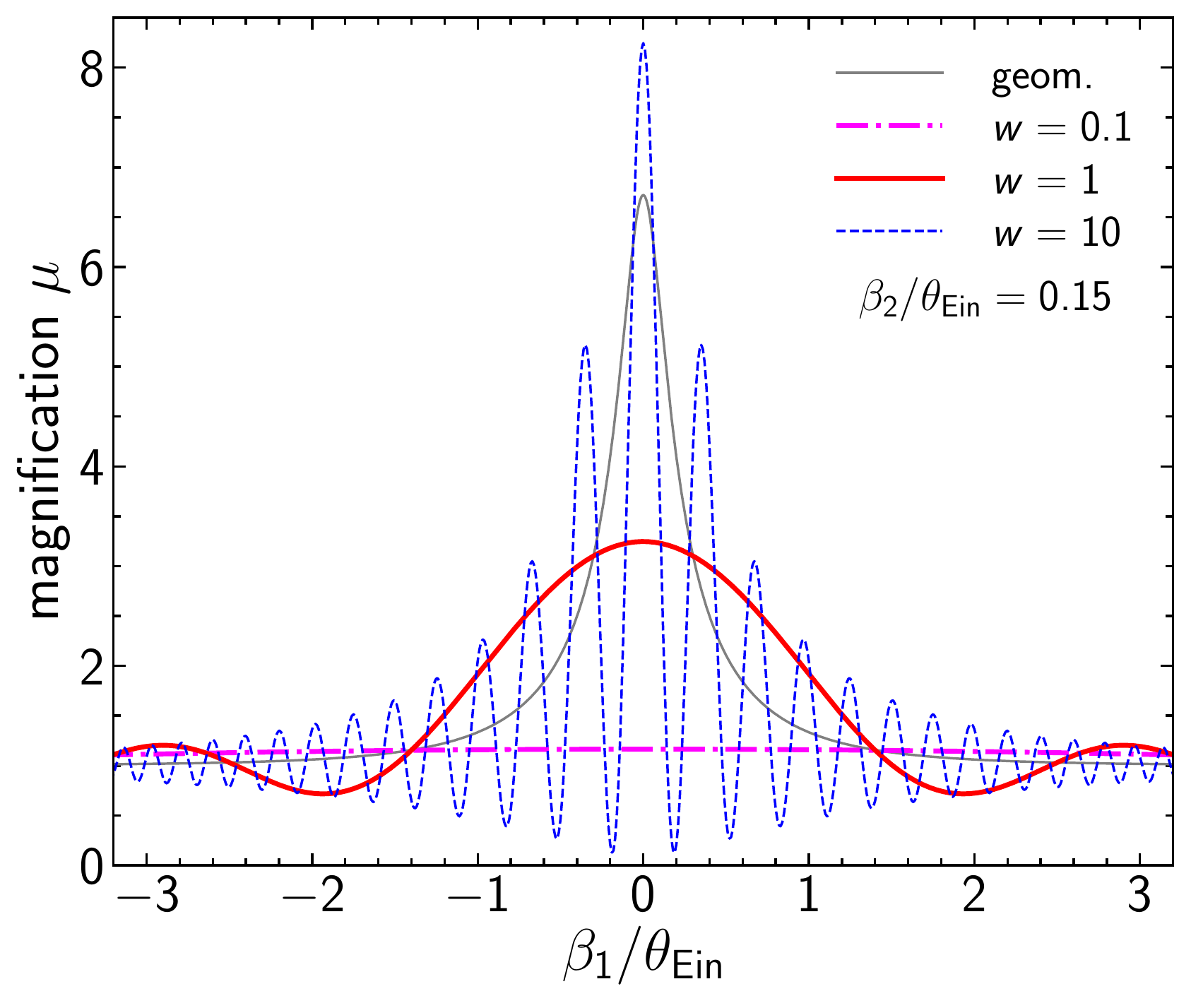}
\caption{The magnification factor for a point mass lens in wave
  optics, which is computed from equation~(\ref{eq:f_amp_point}), as a
  function of the source position. Here we fix $\beta_2/\theta_{\rm
    Ein}=0.15$ and change $\beta_1$ to see how the magnification 
  factor change as a a function of the source position
  $\boldsymbol{\beta}=(\beta_1,\,\beta_2)$, for $w=0.1$ 
  ({\it dash-dotted magenta}), $1$ ({\it thick solid red}), and $10$
  ({\it dashed blue}), where $w$ is the dimensionless parameter
  defined by equation~(\ref{eq:wave_w_point}). The magnification
  factor for the geometric optics case is shown by the thin solid gray line.}
\label{fig:wave_mag}
\end{center}
\end{figure}

Equations~(\ref{eq:amp_wave}) and (\ref{eq:f_amp_point}) suggest that 
the gravitational lensing amplification shows an oscillating behavior
as a function of the source position or the frequency of
waves. Figure~\ref{fig:wave_mag} shows some examples. If observed, it
provides a direct evidence of wave optics effects in action.  As mentioned
above, however, wave optics effects may be suppressed due to the finite
source size (e.g., \cite{2006JCAP...01..023M}). Here we discuss the
finite source size effect using equation~(\ref{eq:amp_wave}), from
which  it is found that the  oscillating behavior comes from
$w\Phi$. For reasonably small $\beta$, equations~(\ref{eq:phi_point})
and (\ref{eq:phi_sis}) imply that $\Phi\sim (\hat{\theta}_j-\hat{\theta}_i)
\sim \hat{\beta} =\beta/\theta_{\rm Ein}$. Therefore the width of
interference oscillations in the source plane is on the order of
$\theta_{\rm Ein}/w$. In order for the interference pattern to be 
observed, the source size in the angular unit,
$\beta_{\rm s}=R_{\rm s}/D_{\rm os}$, should satisfy
$\beta_{\rm s}\lesssim \theta_{\rm Ein}/w$.\footnote{Near
  the fold caustic, the time delay between merging image pairs scales
  as $\Delta t \propto \beta^{3/2}$, where $\beta$ here is the 
  distance from the caustic. Therefore, in this situation this
  condition should be modified as
  $(\beta_{\rm s}/\theta_{\rm Ein})^{3/2}\lesssim 1/w$.}
This condition yields 
\begin{equation}
 R_{\rm s}\lesssim \frac{D_{\rm os}\theta_{\rm Ein}}{w}.
\label{eq:wave_sizelimit}
\end{equation}
In the case of the point mass lens, this condition is expressed as  
\begin{eqnarray}
  R_{\rm s}&\lesssim& 2.24\times 10^{15}\,{\rm km}\left(\frac{1+z_{\rm l}}{1.5}\right)^{-1} \nonumber\\
&&\times  
  \left(\frac{M}{M_\odot}\right)^{-1/2}\left(\frac{f}{\rm Hz}\right)^{-1}
\left(\frac{D_{\rm os}D_{\rm ls}/D_{\rm ol}}{0.949\,{\rm Gpc}}\right)^{1/2},
\label{eq:wave_sizelimit2}
\end{eqnarray}
where distances are normalized to values at the lens redshift
$z_{\rm l}=0.5$ and the source redshift $z_{\rm s}=1.0$. 

To summarize, wave optics effects suppress the gravitational lensing
magnification when $w\lesssim 1$ due to diffraction, where $w$ is
defined in equation~(\ref{eq:wave_w}). On the other hand, when
$w\gtrsim 1$, the gravitational lensing magnification exhibits
oscillating behavior as a function of the source position or the wave
frequency, which can be observed only when the source size $R_{\rm s}$
satisfies the condition given by equation~(\ref{eq:wave_sizelimit}).
We will discuss specific examples in Section~\ref{sec:compact_dm}.

\section{Explosive transients}
\label{sec:transients}

\subsection{Supernovae}
\label{sec:intro_sn}

A supernova is an explosion associated with the death of a star.
Observations of supernovae have a long history, for example some
supernovae that took place in the Milky Way were observed even in the
naked eye and were recorded in the literature. Here we provide a brief 
overview of supernovae. Interested readers are referred to
reviews (e.g., \cite{1997ARA&A..35..309F}) and textbooks (e.g.,
\cite{2017suex.book.....B}) for more details.

Observationally there is a great deal of diversity in properties of
supernovae, including their light curves and spectral features. First,
supernovae are classified based on the presence or absence of hydrogen
lines. Supernovae without hydrogen lines are classified as Type I,
whereas those with hydrogen lines are classified as Type II. Type I
supernovae are further divided  into subclasses based on the presence
or absence of a singly ionized silicon line (SiII) such that those
with the strong silicon line are Type Ia and those with the weak or no
silicon line are Type Ib/c. Type II supernovae are also classified
into e.g., Type IIP, IIL, and IIn, depending on their shapes of the
light curves and/or the presence of absence of narrow line features in
their spectra.

We can classify supernovae on more physical basis, depending on their
explosion mechanisms. Type Ia supernovae are thought to be
thermonuclear explosions of white dwarfs near the Chandrasekhar mass,
$\approx 1.4 M_\odot$. The explosion of a white
dwarf is triggered by the matter accretion from a companion star.
There is a long controversy whether the companion star is a
non-degenerate star such as a red giant or a main sequence star
(single degenerate scenario) or the companion star is also a white
dwarf i.e., a Type Ia supernova is trigger by the merger of two white
dwarfs (double degenerate scenario). See e.g., a review by Maoz {\it
  et al.} \cite{2014ARA&A..52..107M} for more details on this topic. On the
other hand, both Type Ib/c and Type II supernovae are thought to be 
produced by the core collapse of massive stars. There are several
possible mechanisms to trigger the explosion, including the development
of an iron core that exceeds the Chandrasekhar and leads to the
collapse and bounce of the core. After the bounce the outgoing shock
is heated by neutrino emitted from the core, which is thought to be a
key ingredient for the successful explosion. Extensive numerical
simulations to understand the explosion mechanism of core-collapse
supernovae are ongoing (see e.g., \cite{2016PASA...33...48M}).

\begin{figure}
\begin{center}
\includegraphics[width=8.0cm]{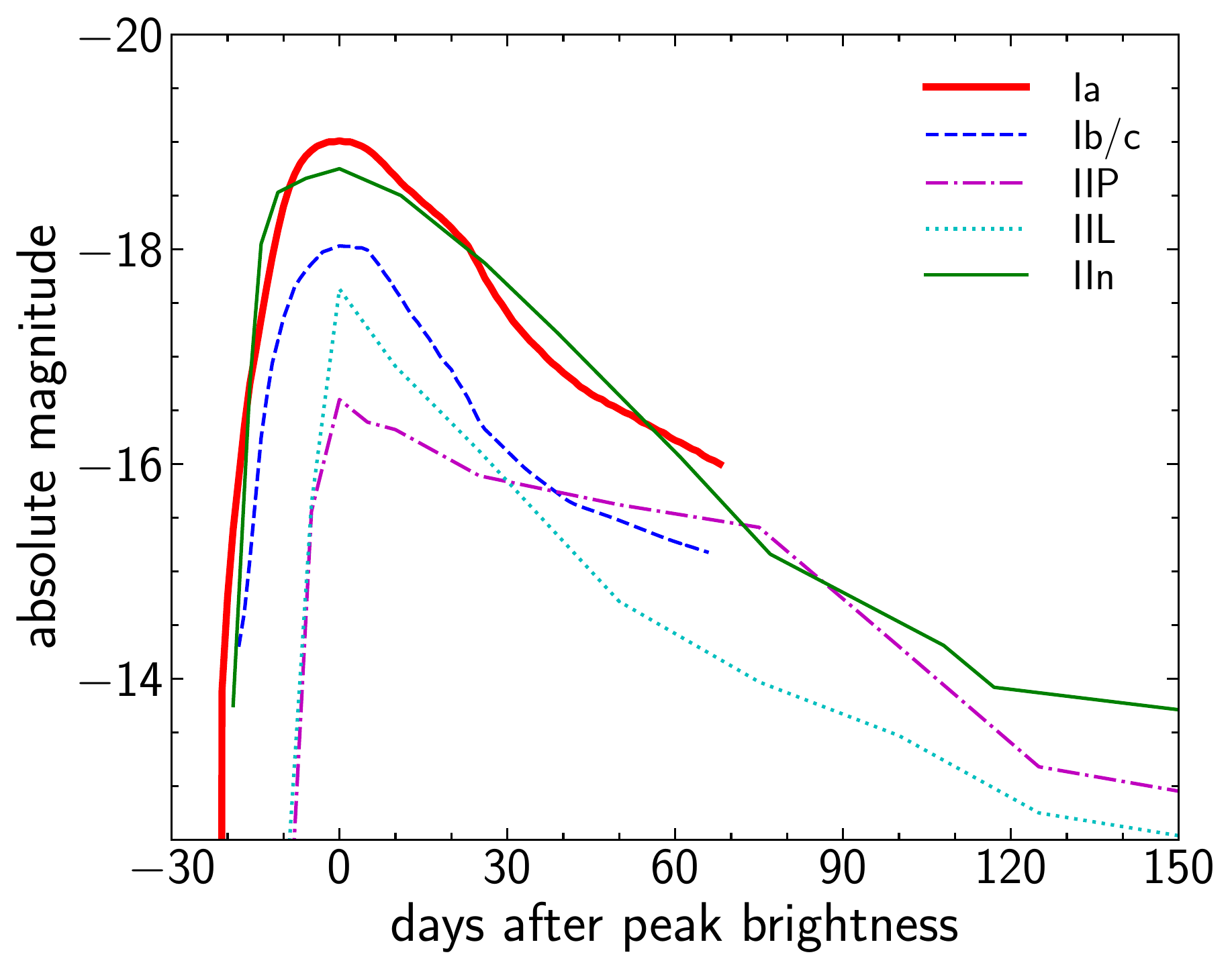}
\caption{Light curves of various types of supernovae. We show template
  light curves of Ia ({\it thick solid red}), Ib/c ({\it dashed
    blue}), IIP ({\it dot-dashed magenta}), IIL ({\it dotted cyan}),
  and IIn ({\it thin solid green}).
  The light curve templates in $V$-band are taken from the webpage
  {\tt https://c3.lbl.gov/nugent/nugent\_templates.html}. Absolute
  magnitudes at the peaks correspond to typical magnitudes for these
  supernova types \cite{2002AJ....123..745R}.}
\label{fig:snlc}
\end{center}
\end{figure}

Figure~\ref{fig:snlc} shows template light curves of various 
supernovae. It is found that supernovae are luminous. Peak
luminosities of luminous supernovae are comparable to galaxy
luminosities, which indicate that we can observe supernovae out to
high redshifts, $z\gtrsim 1$. The Figure also indicates that the
typical time scale of the light curves is a month, if we define the
time scale by the full-width-half-maximum (FWHM) of the light curve.
Their shapes are simple with a rise and a fall, although details are
different for different types of supernovae. We note that these are
template light curves in the supernova rest frame. Observed light
curves of supernovae at cosmological distances are stretched due to
cosmological dilation by a factor of $1+z$, which indicates that we
expect the time scale of a few months in the observer frame
for supernovae at $z\sim 1-2$.

Studies of supernovae are important in several ways. For instance,
supernovae are associated with the death of stars, and therefore their
rates as a function of galaxy type or redshift reflect the cosmic
history of star formation. Supernovae are produced only when masses
of progenitor stars fall in a particular range, from which we can
obtain information on the stellar initial mass function. One of the
most important applications of supernovae is the measurement of the
cosmic expansion. It is known that peak luminosities of Type Ia
supernovae are quite similar, which is particularly true if the
empirical relation between peak luminosities and widths of light curves
is taken into account \cite{1993ApJ...413L.105P}. This ``standardizable
candle'' nature of Type Ia supernovae allows us to measure luminosity
distances to supernovae. Combining the distance measurements with
redshift information, one can constrain the Hubble constant $H_0$ as
well as the cosmic expansion history out to sufficiently high
redshifts. For instance, luminosity distance measurements out to $z\sim
1$ with Type Ia supernovae led to the direct confirmation of the
accelerated expansion of the Universe and hence the significant amount
of dark energy in the Universe
\cite{1998AJ....116.1009R,1999ApJ...517..565P}. 
Type Ia supernovae also play a crucial role in the measurement of
$H_0$ with the so-called distance ladder method (e.g.,
\cite{2001ApJ...553...47F}). 

Because of their importance, a number of supernova surveys have been
conduced, including Supernova Legacy Survey \cite{2011ApJ...737..102S},
Sloan Digital Sky Survey II Supernova Survey
\cite{2018PASP..130f4002S}, Palomar Transient Factory
\cite{2009PASP..121.1395L},
Hubble Space Telescope Cluster Supernova
Survey \cite{2012ApJ...746...85S}, Pan-STARRS Medium Deep Survey
\cite{2018ApJ...859..101S}, All-Sky Automated Survey for Supernovae
\cite{2017MNRAS.471.4966H},
Dark Energy Survey Supernova Program
\cite{2019ApJ...874..150B}, and Hyper Suprime-Cam Transient Survey
\cite{2019arXiv190409697Y}. The total number of supernovae discovered
by now amounts to $O(10^4)$ both for Type Ia supernovae and for
core-collapse supernovae (e.g., \cite{2017ApJ...835...64G}).

These supernova surveys also revealed new classes of supernovae. Among
others, an interesting class of supernovae relevant for
this review article is a superluminous supernova (e.g.,
\cite{2012Sci...337..927G,2018SSRv..214...59M}).
One of the first examples of
this class, SN 2005ap, was discovered by the Texas Supernova Survey in
2005 \cite{2007ApJ...668L..99Q}. Superluminous supernovae, whose
origin is yet to be known but is believed to be associated with
the deaths of very massive stars, have peak
absolute magnitudes less than $-21$, and hence are much more luminous
than normal Type Ia and core-collapse supernovae (see
Figure~\ref{fig:snlc}). Their light curves are also wider, with the
typical time scale of up to $\sim 100$ days in the rest frame rather
than a month. Thanks to their bright luminosities, they can be
observed out to very high redshifts of $z\gtrsim 2$ (e.g.,
\cite{2012Natur.491..228C,2019ApJS..241...16M}).  

Finally, we summarize event rates and sizes of supernovae, which are
important for strong lensing studies. Li {\it et al.}
\cite{2011MNRAS.412.1473L} derived supernova 
rates in the local Universe as 
$R_{\rm SNIa}=(3.0\pm 0.6)\times 10^{4}$~Gpc$^{-3}$yr$^{-1}$
for Type Ia and $R_{\rm
  SNcc}=(7.1\pm1.6)\times10^{4}$~Gpc$^{-3}$yr$^{-1}$ 
for core-collapse (i.e., Type Ib/c and Type II). We note that these
comoving rates
increase toward higher redshifts, mainly due to the increase of the
cosmic star formation rate density from $z\sim 0$ to $\sim 2$.
Quimby {\it et al.} \cite{2013MNRAS.431..912Q} estimated the rate of
superluminous supernovae at $z\sim 0.2$ to
$R_{\rm SLSN}=(1.99^{+1.37}_{-0.86})\times 10^{2}$~Gpc$^{-3}$yr$^{-1}$. The size of
a supernova changes with time because of the dynamical evolution of
the photosphere. In the case of supernovae, the ejecta is expected to
enter the homologous expansion phase, in which the radial expansion
velocity is proportional to the radius, after a few times of the
expansion time scale Initially the photospheric radius increases as
the ejecta expands, and then it decreases as the density of the ejecta
decreases. As a result, the photospheric radius is $\sim 10^{10}$~km
at around the peak of the light curve (e.g., \cite{2018ApJ...868L..24L}). 

\subsection{Gamma-ray bursts}
\label{sec:intro_grb}

Gamma-ray bursts are very energetic explosions that are observed in
the gamma-ray band. Their durations are quite short, $<100$~sec, and
such high energy prompt emission is followed by afterglow emissions
observed from the X-ray to radio wavelength range. Gamma-ray bursts 
were discovered for the first time in 1960s by the Vela satellites
\cite{1973ApJ...182L..85K}, but their origin was totally unknown at
that time. Later the Burst and Transient Source Experiment (BATSE) on
the Compton Gamma Ray Observatory observed many gamma-ray bursts to
show that their distribution on the sky is isotropic, which supports
the extragalactic origin of gamma-ray bursts
\cite{1992Natur.355..143M}.  Detections of the afterglow emissions 
\cite{1997Natur.387..783C} led to identifications of their host
galaxies, which confirm that gamma-ray bursts indeed lie at
cosmological distances \cite{1998Natur.393...35K}. Now gamma-ray
bursts are detected regularly by the 
{\it Swift} satellite \cite{2004ApJ...611.1005G} as well as the Fermi
Gamma-ray Space Telescope \cite{2009ApJ...697.1071A,2009ApJ...702..791M}.
Here we summarize basic properties of gamma-ray bursts, see reviews
\cite{2004RvMP...76.1143P,1999PhR...314..575P,2014ARA&A..52...43B} for
more details.

\begin{figure}
\begin{center}
\includegraphics[width=8.0cm]{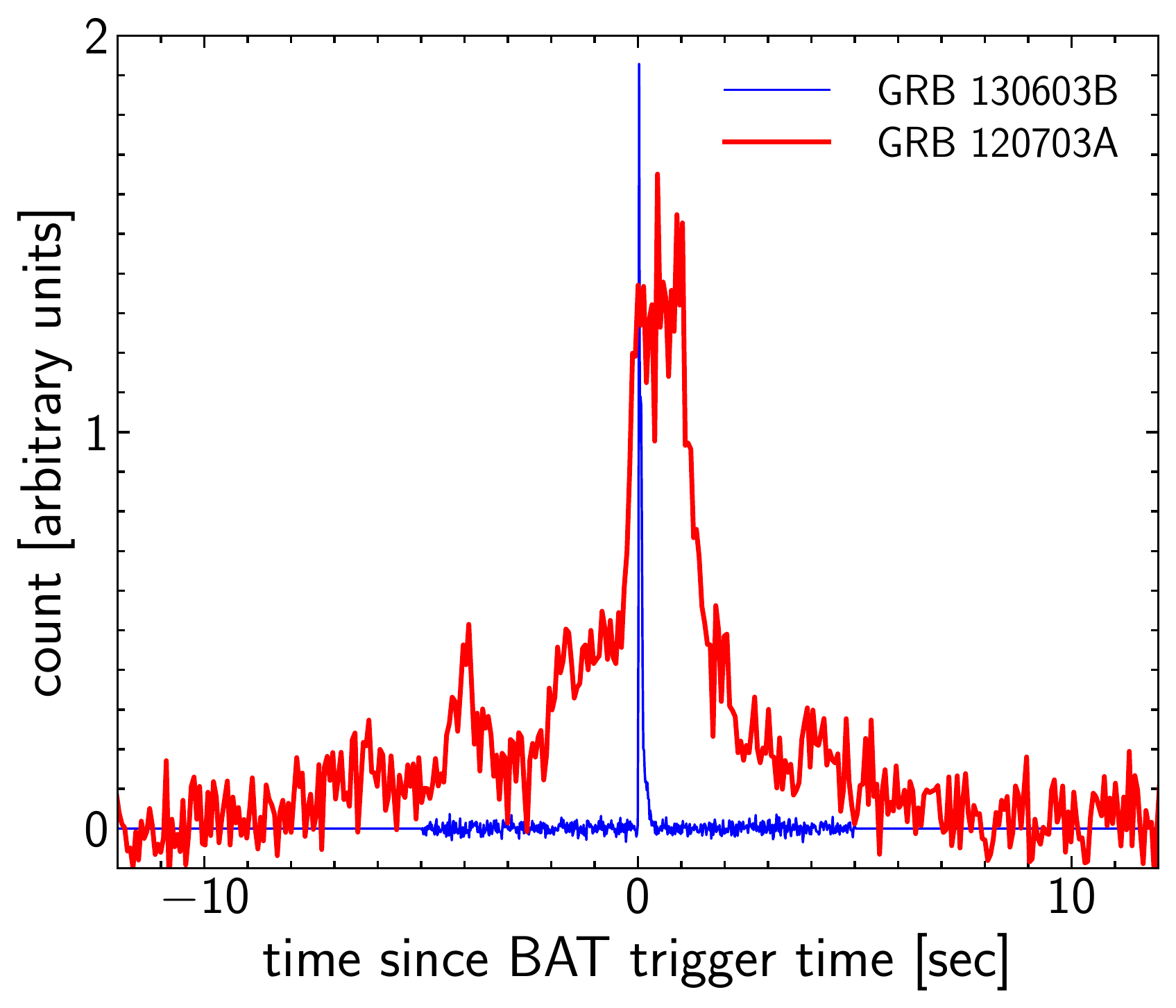}
\caption{Examples of light curves of short (GRB 130603B, {\it
    thin blue}) and long (GRB 120703A, {\it thick red}) gamma-ray bursts
  observed by the Burst Alert Telescope (BAT) on the {\it Swift}
  satellite \cite{2004ApJ...611.1005G}. The light curves are rescaled
  so that their peak counts roughly match.} 
\label{fig:grblc}
\end{center}
\end{figure}

From the analysis of gamma-ray bursts detected by the BATSE, it is
found that gamma-ray bursts are classified into two classes,
long (or long-soft) and short (or short-hard) gamma-ray bursts
\cite{1993ApJ...413L.101K}. They are divided based on the duration of
the emission, such that gamma-ray bursts with their durations longer
and shorter than 2~sec are classified into long and short gamma-ray
bursts, respectively. Figure~\ref{fig:grblc} shows examples of light
curves of both long and short gamma-ray bursts. The clear difference
of the durations between short and long gamma-ray bursts is seen. 
It is also seen that the light curve of the long gamma-ray burst is
complicated with several subpeaks. Indeed shapes of light curves of
different long gamma-ray bursts are quite different with each other.
So far $>5000$ long gamma-ray bursts and $>1000$ short gamma-ray
bursts have been discovered mainly by Compton Gamma Ray Observatory, the
{\it Swift} satellite, and  Fermi Gamma-ray Space Telescope.

Long gamma-ray bursts are thought to be caused by the death of massive
stars, because of the following reasons. First, in most cases host
galaxies of long gamma-ray bursts are young star-forming galaxies
in which many massive stars are recently formed (e.g.,
\cite{2009ApJ...691..182S}). Second, it was found
that some gamma-ray bursts are accompanied by core-collapse supernovae
that are also thought to be caused by the death of massive stars (e.g., 
\cite{1998Natur.395..670G,1999Natur.401..453B}). In this scenario,
gamma-ray emissions can be explained by a relativistic ejecta due to a
strong relativistic jet that is launched after the
core-collapse. However, for this scenario to work, the central engine
that drives the relativistic jet is needed, although the true nature
of the central engine is still yet to be understood (e.g.,
\cite{2018RPPh...81b6901N} for a review).    

The connection of long gamma-ray bursts to star formation 
suggests that observations of long gamma-ray bursts may help
understand the star formation history in the Universe. An advantage of
long gamma-ray bursts is their very high luminosities that allow us to
observe them out to very high redshifts. Indeed, the redshift
distribution of long gamma-ray bursts detected in {\it Swift} has the
median of $z\approx 2$ \cite{2006A&A...447..897J,2014ARA&A..52...43B},
and extends out to $z\sim 9.4$ \cite{2011ApJ...736....7C}. 
Regarding the rate of long gamma-ray bursts, from the {\it Swift}
gamma-ray burst sample Wanderman and Piran \cite{2010MNRAS.406.1944W}
derived the local rate of
$R_{\rm LGRB}=1.3^{+0.6}_{-0.7}$~Gpc$^{-3}$yr$^{-1}$ for
$L>10^{50}$~erg\,s$^{-1}$. The comoving rate increases with redshift as
$(1+z)^{2.1^{+0.5}_{-0.6}}$ at $z<3$, and  
decreases as $(1+z)^{-1.4^{+2.4}_{-1.0}}$ at $z>3$. We note that this
is the rate of events that we can observe i.e., gamma-ray bursts with
the jet orientations aligned with the line-of-sight directions. In
order to derive the true event rate in the Universe we have to apply
for the beaming factor correction, which would make the long gamma-ray
burst rate about two orders of magnitude higher. 

On the other hand, short gamma-ray bursts are thought to have a
different origin, because of their markedly different properties from
those of long gamma-ray-bursts. For example, in many cases host
galaxies of short gamma-ray bursts are elliptical galaxies with little
star formation (e.g., \cite{2005Natur.438..988B}) in contrast to
star-forming host galaxies of long gamma-ray bursts. In addition,
the association of short gamma-ray bursts with supernovae is lacking 
(e.g., \cite{2005Natur.437..845F}). A promising scenario that
explains these properties is that short gamma-ray bursts are caused by
binary mergers of compact objects such as neutron stars and black holes.
This scenario is confirmed by the discovery of gravitational waves
from a binary neutron star merger, GW170817, for which the associated
short gamma-ray burst GRB 170817A was detected (see also
Section~\ref{sec:intro_gw}).  From BATSE, {\it Swift}, and Fermi short
gamma-ray samples, Wanderman and Piran \cite{2015MNRAS.448.3026W} 
derived the local rate of 
$R_{\rm SGRB}=4.1^{+2.3}_{-1.9}$~Gpc$^{-3}$yr$^{-1}$ for $L>5\times
10^{49}$~erg\,s$^{-1}$. The comoving rate rapidly increases with
increasing redshift, at least out to $z\sim 1$.

There have been many proposals to use gamma-ray bursts as
standardizable candles to probe the cosmic expansion history (see e.g.,
\cite{2015NewAR..67....1W} for a review), just like
Type Ia supernovae. Many luminosity correlations that can be used to
standardize gamma-ray bursts are proposed, including the correlation
between the time variability and the luminosity
\cite{2000astro.ph..4176F}, the isotropic energy and the rest-frame
peak energy \cite{2002A&A...390...81A}, the luminosity and the
rest-frame peak energy \cite{2004ApJ...609..935Y}, and the peak energy
and the collimated energy \cite{2004ApJ...616..331G}. Cosmology with
gamma-ray bursts is potentially very powerful as the Hubble diagram
can be extended to very high redshifts out to $z> 8$.

The size of the emission region has also been studied in the
literature (e.g.,
\cite{2014ApJ...794L...8B,2015ApJ...811...93G,2018JCAP...12..005K}).
The size is estimated as 
\begin{equation}
R_{\rm em}\sim \Gamma^2 c\Delta t_{\rm var},
\end{equation}
where $\Gamma=\left\{1-(v/c)^2\right\}^{-1/2}$ is the Lorentz factor
of the ejecta with velocity $v$ and $\Delta t_{\rm var}$ is
the variability time scale. The Lorentz factor is thought to be
typically $\mathcal{O}(100)$. From observed variabilities of light
curves, we have $R_{\rm em}\sim 10^{13}$~cm for short gamma-ray bursts
and $R_{\rm em}\sim 10^{14}$~cm for long gamma-ray bursts, albeit
with large uncertainties. Due to the relativistic effect, the
transverse extent of the emission region $R_{\rm tv}$ differs from
$R_{\rm em}$ by a factor of $\Gamma$ i.e.,  
\begin{equation}
R_{\rm tv}\sim \frac{R_{\rm em}}{\Gamma},
\end{equation}
which suggests that, assuming $\Gamma\sim 300$, $R_{\rm tv}\sim
3\times 10^{10}\,{\rm cm}=3\times 10^5\,{\rm km}$
for short gamma-ray bursts and $R_{\rm tv}\sim
3\times 10^{11}\,{\rm cm}=3\times 10^6\,{\rm km}$ for
long gamma-ray bursts, again with large
uncertainties. Since the size that is relevant for strong lensing is
$R_{\rm tv}$, in what follows we refer to 
$R_{\rm tv}$ as the size of gamma-ray bursts.  We note that these
sizes are for gamma-ray prompt emissions, and sizes of X-ray and
optical afterglows should be three or more orders of magnitude larger 
than the values mentioned above. 

\subsection{Fast radio bursts}
\label{sec:intro_frb}

\begin{figure}
\begin{center}
\includegraphics[width=7.0cm]{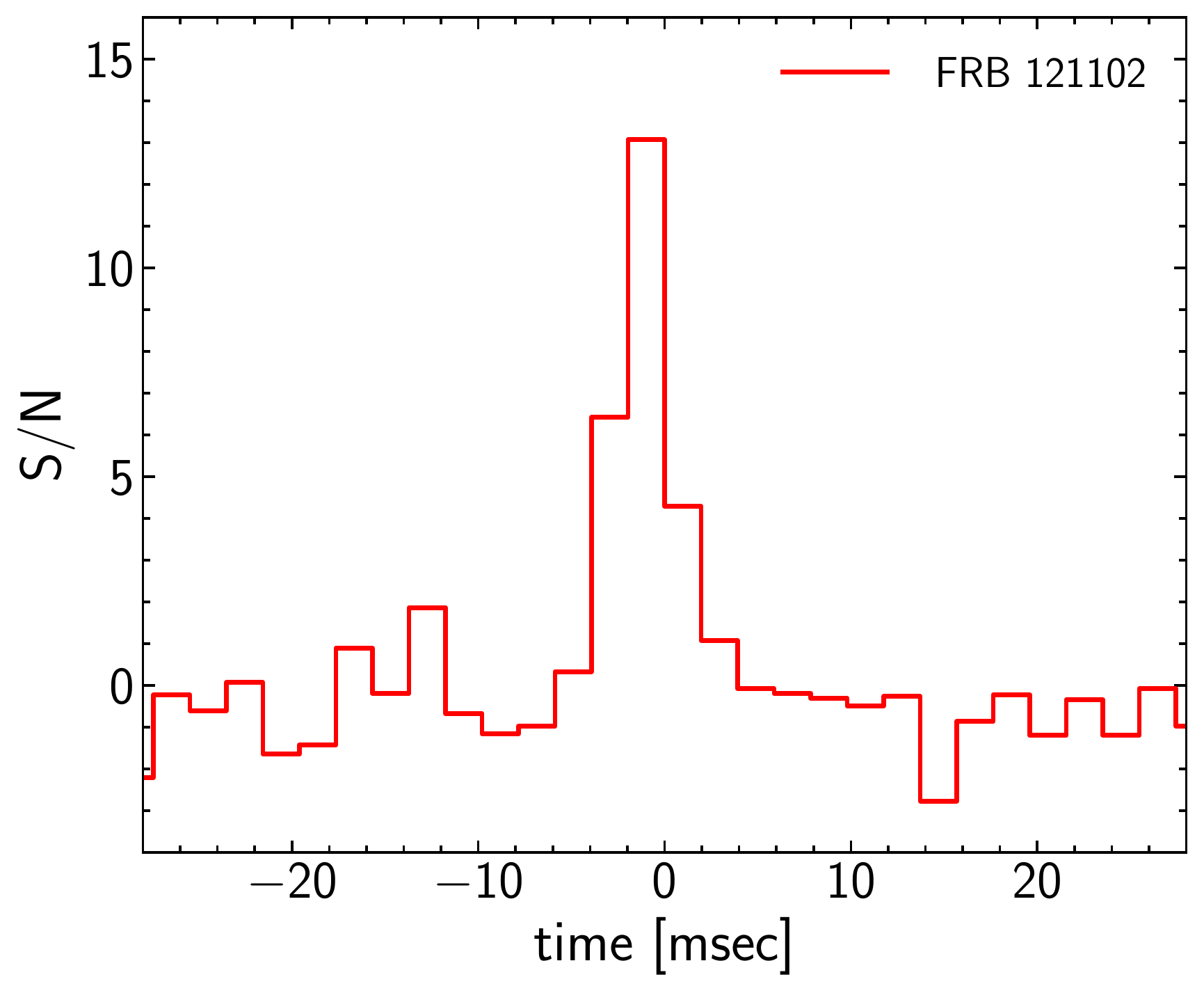}
\caption{An example of light curves of fast radio bursts. Here we show
a dedispersed, averaged pulse profile of FRB 121102 detected with the
Arecibo Observatory \cite{2014ApJ...790..101S}.} 
\label{fig:frblc}
\end{center}
\end{figure}

Fast radio bursts, which are transient radio pulses with the time
scale of a millisecond, are a new class of transients that was
identified relatively recently (see
\cite{2019A&ARv..27....4P,2019arXiv190605878C} for recent reviews).
The first example of fast radio bursts was discovered 
with the Parkes Observatory in 2007 by Lorimer {\it et al.}
\cite{2007Sci...318..777L}. Discoveries of additional four events by
Thornton {\it et al.} \cite{2013Sci...341...53T} support the
astrophysical origin of fast radio bursts. Thanks to Canadian Hydrogen
Intensity Mapping Experiment (CHIME) \cite{2018ApJ...863...48C} and
Australian Square Kilometre  Array Pathfinder (ASKAP)
\cite{2014PASA...31...41H}, the number of known fast radio bursts is
now rapidly increasing, and is reaching $O(100)$
\cite{2018Natur.562..386S,2019Natur.566..230C}.  An example of the
light curve is shown in Figure~\ref{fig:frblc}.

A key quantity that characterizes each fast radio burst is the
dispersion measure. Because of dispersive effects, electromagnetic
waves propagate through a plasma with different speeds at different
frequencies. More specifically, electromagnetic waves with frequency
$\nu$ have the following delay of the arrival time
\begin{equation}
\Delta t =\frac{e^2}{2\pi m_e c}\frac{\rm DM}{\nu^2}\approx
4150~{\rm sec}
\left(\frac{\nu}{\rm MHz}\right)^{-2}\left(\frac{\rm DM}{\rm cm^{-3}\,pc}\right),
\end{equation}
where DM is the dispersion measure, which is essentially 
the column density of free election along the line-of-sight
\begin{equation}
{\rm DM}=\int_0^d n_e dl,
\end{equation}
where $d$ is the distance to the fast radio burst. Since the Universe
is ionized at $z\lesssim 6$, a large contribution from the
intergalactic medium (IGM) to the observed DM after subtracting the
Galactic contribution is expected. A useful approximation that relates
the DM from the IGM and redshift $z$ is \cite{2019A&ARv..27....4P}
\begin{equation}
{\rm DM}_{\rm IGM}\approx 1000\times z~{\rm cm^{-3}\,pc},
\label{eq:dm_approx}
\end{equation}
which is reasonably accurate at least out to $z\sim 2$.

The dispersion measures of fast radio bursts discovered so far are
typically $100-1000$~${\rm cm^{-3}\,pc}$ after subtracting the
Galactic contribution, which suggest their redshifts of $\approx 0.1-1$
according to equation~(\ref{eq:dm_approx}). A complication is that
there may also be contributions from host galaxies and local
environments. For example, if the source is surrounded by a dense
plasma, the contribution of the local environment to DM can be as large
as $\sim 1000~{\rm cm^{-3}\,pc}$ and hence can be comparable or larger
than DM from the IGM. Therefore the redshift estimated by
equation~(\ref{eq:dm_approx}) should be taken as the upper limit of
the source redshift. 

Accurate distances to fast radio bursts are obtained if their host
galaxies are successfully identified. However, identifications of host
galaxies have been challenging due to limited localization
capabilities. The host galaxy was recently identified for the first
time for a 
repeating fast radio burst, which represents a rare class of fast
radio bursts with repeating pulses. So far only ten repeating fast
radio bursts, FRB 121102 \cite{2016Natur.531..202S}, FRB
180814.J0422+73 \cite{2019Natur.566..235C}, and some more new
repeating FRBs \cite{2019arXiv190803507T} have been identified. In the
case of FRB 121102, its host galaxy is identified to be a
low-metallicity, low-mass dwarf galaxy at $z=0.193$
\cite{2017ApJ...834L...7T}, which confirms the extragalactic origin of
fast radio bursts.  

Very recently, host galaxies have been identified for non-repeating
fast radio bursts as well. A luminous galaxy at $z=0.3214$ has been
identified as a host galaxy of the non-repeating fast
radio burst FRB 180924 detected by ASKAP \cite{2019arXiv190611476B}.
A massive galaxy with a relatively low specific star-formation rate at
$z=0.66$ has been identified as a host galaxy of FRB 190523 detected
by the Deep Synoptic Array ten-antenna prototype
\cite{2019arXiv190701542R}. These results highlight the possibility of
the association of fast radio bursts with relatively old stellar
populations. 

The mechanism to produce these fast radio bursts is still unknown. 
Many progenitor models that explain fast radio bursts have
been proposed (see \cite{2019A&ARv..27....4P} for a summary).
The statistical analysis of a large sample of fast radio bursts is a
key for discriminating these different scenarios. Another important
clue will be obtained by identifying many host galaxies. For example
the host galaxy of FRB 121102 implies the connection between fast
radio bursts and massive star formation, although it is also unknown
whether repeating and non-repeating fast radio bursts have the same
origin. 

The volumetric rate of fast radio bursts is also still very uncertain,
although it gives another important clue to the origin. Assuming that
observed fast radio bursts are distributed out to $z\sim 1$, we
crudely obtain $R_{\rm FRB}\sim 2\times 10^3$~Gpc$^{-3}$yr$^{-1}$
\cite{2019A&ARv..27....4P}. The recent study by Lu and Piro
\cite{2019arXiv190300014L} suggests an order of magnitude higher 
rate, $\sim 3\times 10^4$~Gpc$^{-3}$yr$^{-1}$ (see also
\cite{2019arXiv190706619R}). In either case, the high event rate of
fast radio bursts is a great promise for the future.  

The size of the emission region of fast radio bursts is poorly
constrained. The direct upper limit of
$R_{\rm tv}<0.7\,{\rm pc}\sim 2\times 10^{13}\,{\rm km}$ is obtained
from observations of the repeating fast radio burst FRB 121102 with
European VLBI Network \cite{2017ApJ...834L...8M}, although this
constraint is not quite tight. Tighter constraints of the size will
greatly help discriminate different progenitor models.

\subsection{Gravitational waves}
\label{sec:intro_gw}

\begin{figure}
\begin{center}
\includegraphics[width=8.0cm]{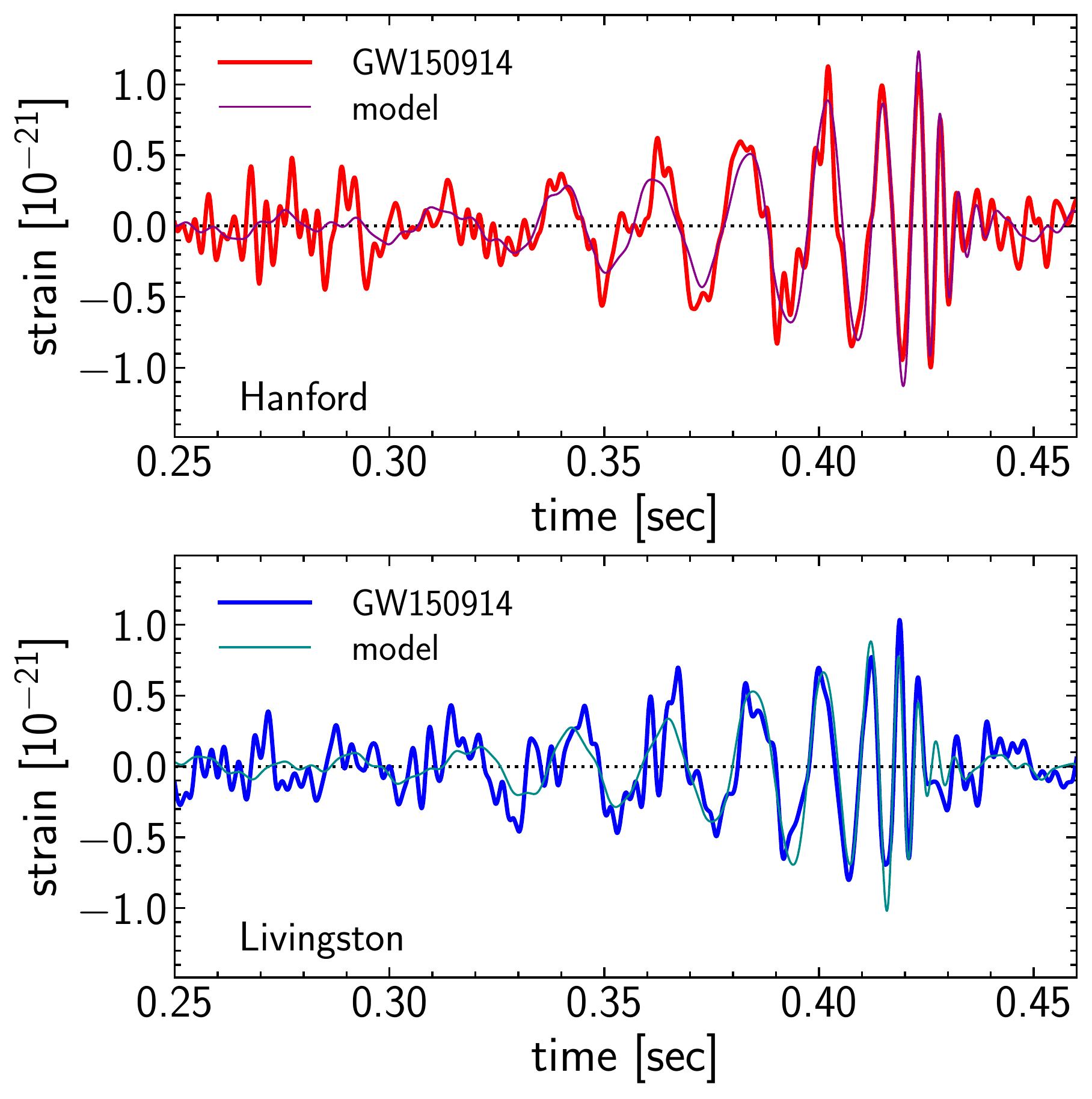}
\caption{The first gravitational wave event GW150914 
  \cite{2016PhRvL.116f1102A} observed by
  Advanced LIGO. This event was detected by both Hanford ({\it upper})
  and Livingston ({\it lower}) detectors. The observed waveform ({\it
    thick}) is plotted together with the best-fitting model ({\it thin}).} 
\label{fig:gw150914}
\end{center}
\end{figure}

The existence of gravitational waves was predicted by Albert Einstein
in 1916 on the basis of General Relativity. Gravitational waves are
essentially the propagation of fluctuations of curvature in spacetime, 
but the strain amplitude is so small that its detection has been quite
challenging. The first direct detection \cite{2016PhRvL.116f1102A} was
made in 2015 by the Advanced Laser Interferometer Gravitational-Wave
Observatory (Advanced LIGO; \cite{2015CQGra..32g4001L}). The event
named GW150914 was produced by a merger of a binary black hole with
masses $\sim 36$~$M_\odot$ and $\sim 
29$~$M_\odot$ located at redshift $z\sim 0.09$. 
Figure~\ref{fig:gw150914} shows the waveform of GW150914, which was
detected both in the Hanford and Livingston detectors. The slight
offset of the arrival times and the relative amplitudes between the
two detectors contain information on the position of the
gravitational wave source on the sky. Since then, the study of
gravitational waves is progressing rapidly. See e.g.,
\cite{2008grwa.book.....M,2018grwa.book.....M} for details of theory
and experiments of gravitational waves.  

Gravitational waves that are detectable with Advanced LIGO are
thought to be produced from mergers of binary black holes, binary
neutron stars, and black hole-neutron star binaries. Since Advanced
LIGO can detect 
gravitational waves in the frequency range $f\sim 10^{1-4}$~Hz, only
gravitational waves at the final inspiraling and merging stages are
observed. After the discovery of GW150914, there are more than 10
gravitational wave observations from binary black hole mergers out to
$z\sim 0.5$ with the total mass ranging from $\sim 20$~$M_\odot$ to 
$\sim 80$~$M_\odot$ 
(e.g., \cite{2018arXiv181112907T} for the summary of the second
observing run, and there are more observations from the third
observing run). The first observation of gravitational waves from a
binary neutron star merger was reported in 2017
\cite{2017PhRvL.119p1101A}.  Currently there is no confirmed
observation of a black hole-neutron star binary merger. 

We can measure various properties of merging binaries from
observations of gravitational waves. One of the most important
quantities that can be accurately constrained from observations of
gravitational waves is the (redshifted) chirp mass
\begin{equation}
\mathcal{M}_z=(1+z)\mathcal{M}=(1+z)\frac{(m_1m_2)^{3/5}}{(m_1+m_2)^{1/5}},
\label{eq:chirpmass}
\end{equation}
where $m_1$ and $m_2$ are masses of two compact objects that
constitute the binary. The chirp mass is constrained very well because
the orbital evolution during inspiraling at a given frequency depends
only on the chirp mass at the leading-order. The degeneracy between
$m_1$ and $m_2$ is broken by the analysis of the waveform around the
merger and ringdown phase. In addition, we can obtain information on
the spin from the analysis of the waveform.

Importantly, we can also measure the luminosity distance $D_{\rm L}$
to the binary from observations of gravitational waves. This is
because the frequency and its time evolution of a merging binary
constrain the chirp mass accurately, which in turn predicts the
amplitude of gravitational waves emitted from the binary. Since the
propagation of gravitational waves decreases the amplitude as 
$D_{\rm L}^{-1}$, the observation of the strain amplitude of
gravitational waves directly constrains $D_{\rm L}$. On the other
hand, the redshift is not directly measured by gravitational wave
observations. Therefore, the redshift of $z\sim 0.09$ for GW150914 was
in fact the value {\it inferred} from the luminosity distance
measurement.  

Redshifts of gravitational wave sources are obtained directly if
we successfully identify their host galaxies. However the
identification of a host galaxy is challenging, mainly because of the
poor angular resolution of gravitational wave observations. In the
case of GW150914, the error circle of the arrival direction has an
area of 600~deg$^2$, which is too wide to pinpoint its host
galaxy. One way to improve the localization accuracy is to detect
gravitational waves with more detectors, as demonstrated by
observations of GW170814 \cite{2017PhRvL.119n1101A} whose arrival
direction was constrained to an area of 60~deg$^2$ thanks to the
detection by Advanced Virgo \cite{2015CQGra..32b4001A} in addition to
two detectors of Advanced LIGO. In the near future KAGRA
\cite{2013PhRvD..88d3007A} and LIGO India will join the observing run,
which allows us to localize gravitational wave sources to a few square
degrees, although this is still insufficient for determining host
galaxies from gravitational wave observations alone in most cases.

Therefore, we usually rely on observations of electromagnetic
counterparts for secure identifications of host galaxies. The search
for electromagnetic counterparts for binary black hole mergers have
been unsuccessful so far, which implies that electromagnetic
counterparts for binary black hole mergers are weak if at all exist.
One the other hand, as already mentioned in
Section~\ref{sec:intro_grb}, binary neutron star mergers are a
prominent candidate of the central engine of short gamma-ray bursts.
Shortly after observations of the first neutron star merger event
GW170817 \cite{2017PhRvL.119p1101A} a likely counterpart in gamma-ray,
GRB 170817A, was discovered by the Fermi satellite
\cite{2017ApJ...848L..13A}. The gamma-ray burst was observed 1.7~sec
after the coalescence. Subsequently, electromagnetic counterparts in
other wavelengths such as X-ray, ultraviolet, optical, infrared, and
radio were identified \cite{2017ApJ...848L..12A}. From these
observations, the host galaxy of GW170817 is identified to NGC 4993
at $z=0.0098$.

Such identification of a host galaxy opens up a new application in
cosmology. As emphasized above, gravitational wave observations
directly measure the luminosity distance to the source. Together with
the redshift information from the host galaxy, one can constrain the
distance-redshift relation and hence the Hubble constant
\cite{1986Natur.323..310S}. This application, which is sometimes
referred to as a standard siren, provides a powerful means of deriving
accurate constraints on the Hubble constant because of the simple and
well understood physics behind the method. In the specific case of
GW170817, the Hubble constant is constrained to
$H_0=70^{+12.0}_{-8.0}\,{\rm km/s/Mpc}$ only from a single event 
\cite{2017Natur.551...85A}.

The current estimate of event rates of compact binary mergers from
gravitational wave observations depends on the prior on the mass
distribution. From the analysis of Advanced LIGO and Advanced
Virgo First and Second Observing Runs \cite{2018arXiv181112907T}, 
the event rate for binary black hole (BBH) mergers is constrained to $R_{\rm
  BBH}\sim 9.7-101$~Gpc$^{-3}$yr$^{-1}$ and that for binary neutron
star (BNS) mergers to $R_{\rm BNS}\sim 110-3840$~Gpc$^{-3}$yr$^{-1}$, which
are obtained by combining results from different priors on the mass
distribution. Since any black hole-neutron star (BHNS) binary merger
was not observed in those observing runs, only the upper limit of 
$R_{\rm BHNS}<610$~Gpc$^{-3}$yr$^{-1}$ is obtained. While the redshift
evolution of the event rates is not yet constrained from the
observations, theoretical models generally predict that the comoving
rates increase toward higher redshifts out to $z\sim 2-10$ (e.g.,
\cite{2016MNRAS.456.1093K,2017MNRAS.471.4702B}). 

The size of gravitational wave sources is effectively the orbital
radius. Since the gravitational wave frequency $f$ is related to the
angular velocity $\Omega$ of the binary orbit as $f=\Omega/\pi$, from
Kepler's law we obtain the size $R$ for a binary system of compact
objects with equal masses $m_1=m_2=m$ as
\begin{equation}
  R =\left(\frac{2Gm}{\pi^2f^2}\right)^{1/3}
  \approx 3000\,{\rm
    km}\left(\frac{m}{M_\odot}\right)^{1/3}\left(\frac{f}{\rm Hz}\right)^{-2/3},
\end{equation}
which indicates that the size is quite small. Setting
$m\sim 10-40\,M_\odot$ and $f\sim 10-1000$~Hz, the size of
gravitational waves from binary black hole mergers detected by
Advanced LIGO is $\sim 20-700$~km, and assuming $m\sim 1.4\,M_\odot$
the size of gravitational waves from binary neutron star mergers
detected by Advanced LIGO is $\sim 10-200$~km.

\begin{table*}[t]
\caption{Summary of explosive transients discussed in this review article. See
  the text in each Section for details and
  references.\label{tab:exp_summary}}     
 \begin{tabular}{@{}cccccccc}
 \hline
   Type
   & Subclass
   & Number
   & $z_{\rm max}$
   & Wavelength
   & Time scale
   & Local rate
   & Size 
   \\
   &
   &
   &
   & ($f$ [Hz])
   & 
   & [Gpc$^{-3}$yr$^{-1}$]
   & [km]
   \\
 \hline
Supernova           & Ia            & $\mathcal{O}$($10^4$)& $\sim 2$     & optical   & $\sim 30$~days  & $\sim 3\times 10^4$  &  $\sim 10^{10}$ \\
(Section~\ref{sec:intro_sn})    &   &                      &              & ($\sim 10^{14-15}$) &        &                      & \\
                    & core-collapse & $\mathcal{O}$($10^4$)& $\sim 2$     & optical   & $\sim 30$~days  & $\sim  7\times 10^4$ &  $\sim 10^{10}$ \\
                    &               &                      &              & ($\sim 10^{14-15}$) &        &                      & \\
                    & superluminous & $\mathcal{O}$($100$) & $\sim 4$     & optical   & $\sim 100$~days & $\sim  200$          &  $\sim 10^{10}$ \\
                    &               &                      &              & ($\sim 10^{14-15}$) &        &                      & \\
 \hline
Gamma-ray burst     & long          & $>5000$              & $\sim 9$     & $\gamma$  &  a few sec      & $\sim 1$             & $\sim 10^{6-7}$\\
(Section~\ref{sec:intro_grb}) &     &                      &              & ($\sim 10^{18-23}$) &        &                      & \\
                    & short         & $>1000$              & $\sim 3$     & $\gamma$  &  $<$sec         & $\sim 1-10$          & $\sim 10^{5-6}$\\
                    &               &                      &              & ($\sim 10^{18-23}$) &        &                      & \\
\hline
Fast radio burst    & $\cdots$      & $\mathcal{O}$($100$) & $\sim 3?$    & radio     & $\sim$msec      & $\sim 10^{3.5-4.5}$   & $<10^{13}$\\
(Section~\ref{sec:intro_frb}) &     &                      &              & ($\sim 10^9$) &             &                      & \\
\hline
Gravitational wave  & BBH           & $>10$                & $\sim 0.5$   & LIGO band & $\lesssim$sec   & $\sim 10-100$        & $\sim 100$\\
(Section~\ref{sec:intro_gw}) &      &                      &              & ($\sim 10^{1-4}$) &          &                      & \\
                    & BNS           & $\geq 1$             & $\sim 0.05?$ & LIGO band & $\lesssim$sec   & $\sim 100-4000$      & $\sim 100$\\
                    &               &                      &              & ($\sim 10^{1-4}$) &          &                      & \\
                    & BHNS          & 0                    & $\cdots$     & LIGO band & $\lesssim$sec   & $< 600$              & $\sim 100$\\
                    &               &                      &              & ($\sim 10^{1-4}$) &          &                      & \\
\hline
 \end{tabular}
\end{table*}

\subsection{Summary of explosive transients}
\label{sec:intro_summary}

Table~\ref{tab:exp_summary} gives a summary of explosive transients
discussed in previous Sections. Some quantities that characterize each
transient and are relevant for discussions of strong lensing are
listed. For comparison, the size of the quasar accretion 
disk depends on the black hole mass, but for typical quasars with
black hole masses $M\sim 10^{8-9}\,M_\odot$ the size of the optical
emission region is found to be $\sim 10^{10-11}$~km and that of the
X-ray emission region is $\sim 10^{9-10}$~km (e.g.,
\cite{2016AN....337..356C}). Therefore the sizes of these explosive
transients are comparable or much smaller than that of quasars. 

\section{Applications}
\label{sec:applications}

\subsection{Introduction}

In this Section, we discuss possible applications of strong
lensing of explosive transients that are introduced in
Section~\ref{sec:transients}. We emphasize advantages of
these new strong lensing events as compared with more traditional
strong lensing such as strong lensing of quasars. 

\subsection{Time delay cosmography}
\label{sec:tdcosmo}

As emphasized in Section~\ref{sec:introduction}, time delay
cosmography, which is made so far mostly using variabilities in lensed
quasars, is becoming more and more important, because of the
apparent tension of $H_0$ between the distance ladder (e.g.,
\cite{2019ApJ...876...85R}) and {\it Planck} cosmic microwave
background measurements (e.g., \cite{2018arXiv180706209P}).
Thus an independent measurement from gravitational lens time delays is  
very important. 
Furthermore, time delays actually measure the time
delay distance, which is a combination of three angular diameter
distances, $D_{\rm ol}D_{\rm os}/D_{\rm ls}$, as shown in
equation~(\ref{eq:tdelay}). The time delay distance depends not only
$H_0$ but also other cosmological parameters such as $\Omega_M$ and
dark energy equation of state parameter $w_{\rm de}$. Since the
dependence of the time delay distance on cosmological parameters
differs considerably from those of other cosmological probes, time
delays provide unique cosmological information that is highly
complementary to other cosmological probes (e.g.,
\cite{2009arXiv0912.0201L,2011PhRvD..84l3529L}).

In addition to the measurements of source and lens redshifts, key
observations that lead to precise measurements of $H_0$ from time
delays include (see e.g., \cite{2016A&ARv..24...11T} for more
discussions); (i) precise time delay measurements, (ii) precise
measurements of image positions, (iii) detailed measurements of a
lensed host galaxy to constrain the lens potential, (iv) the
measurement of the stellar velocity dispersion of the lensing galaxy,
and (v) the proper understanding of the structure along the
line-of-sight. In the future, we will be able to measure $H_0$ at the
high precision by combining many strong lens systems, but in order to
assure the high accuracy we need to keep various systematics under
control. In what follows, we discuss possible systematics and argue
how strong lensing of explosive transients mitigate some of the
systematics. 

Strong lensing of explosive transients can definitely improve the
point (i) above. Since $H_0$ is inversely proportional to the time
delay, ideally we want to measure time delays at a percent level in
order for the measurement errors not to degrade cosmological
constraints. In the case of strong lensing of quasars, due to the
stochastic nature of the quasar light curve, the robust measurement of
time delays requires monitoring of lensed quasar images for many years.
Microlensing due to stars in lensing galaxies, which we will discuss
in more detail later, add additional variability to the light curve,
making the robust measurement even more challenging. As a result,
reliable measurements of quasar time delays require $\sim 10$~yr
monitoring observations, and the resulting accuracy on time delay
measurements is on the order of $\sim 1$~day (e.g., 
\cite{2013A&A...556A..22T,2013A&A...553A.121E,2018A&A...616A.183B}).

In contrast, since light curves of explosive transients introduced in
Section~\ref{sec:transients} are simple, we do not need monitoring
much beyond the time delay. In the case of gamma-ray bursts, fast
radio bursts, and gravitational waves, their time scales of light
curves are less than $\sim 1$~sec, which indicates that time delays
can be measured with an accuracy better than $\sim 1$~sec, much better
than current measurements with lensed quasars. This point has been
discussed in
\cite{2017NatCo...8.1148L,2017MNRAS.472.2906W,2019ApJ...873...37L,2019MNRAS.tmp.1176L}
for gravitational waves, in \cite{2006JCAP...01..012M} for gamma-ray
burst, and in \cite{2018NatCo...9.3833L} for fast radio
bursts. Repeated observations of very precise time delay measurements 
with repeating fast radio burst may allow us to directly measure 
the cosmic expansion \cite{2018ApJ...866..101Z}.
We also expect accurate and robust measurements of time delays
for strong lensing of supernovae, even though the time scale of their
light curves is $\gtrsim 30$~days, because of their simple and
well-known light curves \cite{2001ApJ...556L..71H,2003ApJ...583..584O,2019arXiv190300510H}.

However, one complication that may affect the accuracy and precision
of time delay measurements is microlensing, which refers to flux
variabilities due to stars in lensing galaxies (see
Section~\ref{sec:microlens}). Table~\ref{tab:exp_summary} indicates
that size of the explosive transients tend to be smaller than $R_{\rm
  Ein}$, which suggests the importance of microlensing in strong
lensing of explosive transients (see also Figure~\ref{fig:size_dm}). 

The microlensing effect can be time dependent due to the transverse
motions as well as the change of the size of the emitting region with
respect to time, and therefore can distort the light curve in a
non-trivial manner, which is particularly significant for strong
lensing of supernovae \cite{2006ApJ...653.1391D}. Recent studies discuss 
possible ways to mitigate the effect of microlensing on supernova lensing
\cite{2018ApJ...855...22G,2019A&A...621A..55B,2019ApJ...876..107P,2019ApJ...871..113L},
which indicates that a percent level measurement of time delays is
possible even in the presence of microlensing by taking advantage of
multiband light curves.  

However, an exception is strong lensing of gravitational waves. Thanks
to their long wavelength, microlensing variabilities are suppressed by
wave optics effects. This is obvious from
equation~(\ref{eq:wave_w_point}), as it is found $w\lesssim 0.1$ for
$M\sim 1\,M_\odot$ and $f\lesssim 10^3$~Hz, for which diffraction is
quite effective. The effect of microlensing by stars in lensing
galaxies on strong lensing of gravitational waves have been studied by 
\cite{2019A&A...625A..84D,2019A&A...627A.130D,2019arXiv190311809M}, in
which it is concluded that microlensing can modify the waveform
significantly only for highly magnified sources. In high magnification
regions, the effective Einstein radius of each microlens is enhanced 
by the macro model magnification, which leads to an increase of $w$
for the same mass of the microlens. Put another way,
due to diffraction microlensing by normal stars in lensing galaxies is
not effective for strong lensing of gravitational waves with moderate
magnifications, which is more common. This insensitivity to
microlensing can be seen as an advantage for the application of
gravitational wave lensing for cosmology. 

Strong lensing of explosive transients can also improve points
(iii) and (iv) mentioned above. Quasars are very bright so that they
outshine their host galaxies and sometimes lensing galaxies as well.
Such bright quasar images make detailed measurements of shapes
of lensed host galaxies very difficult. In contrast, for transient
events, we can always use images before the transient event happens or
after it fades away to measure shapes of lensed host galaxies
accurately. This point is emphasized in \cite{2001ApJ...556L..71H} for
strong lensing of supernovae and \cite{2017NatCo...8.1148L} for strong
lensing of gravitational waves. The images without bright lensed
sources also make it easier to conduct deep spectroscopy of lensing
galaxies to measure velocity dispersion profiles including resolved
two-dimensional velocity dispersion maps
\cite{2019arXiv190407237Y}. 

On the other hand, point (ii) above may be challenging in some
cases. This is because of poor angular resolutions of observations
for detecting some of the explosive transients, including gamma-ray
bursts, fast radio bursts, and gravitational waves. One way to obtain
accurate astrometry of lensed images is to identify their counterparts
in other wavelengths, in particular optical. Such optical counterparts
are known to be available at least for gamma-ray bursts and
gravitational waves from binary neutron star mergers, and deep
high-quality observations of multiple images of the optical
counterparts enable us to determine the image position on the order of
milliarcseconds,  which is required for precise time delay cosmography
\cite{2019arXiv190410965B}. For fast radio bursts, very accurate
measurements of image positions may be possible using high-resolution
radio imaging such as VLBI \cite{2017ApJ...834L...8M}, although this
may be practical only for strong lensing of repeating fast radio
bursts \cite{2018NatCo...9.3833L}.

Finally, strong lensing of these explosive transients may provide new
information that is not available for traditional strong lens systems.
One such example is magnification factors that are available for strong
lensing of Type Ia supernovae. For traditional quasar strong lensing,
we cannot measure magnification factors directly because intrinsic
magnitudes of lensed quasars are unknown.  In contrast, the
standardizable candle nature of Type Ia supernovae allows us to
directly measure the magnification factors, which break the mass-sheet
degeneracy and related degeneracies (see Section~\ref{sec:timedelay}). 
The idea to use strong lensing of Type Ia supernovae to break the
$H_0$-slope degeneracy and to obtain accurate $H_0$ measurements has
been proposed in \cite{2003MNRAS.338L..25O}. Accurate measurements of
$H_0$ may be possible also by strong lensing of Type Ia supernovae due
to clusters
\cite{2003ApJ...592...17B,2011A&A...536A..94R,2014ApJ...789...51Z}.
Inversely, we can use strong lensing of Type Ia supernovae to
calibrate their absolute magnitudes \cite{2019arXiv190702693W}.
Again, an obstacle is
microlensing which can change the total magnification of each lens
system considerably in some cases \cite{2006ApJ...653.1391D,2018MNRAS.478.5081F}.
Similarly, the standard siren nature of gravitational waves can add
useful information to time delay cosmography.

\subsection{Test of Fundamental Physics}

The measurements of the propagation speed for different particle types
or energies provide an important means of testing fundamental physics. 
For instance, the violation of weak equivalence principle leads to
different propagation speeds between e.g., photons and neutrinos
(e.g., \cite{1988PhRvL..60..173L}). The
violation of Lorentz invariance, which is predicted by some quantum
gravity theories, results in an energy dependent dispersion to photons
and changes the propagation speed of photons as a function of the
energy (e.g., \cite{2009Natur.462..331A}). Moreover, some modified
gravity theories predict 
the propagation speed of gravitational waves that differs from the
speed of light (e.g., \cite{2016JCAP...03..031L}). 

These effects can be tested by observations of explosive transients
(e.g., \cite{1988PhRvL..60..176K,2009Natur.462..331A,2015PhRvL.115z1101W,2017ApJ...848L..13A}), 
by checking arrival time differences between different particle types
or energies. However the observed arrival time
difference consists of both 
the intrinsic time delay and the time delay caused by the different
propagation speeds. The former is usually unknown or poorly 
constrained, which makes the result somewhat uncertain. One
interesting way to overcome this intrinsic time delay is to make use 
of strong lensing. This is because the contribution of the intrinsic
time delay vanishes if we compare the difference of time delays
between multiple images among different particles or energy.
This idea has been applied to strong lensing of gamma-ray bursts
\cite{2009MNRAS.396..946B} and gravitational waves
\cite{2017PhRvL.118i1101C,2017PhRvL.118i1102F,2017PhRvD..95f3512B,2018EPJC...78..692Y}. 
A caveat is that the difference between the propagation speeds of
gravitational waves and their electromagnetic counterparts might also
be caused by wave optics effects (Section~\ref{sec:waveoptics}) in
gravitational lensing, because the propagation of gravitational waves
is not affected by small intervening matter due to diffraction 
\cite{2017ApJ...835..103T}. 

Another test of fundamental physics includes the time variation of
fundamental constants such as the gravitational constant and the speed
of light. Strong lensing may also help for this type of test as
such time variation changes strong lensing observables such as time
delays (e.g., \cite{2018ApJ...867...50C}). 

\subsection{Compact dark matter}
\label{sec:compact_dm}

There have been long discussions on whether dark matter is composed of
unknown elementary particle or compact objects such as primordial
black holes (PBHs). The possibility of compact dark matter has been
tested with various observations including microlensing in and around
the Milky Way (see \cite{2018CQGra..35f3001S} for a review), from
which constraints on the abundance of compact dark matter are derived
as a function of the mass of compact dark matter. 

Strong lensing of explosive transients helps improve these constraints. 
For instance, strong lensing or microlensing of gamma-ray bursts and
their afterglow emissions have been studied extensively as a means of
testing the compact dark matter scenario
\cite{1987ApJ...317L..51P,1992ApJ...391L..63B,1992ApJ...389L..41M,1993ApJ...402..382M,1995ApJ...452L.111N,1998ApJ...495..597L,1999ApJ...512L..13M,2000MNRAS.319.1163W,2001MNRAS.325.1317K,2005ApJ...618..403B,2018PhRvD..98l3523J}. The
ideas include the search for echo signals in gamma-ray bursts and the
modification of the afterglow light curve due to the size dependence of
microlensing. 

When the mass of compact dark matter is very small, from $\sim
10^{-13}\,M_\odot$ to $\sim 10^{-16}\,M_\odot$ wave optics effects
(Section~\ref{sec:waveoptics}) become important even in gamma-ray. In
this case, the interference between multiple images induces 
an oscillating feature in the photon energy spectrum. The application
of this effect to gamma-ray bursts, which is referred to as
femtolensing, was proposed in \cite{1992ApJ...386L...5G} (see also
\cite{1993ApJ...413L...7S,1995ApJ...442...67U}). This method has been
applied to Fermi  Gamma-ray Burst Monitor data to place useful
constraints on the abundance of compact dark matter in the mass range
mentioned above \cite{2012PhRvD..86d3001B}. However, the finite source
size effect, which has been ignored before, is in fact crucial in this
application \cite{2006JCAP...01..023M}. The recent study by Katz {\it
  et al.} \cite{2018JCAP...12..005K} revisited constraints from
femtolensing taking full account of the finite source size effect to
find that a useful constraint on the abundance of compact dark matter
cannot be placed from the currently available data.

The search for echo signals due to strong lensing is possible also with
other explosive transients. For instance, the possibility of using
fast radio bursts to constraint compact dark matter with $M\gtrsim
20\,M_\odot$ has been proposed in \cite{2016PhRvL.117i1301M} and
subsequently studied in \cite{2018A&A...614A..50W,2018arXiv181211810L}.
For the mass of $\sim 20\,M_\odot$, we expect to observe multiple
bursts separated by a typical time delay of a few milliarcseconds. 
It is found that ongoing experiments such as CHIME can place
meaningful constraints on the abundance of compact dark matter 
in that mass range. Wave optics effects in strong lensing of fast radio
bursts and its application to the compact dark matter search were
discussed in \cite{2014ApJ...797...71Z} and also noted in
\cite{2016PhRvL.117i1301M}.  

Although the similar search is possible with strong lensing of
gravitational waves, their long wavelengths and compact sizes indicate
that wave optics effects definitely play an important role 
(Section~\ref{sec:waveoptics}). Again,
equation~(\ref{eq:wave_w_point}) indicates that we need compact dark
matter with the mass $M\gtrsim 10-100\,M_\odot$ to avoid 
diffraction and to observe strong lensing magnifications. When the
mass is near the threshold, the signal-to-noise ratio of lensed
waveforms shows an oscillatory behavior as the frequency sweeps up due
to wave optics effects, which can be regarded as a smoking gun signature
of strong lensing  \cite{1998PhRvL..80.1138N}. Strong lensing of
gravitational waves by compact objects is recently revisited after the
first direct observation of gravitational waves from binary black hole
mergers, including the rate estimate and expected constraints on the
abundance of compact dark matter 
\cite{2018PhRvD..98h3005L,2018PhRvD..98j3022C,2019PhRvL.122d1103J,2019ApJ...875..139L,2018arXiv180906511H}.

The compact dark matter scenario can be tested with strong lensing of
supernovae as well. In particular, strong lensing of Type Ia
supernovae by compact dark matter produces a non-Gaussian tail in
their apparent magnitude distribution for a given redshift, from which
useful constraints on compact dark matter for a wide mass range
$M\gtrsim 0.01\,M_\odot$ are obtained \cite{2018PhRvL.121n1101Z}.

\begin{figure}
\begin{center}
\includegraphics[width=8.0cm]{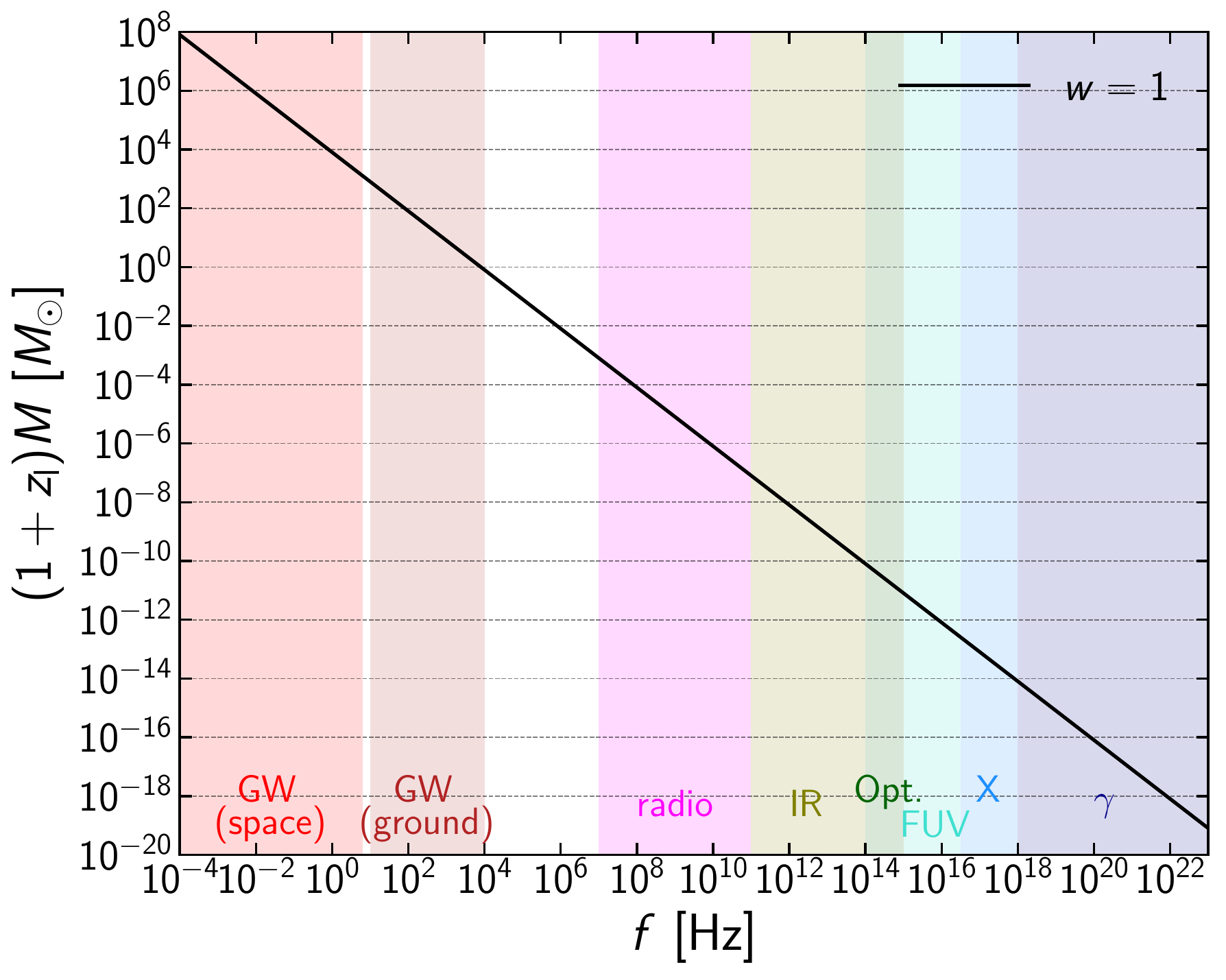}
\caption{The relation between the frequency $f$ and the (redshifted)
  mass $(1+z_{\rm l})M$ of a point mass lens for the dimensionless parameter 
  $w=1$, where $w$ is defined in equation~(\ref{eq:wave_w_point}). The
  region below the solid line corresponds to the case that the
  gravitational lensing magnification is significantly suppressed due
  to the diffraction, which is one of wave optics effects. }  
\label{fig:wave_w}
\end{center}
\end{figure}

In testing the compact dark matter scenario with strong lensing,
wave optics effects and the finite source size effect sometimes become
very important. To guide future studies along this line, we revisit
these effects introduced in Section~\ref{sec:waveoptics} and discuss
them more quantitatively. 

First, an important parameter that controls wave optics effects is the
dimensionless parameter $w$ defined in equation~(\ref{eq:wave_w_point}).
When $w<1$, diffraction originating from wave optics effects
becomes so effective that the gravitational lensing magnification is
highly suppressed i.e., $\mu\sim 1$ irrespective of the impact
parameter. In Figure~\ref{fig:wave_w}, we show the relation between
the frequency $f$ and the (redshifted) lens mass $(1+z_{\rm l})M$ that
satisfy $w=1$ for the case of a point mass lens. We note that the
similar relation holds for other lens mass models, once $M$ is replaced
to the enclosed mass within the Einstein radius (see
equation~\ref{eq:wave_w_mein}). The region below the line in
Figure~\ref{fig:wave_w} corresponds to $w<1$, and hence to
diffraction. Figure~\ref{fig:wave_w} clearly demonstrates that wave 
optics effects are particularly important for gravitational waves. 

\begin{figure}
\begin{center}
\includegraphics[width=7.5cm]{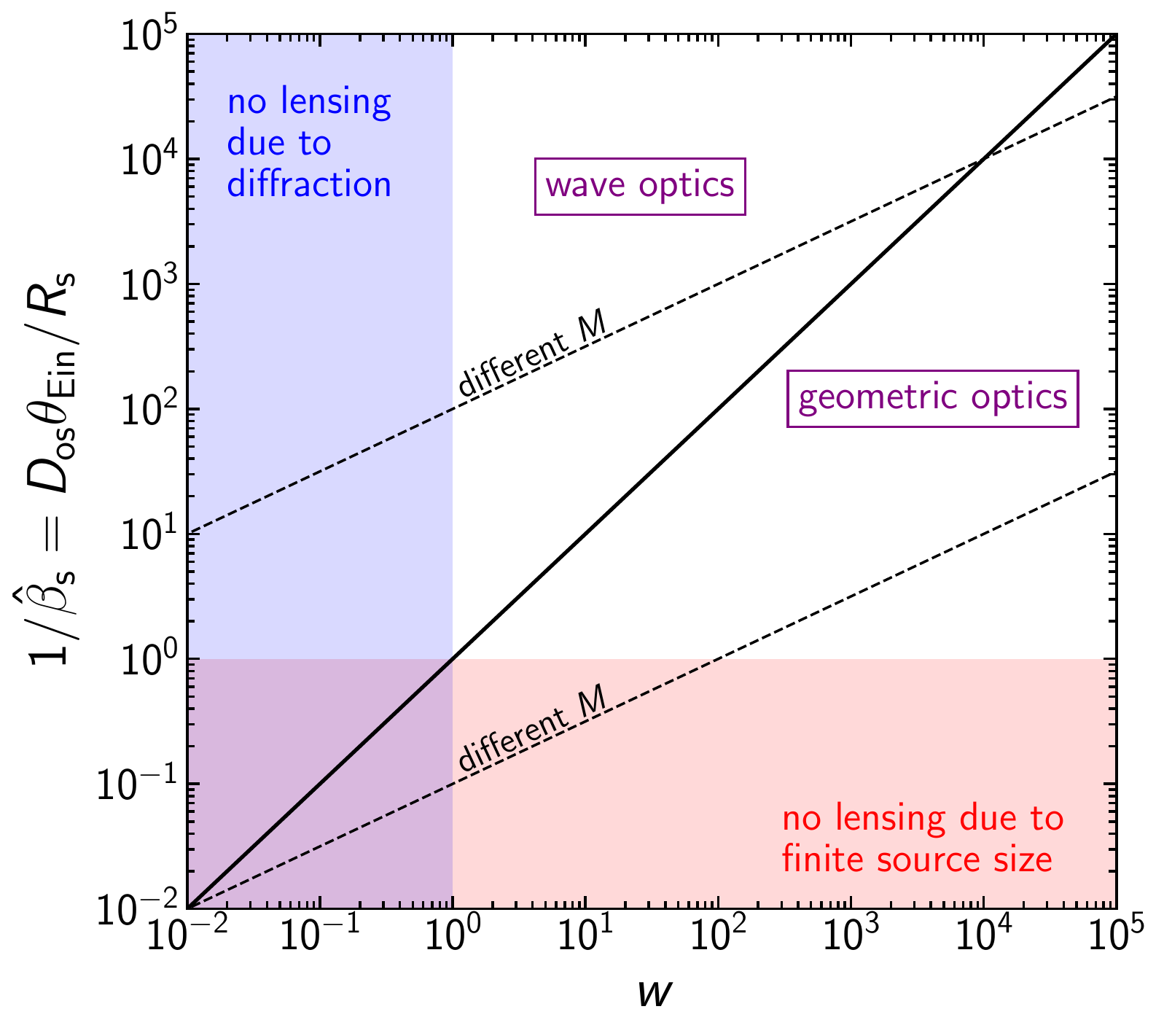}
\caption{The schematic illustration of regions relevant for geometric
  optics and wave optics. The solid line shows $w=1/\hat{\beta}_{\rm
    s}$, where $w$ is defined in equation~(\ref{eq:wave_w_point}) and
  $\hat{\beta}_{\rm s}=R_{\rm s}/(D_{\rm os}\theta_{\rm Ein})$ is the
  source size normalized by the Einstein radius. The region above the
  solid line corresponds to the situation  
  where the interference pattern due to wave optics effects may be
  observed, whereas the region below the solid line corresponds to the
  situation that the geometric optics approximation is relevant. The
  shaded regions show $w<1$ and $1/\hat{\beta}_{\rm s}<1$, for which
  the gravitational lensing magnification is significantly suppressed
  due to diffraction and the finite source size effect,
  respectively. The dotted lines show the direction along which
  parameter values change by changing the lens mass $M$.}  
\label{fig:wave_bet}
\end{center}
\end{figure}

As discussed in Section~\ref{sec:waveoptics}, in order for the
interference pattern due to wave optics effects to be observed, the
source must be sufficiently compact. This condition
(equation~\ref{eq:wave_sizelimit}) is given as $w\lesssim
1/\hat{\beta}_{\rm s}$, where $\hat{\beta}_{\rm s}=R_{\rm s}/(D_{\rm
  os}\theta_{\rm Ein})$ is the source size $R_{\rm _s}$ normalized by
the Einstein radius. We illustrate this condition in
Figure~\ref{fig:wave_bet}. The region above the line 
$w=1/\hat{\beta}_{\rm s}$ corresponds to the situation where the
interference pattern due to wave optics effects may be observed. We note
that this is just a necessity condition, and in order for the
interference pattern to be observe other conditions such as the
frequency band should also be met \cite{1999PThPS.133..137N}.
We also note that in regions with $w<1$ and $1/\hat{\beta}_{\rm s}<1$
the gravitational lensing magnification is significantly suppressed
due to diffraction and the finite source size effect,
respectively. Therefore in these regions we do not observe any
gravitational lensing effect.

Figure~\ref{fig:wave_bet} has several important implications. Since
$w$ and $1/\hat{\beta}_{\rm s}$ depend on the mass $M$ of a point
mass lens as $w\propto M$ and $1/\hat{\beta}_{\rm s}\propto \sqrt{M}$,
respectively, parameter values change along the direction indicated by
the dotted lines. This indicates that in the limit $M\rightarrow
\infty$ the parameter values always fall in the geometric optics
region, which is one of the reasons why the geometric optics
approximation is valid in most astronomical situations. 
Figure~\ref{fig:wave_bet} also suggests that the interference pattern
due to wave optics effects may be observed only when $1/\hat{\beta}_{\rm
  s}>1$ at $w=1$, as in the case of the upper dotted line in
Figure~\ref{fig:wave_bet}. In contrast, in the case of the lower
dotted line in Figure~\ref{fig:wave_bet}, the geometric optics
approximation is valid in all the parameter range of interest.
Even though at the small mass end $w$ becomes smaller than unity,
at $w\sim 1$ the magnification is already strongly suppressed by the
finite source size effect ($1/\hat{\beta}_{\rm s} \ll 1$), suggesting
that the effect of diffraction is unobservable.
From equations~(\ref{eq:wave_w_point}) and (\ref{eq:wave_sizelimit2}),
the necessity condition that the interference pattern is observed for
some lens masses is written as
\begin{eqnarray}
  R_{\rm s}&\lesssim& 3.05\times 10^{13}\,{\rm km}\left(\frac{1+z_{\rm l}}{1.5}\right)^{-1/2} \left(\frac{f}{\rm Hz}\right)^{-1/2}\nonumber\\
&&\times  
  \left(\frac{D_{\rm os}D_{\rm ls}/D_{\rm ol}}{0.949\,{\rm Gpc}}\right)^{1/2},
\label{eq:wave_sizelimit3}
\end{eqnarray}
<
where distances are again normalized to values at the lens redshift
$z_{\rm l}=0.5$ and the source redshift $z_{\rm s}=1.0$.

\begin{figure}
\begin{center}
\includegraphics[width=8.0cm]{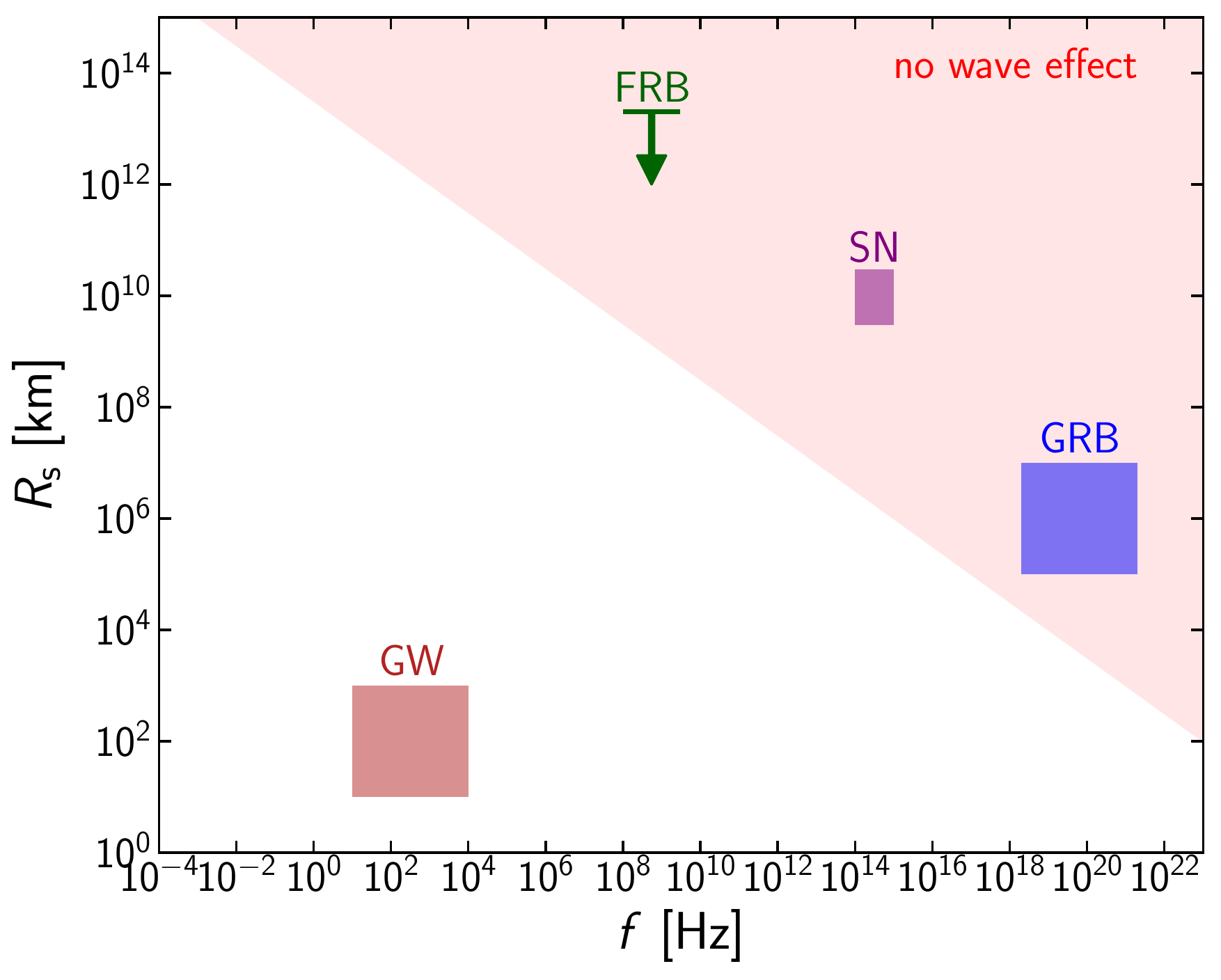}
\caption{Sizes and observed frequency of various explosive
  transients (see also Table~\ref{tab:exp_summary}). The shaded region
 in the upper right corner does not satisfy the condition given by
 equation~(\ref{eq:wave_sizelimit3}), which means that wave optics
 effects are never observed in this region irrespective of the lens 
 mass.  We note that the boundary depends on redshifts, and in this
 example we assume $z_{\rm l}=0.5$ and $z_{\rm s}=1.0$.}  
\label{fig:wave_rs}
\end{center}
\end{figure}

We can check whether the condition given by
equation~(\ref{eq:wave_sizelimit3}) is satisfied for explosive
transients summarized in Table~\ref{tab:exp_summary}. 
The result summarized in Figure~\ref{fig:wave_rs} suggests that
gravitational waves indeed satisfy the condition, and therefore are
ideal site to search for wave optics effects in strong lensing. Based on
the current understanding of their sizes, gamma-ray bursts do not
satisfy the condition, and therefore the so-called femtolensing does
not occur efficiently. Another interesting target to search for wave
optics effects in strong lensing is fast radio bursts for which sizes are
poorly constrained. If the size of fast radio bursts is sufficiently
compact, we may be able to detect the interference pattern in strong
lensing of fast radio bursts for lens masses of 
$M\gtrsim 10^{-5}\,M_\odot$ \cite{2014ApJ...797...71Z,2016PhRvL.117i1301M,2017ApJ...850..159E}.

\subsection{Structure of dark matter and galaxies}

Normal lensing objects such as galaxies and clusters consist of both
dark and luminous matter. Precise measurements of the dark matter
distribution in galaxies and clusters serve as an important test of
dark matter scenario as well as galaxy formation models. Gravitational
lensing is unique in that it probes the total projected mass of the
lensing galaxy robustly. While distributions of dark matter and baryon in
lensing objects have been studied in detail using lensed galaxies and
quasars (e.g., \cite{2010ARA&A..48...87T,2011A&ARv..19...47K} for
reviews), strong lensing of explosive transients can shed new light on
these applications. 

For instance, strong lensing of various transients can be discovered
by monitoring massive clusters of galaxies, which are known to be
efficient lenses. Time delays obtained from measurements of multiple
images of explosive transients break degeneracies in mass models
reconstructed from multiple image positions of strongly lensed
galaxies \cite{2019A&A...621A..91W}. In Section~\ref{sec:snrefsdal}, 
we present a specific example of this application in the case of a
strongly lensed core-collapse supernova. If the background sources are
standardizable candles such as Type Ia supernovae, we can directly 
measure magnification factors that break the mass-sheet degeneracy
\cite{1998MNRAS.296..763K}, as is clear from
equation~(\ref{eq:mass_sheet_mu}), and other mass model degeneracies. 
This application is possible even when background sources are not
multiply imaged (e.g., 
\cite{2014MNRAS.440.2742N,2014ApJ...786....9P,2015ApJ...811...70R}).

As discussed in Section~\ref{sec:microlens}, strong lensing provides
an important means of studying substructures in lensing galaxies,
which serve as a critical test of the CDM model. Gravitational lensing
of gravitational waves offers an alternative approach, as it is
insensitive to microlensing due to wave optics effects as discussed in
Section~\ref{sec:tdcosmo}.  

Substructures can be probed also by perturbations on time delays
between merging pairs of multiple images (see
Section~\ref{sec:microlens}). However, the smaller time
delays suggest that their precise measurements have been difficult for
quasar lenses.  Strong lensing of explosive transients, on the other
hand, can improve time delay measurements significantly due to the
short time scale of their light curves, leading to much more accurate
estimates of the effect of substructures on time delays. This point
was discussed in \cite{2018ApJ...867...69L} for strong lensing of
gravitational waves.  

Finally, using gravitational waves we may be able to detect
substructures more directly. Lensing by substructures with masses
$\sim 10^{3-6}\,M_\odot$ can induce interference pattern in
waveforms, as in the case of microlensing by compact dark matter
discussed in Section~\ref{sec:compact_dm}. This possibility has been
explored in \cite{2018PhRvD..98j4029D}.

\subsection{The nature of explosive transients}
\label{sec:nat_trans}

As discussed in Section~\ref{sec:transients}, the true nature of the
explosive transients that are considered in this review article is yet
to be fully understood. Strong lensing may help reveal their true
nature by taking advantage of its magnifying power as well as its
power to resolve fine structures of sources.

First, the result in Section~\ref{sec:lensrate} indicates that the
strong lensing probability is a steep function of the source
redshift. Therefore, we can constrain the redshift distribution of
explosive transients from their strong lensing probabilities. 
The idea was used in \cite{1999ApJ...510...54H} to constrain the
redshift distribution of gamma-ray bursts. The similar idea was
proposed for fast radio bursts in \cite{2014SCPMA..57.1390L}.

Since strong lensing magnifies background sources, it enables us to
observe very distant events that cannot be observed without
the gravitational lensing magnification. Therefore, we can constrain
the supernova rate at very high redshifts by observations of
strongly lensed supernovae at such high redshifts
\cite{2003ApJ...583..584O,2007APh....27..213A,2018arXiv180502662R}.
The search of lensed high-redshift supernovae can be conducted
efficiently by monitoring massive clusters of galaxies
\cite{2013MNRAS.435L..33P,2013arXiv1312.6330W,2016A&A...594A..54P,2018ARep...62..917P,2019PASJ...71...59M,2019PASJ...71...60W}.

The example above immediately suggests that the gravitational lensing
magnification modifies the observed distribution of explosive
transients. This may be particularly important for gravitational waves
from binary black hole mergers for which redshifts are not directly
measured in most cases. Instead, as discussed in
Section~\ref{sec:intro_gw}, from gravitational wave observations one
can measure the luminosity distance to the source. However, in
presence of gravitational lensing magnification $\mu$, the observed
luminosity distance is modified as 
\begin{equation}
D_{\rm L}^{\rm obs}=\frac{\bar{D}_{\rm L}}{\sqrt{\mu}},
\end{equation}
where $\bar{D}_{\rm L}$ is the luminosity distance to the source in
absence of gravitational lensing (i.e., the luminosity distance to the
source redshift computed assuming a homogeneous and isotropic
Universe) and $D_{\rm L}^{\rm obs}$ is the luminosity 
distance measured from observations of gravitational waves. 
Therefore, for highly magnified events $\mu\gg 1$, the redshift
inferred from the luminosity distance is biased low. The bias in the
estimated redshift directly affects the estimate of the chirp mass 
via equation~(\ref{eq:chirpmass}). Indeed it is pointed out that
strong lensing magnification produces an apparent tail in the high
mass end of the observed chirp mass distribution
\cite{2017PhRvD..95d4011D}. Furthermore, strong lensing of
gravitational waves produce multiple images, some of which are {\it
  demagnified}. Such demagnified images can be observed as apparently
very high redshift events, and hence produce a tail at the high end of
the observed redshift distribution \cite{2018MNRAS.480.3842O}. These
examples highlight the critical importance of gravitational lensing
for the interpretation of observed distributions of gravitational
waves. 

When multiple images of explosive transients are produced, in a sense
we observe the transients multiple times with some time differences. 
If we can predict the appearance of trailing images, it opens up
interesting applications such as the detailed monitoring of early
light curves. This possibility was noted in
\cite{2003ApJ...583..584O} for strong lensing of supernovae, and was
explored in detail in \cite{2018MNRAS.474.2612S}. A particularly
interesting feature in the early light curves of supernovae is the
so-called shock breakout, which is a luminous emission with very
short time scale.

Strong lensing can also be used to resolve fine structures of sources
by e.g., taking advantage of the size dependence of microlensing as
shown in Figure~\ref{fig:size_dm}. The ideas to resolve the jet
structure of gamma-ray sources with strong lensing have been explored
in
\cite{2000PhLA..265..168B,2001ApJ...558..643G,2001ApJ...561..703I,2009MNRAS.397.1084P,2015ApJ...809..100B,2016ApJ...821...58B}. 
For strong lensing of repeating fast radio bursts, one can measure the
change of time delays between multiple images, from which the motion of
fast radio burst sources is measured \cite{2017ApJ...847...19D}.

\section{Past observations and future prospects}
\label{sec:observations}

\subsection{Past observations}

\subsubsection{Strong lensing of supernovae: Before discoveries.}

The expected event rates of strongly lensed supernovae in various
supernova surveys have been computed 
\cite{2000ApJ...531..676W,2000ApJ...532..679P,2001ApJ...556L..71H,2002A&A...393...25G,2002A&A...392..757G,2013MNRAS.429.2392K},
which suggest that future surveys that are aimed at finding many
supernovae at $z\gtrsim 1$ should also be able to find strongly lensed
supernovae. One of the most comprehensive predictions before the first
discoveries has been made in \cite{2010MNRAS.405.2579O}, in which it
was argued that Pan-STARRS1
\cite{2002SPIE.4836..154K,2016arXiv161205560C} can find
$\mathcal{O}(1)$ strongly lensed supernovae, whereas Large Synoptic
Survey Telescope (LSST) \cite{2009arXiv0912.0201L} can find more than
100 strongly lensed supernovae.

We can efficiently search for strongly lensed supernovae by monitoring
plausible sites, such as galaxy-galaxy strong lens systems
\cite{2018ApJ...864...91S} and massive clusters 
\cite{2000MNRAS.319..549S,2000A&A...363..349S,2009A&A...507...61S,2018A&A...614A.103P}. 
The latter search led to discoveries of some supernovae behind
clusters, which are magnified but not multiply imaged 
\cite{2009A&A...507...71G,2011ApJ...742L...7A,2014ApJ...786....9P,2014MNRAS.440.2742N,2015ApJ...811...70R,2018ApJ...866...65R}. Giraud
\cite{1992A&A...259L..49G} reported a possible strongly lensed
variable source in a pair of arclets in the cluster Cl 0302+1658, but
interpreted it as an active galactic nucleus rather than a supernova.

\subsubsection{Discovery of PS1-10afx.}
\label{sec:ps1-10afx}

Pan-STARRS1 Medium Deep Survey (see \cite{2016arXiv161205560C}) is a
time-domain survey with a typical cadence of 3 days. The total survey
area is $\sim 70$~deg$^2$ and the typical 5$\sigma$ depth of nightly
stacks is $23$~mag. PS1-10afx is a new peculiar transient from
Pan-STARRS1 Medium Deep Survey reported by Chornock {\it et al.}
\cite{2013ApJ...767..162C}. PS1-10afx turned out to be a very bright
supernova at $z=1.388$ with an unusually fast light curve and a red
color, from which it was concluded that it is a new type of a
hydrogen-deficient superluminous supernova.

However, Quimby {\it et al.} \cite{2013ApJ...768L..20Q} re-examined
the photometric and spectroscopic data of PS1-10afx and proposed a new
interpretation: PS1-10afx is a normal Type Ia supernova that is
magnified by a factor of $\sim 31$ due to strong gravitational
lensing. In this case, the magnification factor can be estimated
directly thanks to the standardizable nature of a Type Ia
supernova. The lack of any signature of multiple images in both the
supernova images and the light curve is easily explained by the small
image separation between multiple images, $\theta<0.4''$. This
scenario, however, requires the presence of a foreground galaxy that
acts as a lens, which was not clearly seen in the follow-up images of
the supernova host galaxy taken after PS1-10afx faded away.

Quimby {\it et al.} \cite{2014Sci...344..396Q} presented a new
evidence that supports the lensing interpretation of PS1-10afx. They
obtained a deep spectrum of the host galaxy with Keck telescope and
detected a foreground galaxy at $z=1.117$ in the spectrum of the host
galaxy at $z=1.388$. This indicates that there are two galaxies that
are superposed and blended in the ground-based images. The analysis
indicates that the foreground galaxy well explains the small image
separation and time delay that are need to be compatible with the
observed property of PS1-10afx. In addition, the discovery of a lensed
Type Ia supernova from Pan-STARRS1 Medium Deep Survey is in good
agreement with the expected rate \cite{2010MNRAS.405.2579O} that is
extended to include events with unresolved multiple images. 
The comparison of theoretical expectations suggests that PS1-10afx is
likely to consist of four multiple images, although these images
were not resolved. 

\subsubsection{Discovery of SN Refsdal.}
\label{sec:snrefsdal}

\begin{figure}
\begin{center}
\includegraphics[width=8.0cm]{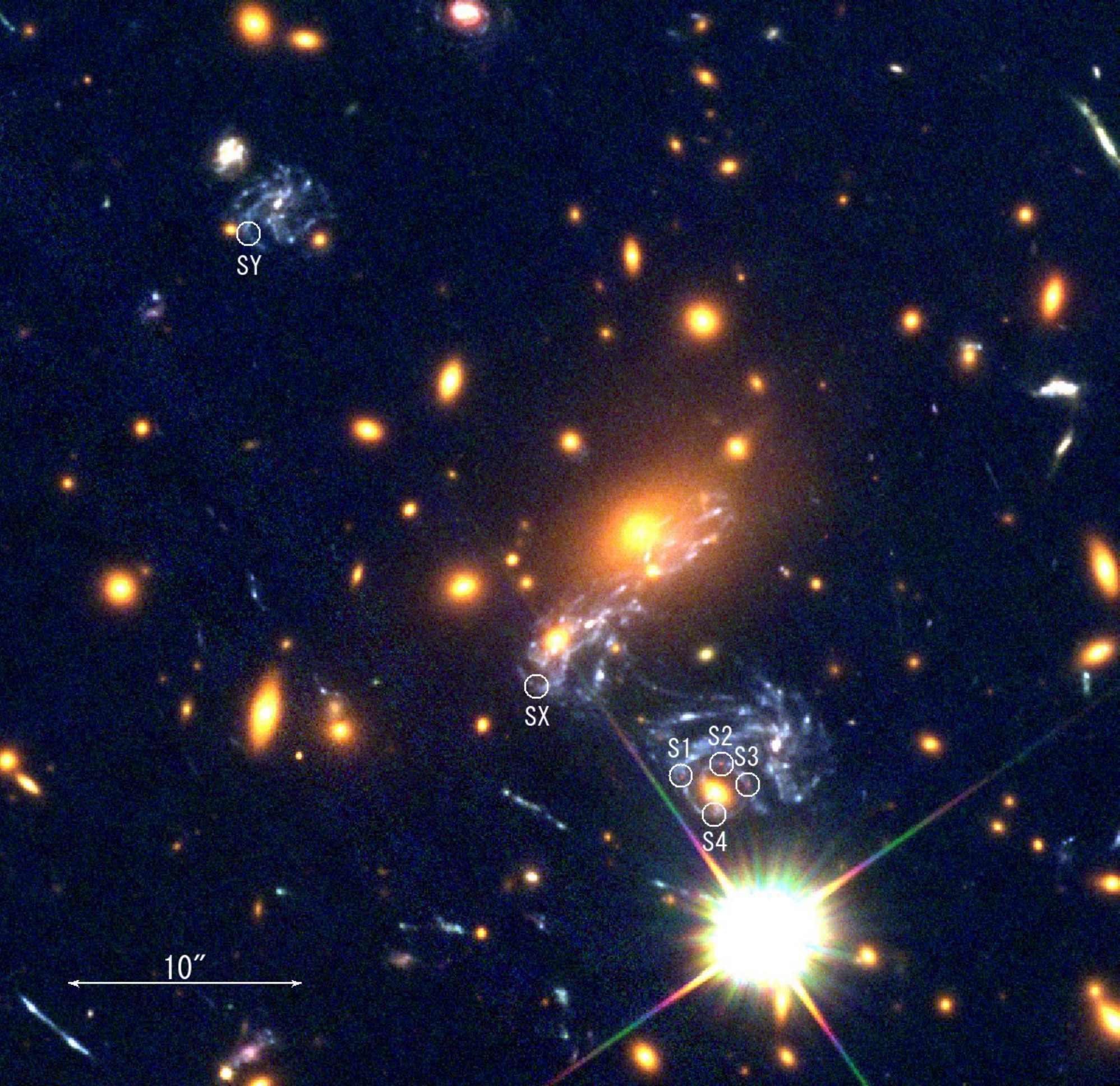}
\caption{Locations of 6 multiple images of the strongly lensed
  Type II supernova SN Refsdal \cite{2015Sci...347.1123K} at the core
  of the massive cluster MACS J1149.6+2223. The image shows a
  color-composite {\it Hubble Space Telescope} image taken in the {\it
    Hubble} Frontier Fields program
  \cite{2017ApJ...837...97L}. Originally the 4 multiple images S1--S4
  are detected, and about 1 year after the discoveries of S1--S4 the
  appearance of the new image SX was observed
  \cite{2016ApJ...819L...8K}. The image SY is never observed but
  predicted to have appeared $\gtrsim 10$~years before the appearance 
  of S1--S4. }  
\label{fig:snrefsdal}
\end{center}
\end{figure}

Kelly {\it et al.} \cite{2015Sci...347.1123K} reported the discovery
of SN Refsdal at $z=1.49$, which is the first strongly lensed
supernova discovered with resolved multiple images and time delay
measurements. It was discovered during {\it Hubble Space Telescope}
observations of the cluster MACS J1149.6+2223 at $z=0.54$, one of six
clusters targeted by the {\it Hubble} Frontier Fields program
\cite{2017ApJ...837...97L} that is aimed at studying distant Universe
with help of gravitational lensing magnifications due to massive
clusters of galaxies. Specifically, SN Refsdal was discovered by the
Grism Lens-Amplified Survey from Space program
\cite{2015ApJ...812..114T}, a follow-up program to acquire
near-infrared grism spectra of massive galaxy clusters including the
{\it Hubble} Frontier Fields clusters. 

Figure~\ref{fig:snrefsdal} shows locations of multiple images of SN
Refsdal. Originally the 4 images S1--S4 that were produced around an
elliptical member galaxy of the cluster were reported in
\cite{2015Sci...347.1123K}. The host galaxy of the supernova is a
face-on spiral galaxy at $z=1.49$ that are multiply imaged by the
foreground cluster as shown in Figure~\ref{fig:snrefsdal}, which
immediately suggests that additional multiple images in addition to the
observed 4 images should exist. This possibility was noted in
\cite{2015Sci...347.1123K} with estimated time delays on the order of
years. 

Soon after the discover was reported, many predictions of expected
time delays between multiple images of SN Refsdal have been made 
\cite{2015MNRAS.449L..86O,2015ApJ...800L..26S,2016ApJ...817...60T,2016MNRAS.456..356D,2016ApJ...819..114K,2016MNRAS.457.2029J,2016ApJ...822...78G}.
Thanks to deep imaging of the {\it Hubble} Frontier Fields program,
there are more than 100 multiple images of background galaxies
identified for this cluster, which allow us to reconstruct the mass
distribution in a reliable manner. These predictions agree in that
there are two additional images in addition to the observed image
S1--S4. Although one of the images, SY, is predicted to have appeared
$\gtrsim 10$~years before the appearance of S1--S4 and hence cannot be
confirmed by future observations, the other image SX is predicted to
appear in the future, which is a falsifiable prediction
with future monitoring observations (see Figure~\ref{fig:snrefsdal}
for the locations of SX and SY on the sky). However, there was a
considerable scatter in the predictions of the appearance of SX,
ranging from about half year to 2 years from the appearance of
S1--S4. The large difference of predictions of time delays despite a
large number of multiple images is partly due to the complex nature of
the cluster mass distribution. This, in turn, implies that the
observation of the reappearance of SX provides a unique opportunity to
check and improve our understanding of the cluster mass distribution
that is dominated by dark matter.

Since the images S1--S4 were discovered in 2014 October, SX has been
expected to appear sometime in 2015--2016. Monitoring follow-up
observations of this cluster with {\it Hubble Space Telescope} indeed
detected the new image SX at the position exactly predicted by mass
models \cite{2016ApJ...819L...8K}. From the observation, the time
delay between S1 and SX is measured to $\sim 350$~days, which is in
excellent agreement with several model predictions, in particular
those made with {\tt glafic}
\cite{2010PASJ...62.1017O,2015MNRAS.449L..86O,2016ApJ...819..114K} 
and {\tt GLEE} 
\cite{2010A&A...524A..94S,2012ApJ...750...10S,2016ApJ...822...78G}.
These successful predictions of the appearance of the image SX support 
the validity of strong lensing mass reconstruction techniques adopted
so far (see also \cite{2017MNRAS.472.3177M} for another validation
using simulated clusters). From the follow-up monitoring observations,
time delays between images S1--S4 were also measured 
\cite{2016ApJ...820...50R} and were found to agree with model 
predictions reasonably well (see also \cite{2016ApJ...817...60T}). 
Based on the observed light curve and
spectrum, SN Refsdal was classified as an SN 1987A-like Type II
supernova \cite{2016ApJ...831..205K}, and therefore the magnification
factor was not directly measured. The total magnification of all the 6
images is predicted to be $\sim 74$ by a best-fit mass model of
\cite{2016ApJ...819..114K}.

If mass distributions of clusters are well understood and the
systematics inherent to strong lens mass reconstructions are kept under
control, we may be able to use SN Refsdal-like events to constrain
$H_0$. Estimates of the constraining power using SN Refsdal indicates
that we can constrain $H_0$ from a single SN Refsdal-like event with
$\sim 10$\% accuracy or even better
\cite{2018ApJ...853L..31V,2018ApJ...860...94G}, although the
accuracy may be degraded by a factor of a few or more if we relax
prior assumptions on the cluster mass distribution
\cite{2019MNRAS.482.5666W}.

\subsubsection{Discovery of iPTF16geu.}
\label{sec:iptf16geu}

\begin{figure}
\begin{center}
\includegraphics[width=7.5cm]{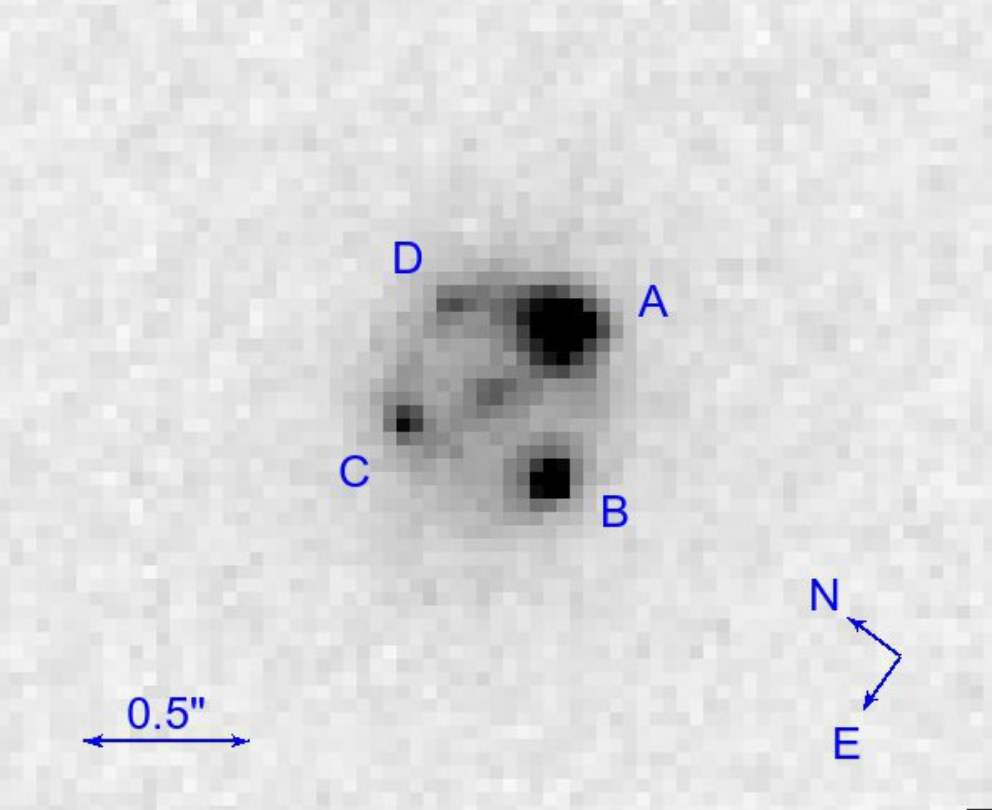}
\caption{The {\it Hubble Space Telescope} F814W image of the strongly
  lensed Type Ia supernova iPTF16geu
  \cite{2017Sci...356..291G}. The 4 supernova images are marked by
  A--D. }  
\label{fig:iptf16geu}
\end{center}
\end{figure}

\begin{table*}
\begin{center}
\caption{Summary of strongly lensed supernovae discussed in this
  review article. See the text in each Section for details and
  references. $N_{\rm img}$ indicates the number of multiple images,
  $m_{\rm peak}$ is an observed peak magnitude (the total magnitude for
  PS1-10afx and iPTF16geu, and the magnitude of the brightest image
  for SN Refsdal), $\mu_{\rm tot}$ is the total magnification factor
  of all the multiple images, which is directly measured from the
  observation for Type Ia, $\theta_{\rm max}$ is the maximum image
  separation between any multiple image pairs, and $\Delta t_{\rm
    max}$ is the maximum time delay between any multiple image
  pairs. Note that the values listed here can be either observed or
  model predicted ones. \label{tab:lensed_sne}}        
 \begin{tabular}{@{}ccccccccc}
 \hline
   Name
   & Type
   & $z_{\rm s}$
   & $z_{\rm l}$
   & $N_{\rm img}$
   & $m_{\rm peak}$
   & $\mu_{\rm tot}$
   & $\theta_{\rm max}$
   & $\Delta t_{\rm max}$
   \\
 \hline
PS1-10afx  (Section~\ref{sec:ps1-10afx}) & Ia & $1.388$ & $1.117$ & 4? & $i\sim 22$  & $\sim 31$ & $<0.4''$    & $<4$~days \\
SN Refsdal (Section~\ref{sec:snrefsdal}) & II & $1.49$  & $0.54$  & 6  & $i\sim 27$  & $\sim 74$ & $\sim 32''$ & $\sim 6000$~days \\
iPTF16geu  (Section~\ref{sec:iptf16geu}) & Ia & $0.409$ & $0.216$ & 4  & $i\sim 19$  & $\sim 52$ & $\sim 0.6''$& $\lesssim 1$~days \\
\hline
 \end{tabular}
\end{center}
\end{table*}

Goobar {\it et al.} \cite{2017Sci...356..291G} reported the discovery
of iPTF16geu, which is the first strongly lensed Type Ia supernova
with resolved multiple images, from the intermediate Palomar Transient
Factory \cite{2013ATel.4807....1K} that is a massive time-domain
survey with the limiting magnitude of $R\sim 20.5$ using a camera
covering the 7.26~deg$^2$ field-of-view on the 48-inch
Oschin telescope at Palomar Observatory.  Figure~\ref{fig:iptf16geu}
shows the follow-up {\it Hubble Space Telescope} image of iPTF16geu, in
which a Type Ia supernova at $z=0.409$ is strongly lensed into 4
multiple images due to a foreground galaxy at $z=0.216$. As in the
case of PS1-10afx, the standardizable nature of a Type Ia supernova
enables the direct measurement of the total magnification of iPTF16geu
to $\sim 52$. Since the foreground galaxy is a relatively low mass
galaxy with the velocity dispersion of $\sim 160\,{\rm km\,s^{-1}}$,
the maximum image separation between multiple images is small,
$\sim 0.6''$.  A gravitationally lensed host galaxy is clearly visible 
particularly in follow-up Keck near-infrared images.

Mass modeling of iPTF16geu has been conducted in More {\it et al.}
\cite{2017ApJ...835L..25M}. Although the supernova image positions and
lensed host galaxy are fitted well by a simple model that consists of
Singular Isothermal Ellipsoid plus an external shear, it was found
that flux ratios between the multiple images predicted by mass models
differ considerably from observed flux ratios. These anomalous flux
ratios have been attributed to microlensing in
\cite{2017ApJ...835L..25M}. It was also found that predicted time
delays between the multiple images are less than a day, making
measurements of time delays in observations very challenging. 

The effect of microlensing on iPTF16geu has been revisited by
\cite{2017arXiv171107919Y,2018MNRAS.478.5081F}, who conclude that
microlensing alone cannot explain the anomalous flux ratios. 
Possible explanations include the too simplistic assumption on the
macro mass model and an additional perturbation on the flux ratios by
substructures in the lensing galaxy. On the other hand, recent mass
modeling by M{\"o}rtsell {\it et al.} \cite{2019arXiv190706609M}
concluded that the anomalous flux ratio can be reconciled with
microlensing if the radial density profile of the lensing galaxy is
shallower than the isothermal model.

The observed light curves of iPTF16geu have been analyzed by Dhawan
{\it et al.} \cite{2019arXiv190706756D} to confirm very short time
delays between images, $\lesssim 1$~day. Specifically, time delays
with respect to the brightest image are measured to $-0.23\pm 0.99$,
$-1.43\pm 0.74$, and $1.36\pm 1.07$~days. They also studied the dust
extinction of multiple images, and derived the extinction-corrected
total magnification factor of $\mu=67.8^{+2.6}_{-2.9}$. In a companion
paper by Johansson {\it et al.} (in prep.), they provide the first
time delay measurement based on spectroscopic measurements. 

The high total magnification of $\sim 52$ even before the extinction 
correction is partly explained by the selection effect, which will be
discussed in detail in Section~\ref{sec:rate_exp}. However, it has
been found from detailed comparisons with theoretically expected
distributions that the observed magnification is higher than expected
given its redshift, even if we take account of the selection effect
\cite{2017Sci...356..291G,2017ApJ...835L..25M,2019ApJS..243....6G}. 
This issue may be related with the anomalous flux ratios mentioned
above. It is of great importance to understand the cause of the
anomalous flux ratios and the high total magnification for the future
use of strongly lensed Type Ia supernovae for cosmology.  

\subsubsection{Implications of the first discoveries for search methods.}

Table~\ref{tab:lensed_sne} summarizes properties of the three strongly
lensed supernovae presented in this review article. Two out of the three 
events have very small image separations such that they are barely
resolved in ground-based imaging observations. Such unresolved events
were not included in the calculation of \cite{2010MNRAS.405.2579O}. 
If we can identify these unresolved strong lensing events from the
survey data in a timely manner, we may be able to increase the number
of strongly lensed supernovae discovered in future time-domain surveys
to enhance their power for cosmological and astrophysical studies.

Based on the discovery of PS1-10afx, Quimby {\it et al.}
\cite{2014Sci...344..396Q} proposed a new method to identify strongly
lensed Type Ia supernovae, utilizing a color-magnitude diagram of
supernovae. Specifically, it was found that strongly lensed Type Ia
supernovae are well separated from unlensed supernovae in $i$-band
magnitude versus $r-i$ color diagram, which allows us to identify
unresolved strong lensing candidates relatively securely. Rapid
follow-up observations of these candidates may lead to measurements of
time delays for these strong lensing events. It was argued that this 
approach can significantly increase the number of strongly lensed
supernovae discovered by LSST.

Goldstein and Nugent \cite{2017ApJ...834L...5G} proposed a slightly
different approach, in which strongly lensed Type Ia supernova
candidates are identified by identifying supernovae near elliptical
galaxies whose absolute magnitudes computed from the redshifts of the
elliptical galaxies are brighter than those of Type Ia
supernovae. This search method is based on the fact that lensing
galaxies are dominated by elliptical galaxies. This method also
enables rapid identifications of unresolved strong lensing events and
potentially increases the number of strongly lensed supernovae
discovered by LSST (see also \cite{2019ApJS..243....6G}).  
 
Even if multiple images of strongly lensed supernova are barely
resolved, we may still be able to see its signature by carefully
checking the morphology of the supernova image to see if it is really
consistent with the Point Spread Function. The possibility of finding
strong lensed supernovae by checking the ellipticity of the supernova
image is discussed in \cite{2018RNAAS...2d.186L}.

Recent work by Wojtak {\it et al.} \cite{2019MNRAS.487.3342W} explored
how effective such new strategy to find unresolved strongly lensed
supernovae is in ongoing and future time-domain surveys. It was found
that finding unresolved strongly lensed supernovae increases the
number of strongly lensed supernovae drastically for shallow surveys
such as Palomar Transient Factory, whereas the increase of the number
is modest for deep time-domain surveys such as LSST.  

\subsubsection{Strong lensing of gamma-ray burst.}

One of the most comprehensive discussions on the detectability of
strong lensing of gamma-ray bursts has been presented in
\cite{2001ApJ...548..522P}. Although the {\it Swift} satellite may be
able to detect strongly lensed gamma-ray bursts, it is argued that
detecting multiple image pairs is unlikely because of its inefficient
duty cycle and the limited sky coverage, $\sim 50\%$ (see also
\cite{2011MNRAS.414..209W}). Fermi Gamma-ray Burst Monitor has more
sky coverage and therefore may be suited to search for multiple image
pairs in this regard.  

Despite some explicit attempts to search for lensed image pairs in the
gamma-ray burst catalogs for a wide range of time delays 
\cite{1994ApJ...432..478N,2014SCPMA..57.1592L,2006ApJ...650..252H,2019ApJ...871..121H},
no secure candidate of multiply imaged gamma-ray bursts has been
identified so far. The latest search by Hurley {\it et al.}
\cite{2019ApJ...871..121H} makes use of the gamma-ray burst sample
detected by {\it Konus}-{\it Wind} \cite{1995SSRv...71..265A}, which
has the high duty cycle and 
large sky coverage, to search for lensed image pairs. Based on the
absence of any candidate of strongly lensed gamma-ray burst, an upper
limit of the lensing probability of 0.0033 is placed. A caution is
that microlensing can distort light curves of strongly lensed
gamma-ray bursts (e.g., \cite{1997MNRAS.286L..11W}), which may affect
the efficiency of searching for multiple image pairs based on the
similarity of the light curves.   

\subsubsection{Strong lensing of fast radio burst.}

While some estimates of expected event rates of strongly lensed fast
radio bursts have been presented in the literature (e.g.,
\cite{2014SCPMA..57.1390L,2018NatCo...9.3833L}), so far no systematic
search for strong lensing of fast radio bursts has been made. Since 
the number of observed fast radio bursts is very rapidly increasing,
the future search in real catalogs will be interesting. 

\subsubsection{Strong lensing of gravitational waves.}

The expected rates of strongly lensed gravitational waves have been
computed both for ground based experiments
(e.g., \cite{2013JCAP...10..022P,2014JCAP...10..080B,2015JCAP...12..006D,2018MNRAS.476.2220L,2018PhRvD..97b3012N,2018MNRAS.480.3842O,2019arXiv190706841B})
and space based experiments 
(e.g., \cite{2003ApJ...595.1039T,2010PhRvL.105y1101S,2011MNRAS.415.2773S,2018MNRAS.480.3842O}). 
These calculations suggest that a large number of strongly lensed
gravitational waves from compact binary mergers will be discovered in
future third-generation ground-based experiments as well as 
future space-based gravitational wave experiments.

The expected rates of strongly lensed gravitational waves in the
previous and ongoing Advanced LIGO observing runs are predicted to be
small (e.g., \cite{2018PhRvD..97b3012N,2018MNRAS.476.2220L,2018MNRAS.480.3842O,2019arXiv190706841B}). 
However, Broadhurst {\it et al.}
\cite{2018arXiv180205273B,2019arXiv190103190B} made an interesting
claim that roughly half of gravitational waves from binary black hole
mergers detected by Advanced LIGO are in fact strongly lensed ones. As
discussed in Section~\ref{sec:nat_trans}, estimated redshifts and
chirp masses of highly magnified gravitational wave events are biased
if gravitational lensing is not taken into account, such that highly
magnified high redshift events are observed as low redshift events
with very large chirp masses. Therefore, binary black holes with
relatively high masses of $\sim 30\,M_\odot$, if interpreted as highly
magnified events, are in fact binary black hole systems with moderate
masses, $\sim 10\,M_\odot$. In order for such events to contribute to
the current observation, the redshift evolution of the event rate must
be very strong such that the event rate at $z\sim 1-2$ is several
orders of magnitude higher than the local event rate. 

For highly magnified strong lensing events, we expect a pair of images
with similar waveforms \cite{2017arXiv170204724D}, which are observed
with a typical time difference of less than a day
\cite{2018MNRAS.480.3842O}. While the absence of such pair events in
Advanced LIGO observations may disfavor the lensing scenario mentioned
above, it is possible that such counterimages are missed due to the
relatively low duty cycle (``glitches'' in the data stream) as well as
the effect of the Earth rotation that changes the sensitivity to a
source located in a given position on the sky as a function of time
\cite{2019ApJ...874..139Y}. These issues are mitigated by increasing
the number of detectors in operation.

Gravitational lensing also rotates the polarization plane of
gravitational waves, which results in the modification of the antenna
pattern function. This effect, however, appears to be negligibly small
in most situations given the small deflection angles due to
gravitational lensing \cite{2019arXiv190707486H}.

There are some attempts to explicitly search for strongly lensed
gravitational wave events among sub-threshold signals, but no
promising candidate of strong lensing events is identified
\cite{2019ApJ...874L...2H,2019arXiv190406020L}. Since strong lensing
of gravitational waves may be produced by foreground galaxy clusters,
the search of strong lensing events can also be conduced around known
massive clusters within error circles of observed gravitational wave
events \cite{2018MNRAS.475.3823S,2019MNRAS.485.5180S}. 

\subsection{Expected event rates}
\label{sec:rate_exp}

Here we present some discussions on expected event rates of strongly
lensed explosive transients. Our strategy here is to provide simple
and concise estimates of strong lensing rates for various explosive
transients in a unified manner. Since we do not fully take account of
the luminosity distributions and selection functions, these estimates
are not very accurate, but a reward of this simple approach is that
the parameter dependence is clearer that leads to the better
understanding of differences of expected strong lensing rates in
different setups. Our approach here is also complementary to detailed
calculations of event rates of strong lensing taking full account of
the luminosity distributions and the selection effects, e.g.,
\cite{2010MNRAS.405.2579O,2019ApJS..243....6G,2019MNRAS.487.3342W} for
strong lensing of supernovae and
\cite{2018MNRAS.476.2220L,2018MNRAS.480.3842O} for strong lensing of
gravitational waves.

We start with the strong lensing probabilities derived in
Section~\ref{sec:lensrate}. The strong lensing probability as a
function of the source redshift, $P_{\rm sl}(z_{\rm s})$ defined by
equation~(\ref{eq:p_sl}), is computed 
following the Monte-Carlo approach \cite{2018MNRAS.480.3842O} assuming
single galaxies as lensing objects. The galaxy mass distribution is
modeled by a Singular Isothermal Ellipsoid plus an external shear (see
Section~\ref{sec:lensrate} for more details). Since groups and
clusters of galaxies are not included in the calculation, these strong
lensing probabilities are likely to be underestimated, although their
contribution of the strong lensing probability to the total strong
lensing probability is thought to be subdominant as discussed in
Section~\ref{sec:lensrate}, especially for strong lensing of explosive
transients whose sizes are compact.  

Given the strong lensing probabilities, we can compute the expected
observed rate of strongly lensed explosive transients at $z<z_{\rm max}$ as
\begin{equation}
R_{\rm sl}(<z_{\rm max})=\Omega_{\rm sky}\int_0^{z_{\rm max}} dz_{\rm
  s} \frac{d^2V}{dz_{\rm s}d\Omega}\frac{R(z_{\rm s})}{1+z_{\rm
    s}}P_{\rm sl}(z_{\rm s}), 
\label{eq:r_sl}
\end{equation}
where $R(z)$ is the comoving rate density of explosive transients as a
function of redshift and $\Omega_{\rm sky}$ is the sky area of the
survey. The factor $(1+z)^{-1}$ takes account of the time dilation
effect, since $R(z)$ is usually defined as the event rate in the rest
frame of the transients. 

We compute $R_{\rm sl}$ in the following setup. We compute expected
observed rates in all sky by setting $\Omega_{\rm sky}=4\pi$. Since strong
lensing events with sufficiently magnified are of more interest and
total magnification factors tend to be high for known strongly lensed
supernovae (see Table~\ref{tab:lensed_sne}), we focus on strong
lensing events with $\mu_{\rm tot}\gtrsim 10$ by setting the
selection bias, $B=1/25=0.04$
(see also the discussion in Section~\ref{sec:lensrate}). For
simplicity, the event rate of explosive transients is assumed to have
the following redshift dependence
\begin{equation}
R(z)=R^{\rm loc}(1+z)^{\alpha_z},
\label{eq:rate_alphaz}
\end{equation}
where $R^{\rm loc}$ is the local event rate and $\alpha_z$
parametrizes the redshift evolution. See Table~\ref{tab:exp_summary}
for the current estimates of $R^{\rm loc}$ for various explosive
transients. In many cases, the redshift evolution of event rates of
explosive transients traces the global star formation history of the
Universe, for which the rate increases toward higher redshifts out to
$z\sim 2$, with the slope corresponding to $\alpha_z\approx 2$.  With
these assumptions, the observed rate (\ref{eq:r_sl}) is rewritten as
\begin{eqnarray}
R_{\rm sl}(<z_{\rm max})&=&R_{\rm sl}^{\rm norm}(<z_{\rm max}; \alpha_z)
\left(\frac{\Omega_{\rm sky}}{4\pi}\right)\left(\frac{B}{0.04}\right)\nonumber\\
&&\times \left(\frac{R^{\rm loc}}{1~{\rm Gpc^{-3}yr^{-1}}}\right),
\label{eq:r_sl_new}
\end{eqnarray}
where $R_{\rm sl}^{\rm norm}(<z_{\rm max}; \alpha_z)$ is calculated by
inserting the fiducial values to equation~(\ref{eq:r_sl}) and adopting
an approximation given by equation~(\ref{eq:p_sl_approx}).

\begin{figure}
\begin{center}
\includegraphics[width=8.0cm]{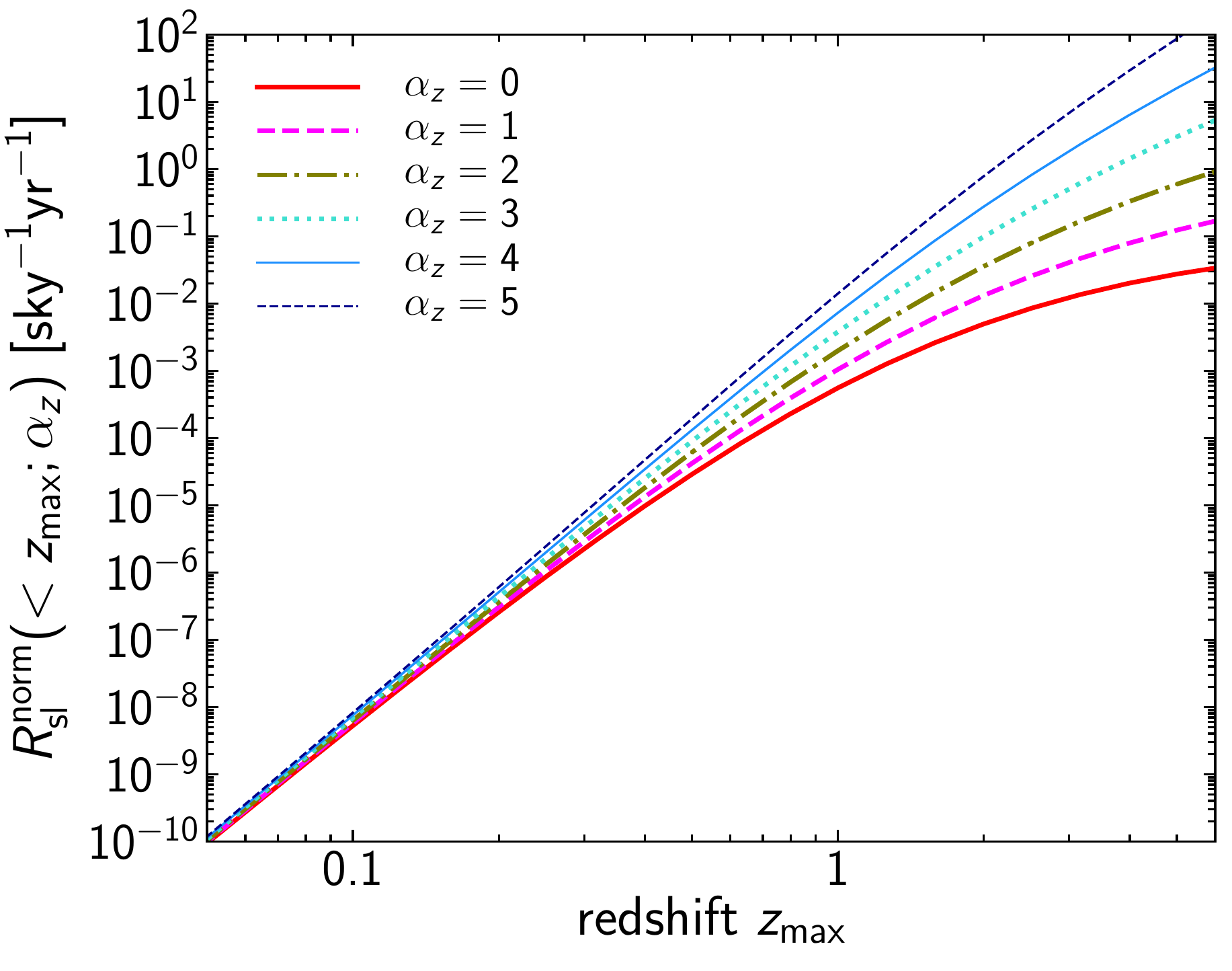}
\caption{The normalization of the observed strong lensing
  rate, $R_{\rm sl}^{\rm norm}(<z_{\rm max}; \alpha_z)$ defined in
  equation~(\ref{eq:r_sl_new}), as a function of the maximum redshift
  $z_{\rm max}$. We show $R_{\rm sl}^{\rm norm}$ for several different
redshift evolution parameter $\alpha_z$ that is introduced in
equation~(\ref{eq:rate_alphaz}). The fitting form of $R_{\rm
  sl}^{\rm norm}$ is given by equation~(\ref{eq:r_sl_norm}).}
\label{fig:slrate}
\end{center}
\end{figure}

Figure~\ref{fig:slrate} shows $R_{\rm sl}^{\rm norm}(<z_{\rm max}; \alpha_z)$  
for several different choices of $\alpha_z$. It is found that the
expected observed strong lensing rate is a steep function of $z_{\rm
  max}$. At low redshift $z_{\rm max}\ll 1$, we roughly have 
$R_{\rm sl}^{\rm norm}(<z_{\rm max}; \alpha_z)\propto z_{\rm max}^6$, in
contrast to the unlensed event rate which is proportional to the
volume at low redshifts i.e., $\propto z_{\rm max}^3$.

We find that the results shown in Figure~\ref{fig:slrate} are fitted
by the following form 
\begin{equation}
R_{\rm sl}^{\rm norm}(<z_{\rm max}; \alpha_z)\approx
\frac{a_1z_{\rm max}^{a_2}(1+z_{\rm max})^{\alpha_z}}{1+a_3z_{\rm max}^{a_4}},
\label{eq:r_sl_norm}
\end{equation}
\begin{equation}
a_1=3\times 10^{-3},
\end{equation}
\begin{equation}
a_2=5.8,
\end{equation}
\begin{equation}
a_3=4.6+0.35\,\alpha_z
\end{equation}
\begin{equation}
a_4=3.1+0.1\,\alpha_z^{0.6}.
\end{equation}
This fitting form is derived in the range $0\leq\alpha_z\leq5$ and
$z_{\rm max}<5$.

\begin{table*}[t]
\caption{Expected observed rates of strong lensing of various transients
  computed using equation~(\ref{eq:r_sl_new}). See also
  Table~\ref{tab:exp_summary} for the summary of properties of these
  transients. The columns $R^{\rm loc}$ and $\alpha_z$ show fiducial
  values of the local event rate and the redshift evolution parameter
  adopted in the calculation. Expected observed rates $R_{\rm sl}$ for
  $\Omega_{\rm sky}=4\pi$ and $B=0.04$ (corresponding to $\mu_{\rm
    tot}\gtrsim 10$) within the maximum redshift $z_{\rm max}=0.5$,
  $1$, $2$, and $3$ are shown.\label{tab:num_summary}}     
 \begin{tabular}{@{}cccccccc}
 \hline
   Type
   & Subclass
   & $R^{\rm loc}$
   & $\alpha_z$
   & $R_{\rm sl}(<0.5)$ 
   & $R_{\rm sl}(<1)$ 
   & $R_{\rm sl}(<2)$ 
   & $R_{\rm sl}(<3)$ 
   \\
   &
   & [Gpc$^{-3}$yr$^{-1}$]
   &
   & [sky$^{-1}$yr$^{-1}$]
   & [sky$^{-1}$yr$^{-1}$]
   & [sky$^{-1}$yr$^{-1}$]
   & [sky$^{-1}$yr$^{-1}$]
   \\
 \hline
Supernova           & Ia            & $3\times 10^4$ & 1 & 1.6 & 30  & 320 & 1300 \\
                    & core-collapse & $7\times 10^4$ & 2 & 5.4 & 130 & 2000 & 10000\\
                    & superluminous & $200$          & 2 & 0.02 & 0.38 & 5.8 & 29 \\
 \hline
Gamma-ray burst     & long          & $1$            & 2 & $<0.01$ & $<0.01$ & 0.03 & 0.15\\
                    & short         & $3$            & 1 & $<0.01$ & $<0.01$ & 0.03 & 0.13 \\
\hline
Fast radio burst    & $\cdots$      & $10^4$         & 2 & 0.78 & 19 & 290 & 1500 \\
\hline
Gravitational wave  & BBH           & $30$           & 2 & $<0.01$ & 0.06 & 0.88 & 4.4\\
                    & BNS           & $600$          & 1 & 0.03 & 0.61 & 6.5 & 25 \\
                    & BHNS          & $10$           & 1 & $<0.01$ & 0.01 & 0.11 & 0.4\\
\hline
 \end{tabular}
\end{table*}

We use equations~(\ref{eq:r_sl}) and (\ref{eq:r_sl_norm}) to compute
expected event rates of strong lensing of various explosive transients
as a function of $z_{\rm max}$. For each explosive transient listed in
Table~\ref{tab:exp_summary}, we choose a fiducial value of $R^{\rm
  loc}$ that is consistent with the current estimates. We also choose 
the redshift evolution parameter defined in
equation~(\ref{eq:rate_alphaz}) to $\alpha_z=1$ or $2$ so that it is
broadly consistent with the current estimates. 
Table~\ref{tab:num_summary} summarizes our fiducial choices of $R^{\rm
  loc}$ and $\alpha_z$, and resulting expected observed rates for $z_{\rm
  max}=0.5$, $1$, $2$, and $3$. Since we adopt $B=0.04$, these
correspond to rates of strong lensing events with  
$\mu_{\rm tot}\gtrsim 10$. We note that these predictions can be
easily modified to those for other parameter sets by using
equation~(\ref{eq:r_sl}). 

Results in Table~\ref{tab:num_summary} do not take account of the
observability. We provide a rough estimate of $z_{\rm max}$ for each
survey as follows. For each survey, we first estimate the redshift
$z_{\rm lim}$, out to which normal unlensed events are largely
detected. Since we consider strong lensing events $\mu_{\rm tot}\gtrsim
10$, we assume that each multiple image is magnified by a factor of
$\sim 4$ or so. The magnification factor of $4$, for instance,
indicates that the event is detected out to a factor of $\sqrt{4}=2$
larger luminosity distance. Therefore, for each survey and explosive
transient with $z_{\rm lim}$, we assume that the strong lensing events
are detected out to $z_{\rm max}$ that satisfies
\begin{equation}
D_{\rm L}(z_{\rm max})=2\,D_{\rm L}(z_{\rm lim}).
\label{eq:def_zmax}
\end{equation}
At sufficiently low redshifts, this relation implies $z_{\rm
  max}\approx 2 z_{\rm lim}$.

\begin{figure}
\begin{center}
\includegraphics[width=8.0cm]{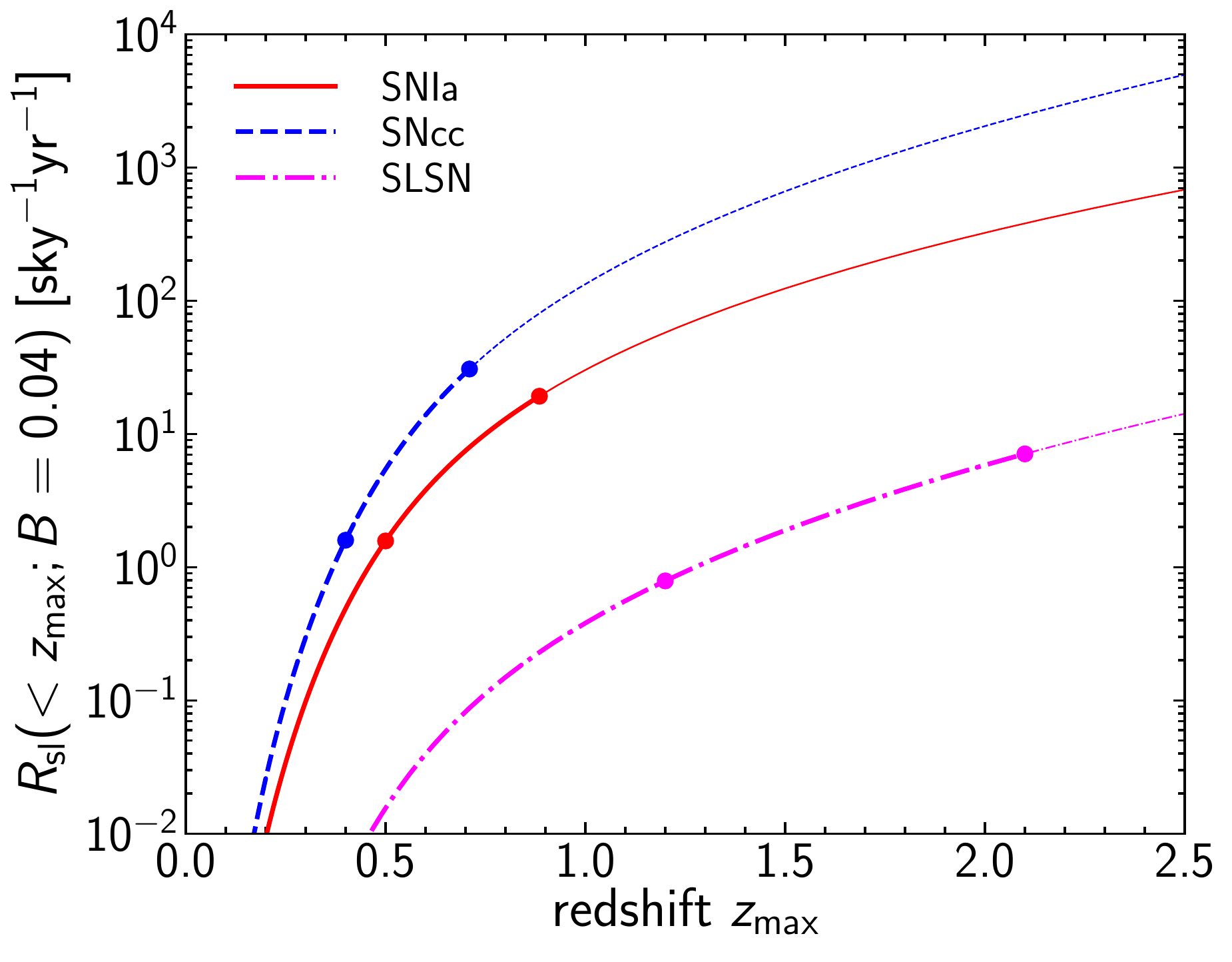}
\caption{Expected observed rates of strong lensing of supernovae as a
  function of the maximum redshift $z_{\rm max}$ computed using
  equation~(\ref{eq:r_sl}). See Table~\ref{tab:num_summary} for the
  fiducial values adopted in the calculation. We show all-sky
  ($\Omega_{\rm sky}=4\pi$) rates with $B=0.04$ that corresponds to
  strong lensing events with $\mu_{\rm tot}\gtrsim 10$. For each
  supernova type, a rough estimate of $z_{\rm lim}$ and the
  corresponding $z_{\rm max}$ (see equation~\ref{eq:def_zmax}) for
  LSST \cite{2009arXiv0912.0201L} are marked by left and right
  circles, respectively. Thick lines indicate the rough redshift
  ranges we can observe with LSST. } 
\label{fig:rates_sn}
\end{center}
\end{figure}

Figure~\ref{fig:rates_sn} shows expected observed rates of strong
lensing of supernovae along with rough estimates of $z_{\rm max}$ for LSST
\cite{2009arXiv0912.0201L}. The LSST monitors a half sky, but the
survey is conduced for 10~years. Therefore in their survey duration we
expected to discover the significant number of strongly lensed Type
Ia and core-collapse supernovae, which is consistent with more
detailed estimates (e.g., 
\cite{2010MNRAS.405.2579O,2019ApJS..243....6G,2019MNRAS.487.3342W}).
The calculation also suggests that we may be able to discover strongly
lensed superluminous supernovae. These are results for the wide survey
of LSST, whereas LSST is also planning to conduct deep drilling fields
survey where strongly lensed supernovae at higher redshifts may be
discovered. 

\begin{figure}
\begin{center}
\includegraphics[width=8.0cm]{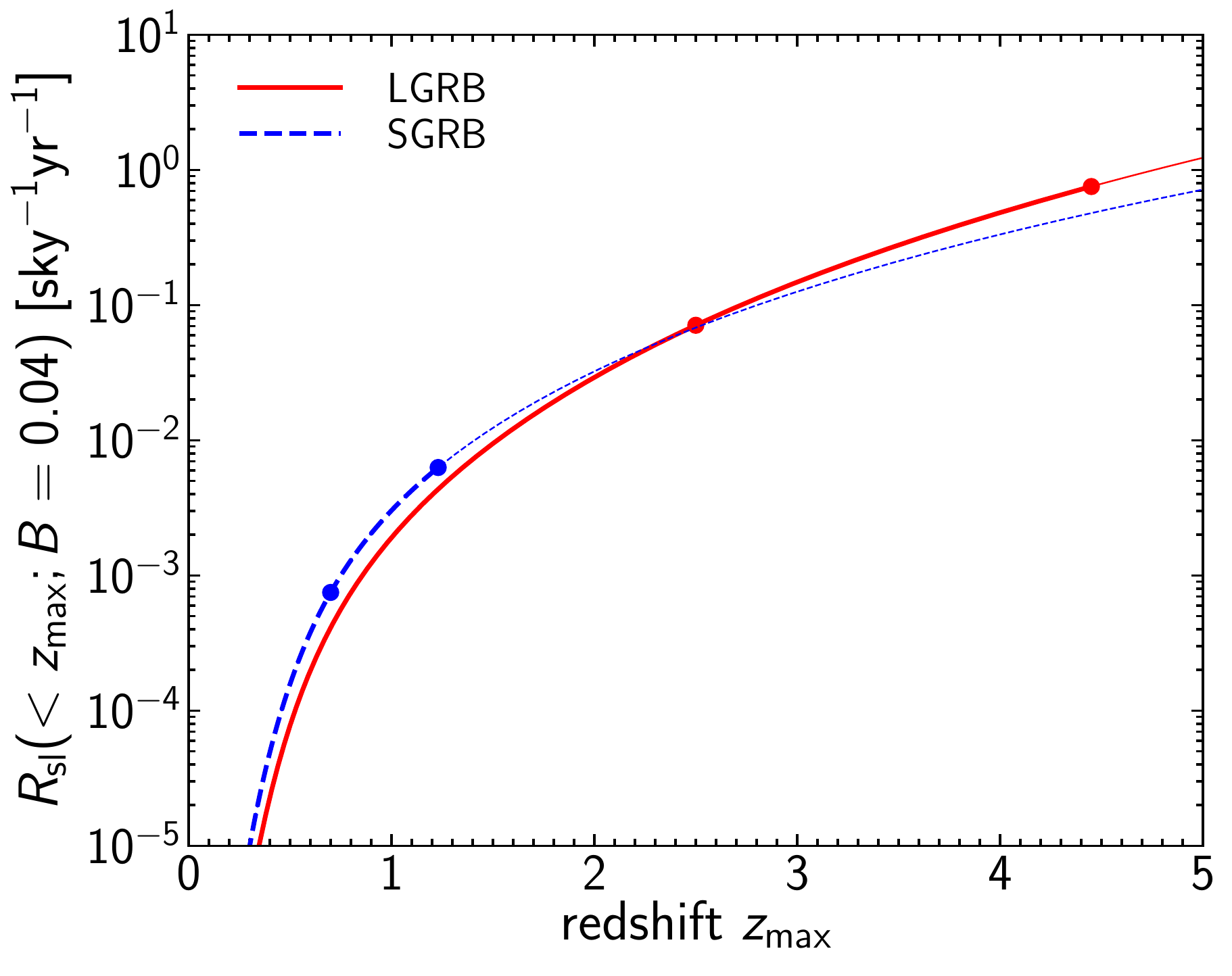}
\caption{Same as Figure~\ref{fig:rates_sn}, but for strong lensing of
  gamma-ray bursts and rough estimates of $z_{\rm lim}$ for {\it
    Swift} \cite{2014ARA&A..52...43B}.}
\label{fig:rates_grb}
\end{center}
\end{figure}

Figure~\ref{fig:rates_grb} shows expected observed rates of strong
lensing of gamma-ray bursts along with rough estimates of $z_{\rm
  max}$ for {\it Swift} \cite{2014ARA&A..52...43B}. Although the event rates of
gamma-ray bursts are low, thanks to the high mean redshift strong
lensing of long gamma-ray bursts can in principle be observed,
although one limitation is its inefficient duty cycle as discussed in
\cite{2001ApJ...548..522P}. 

\begin{figure}
\begin{center}
\includegraphics[width=8.0cm]{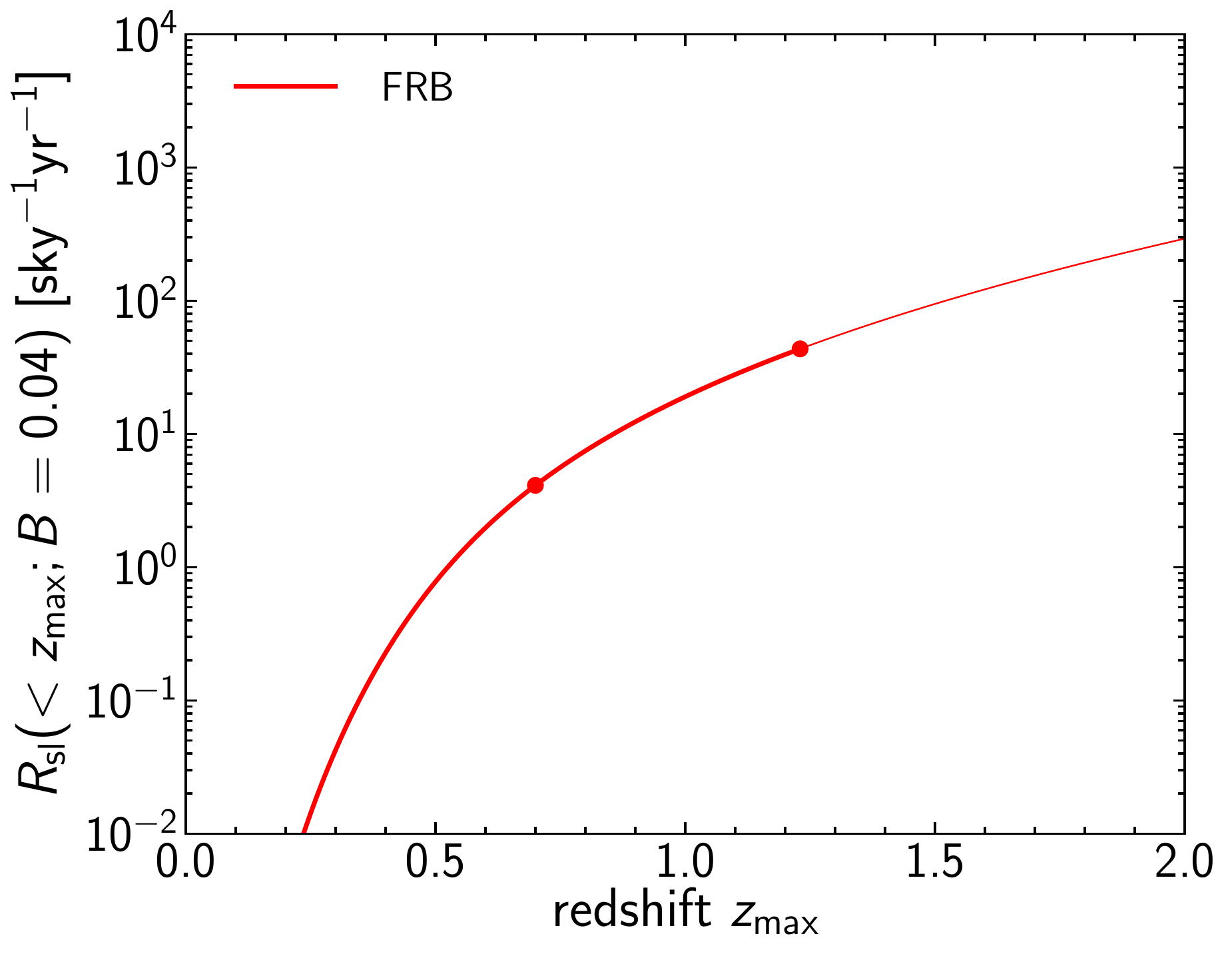}
\caption{Same as Figure~\ref{fig:rates_sn}, but for strong lensing of
  fast radio bursts and a rough estimate of $z_{\rm lim}$ for CHIME
  \cite{2019Natur.566..230C}.} 
\label{fig:rates_frb}
\end{center}
\end{figure}

Figure~\ref{fig:rates_frb} shows expected observed rates of strong
lensing of fast radio bursts along with a rough estimate of 
$z_{\rm lim}$ for CHIME \cite{2019Natur.566..230C}. Thanks to the high
event rate, the expected rate of strong lensing is also high, but
CHIME observes the sky for the area of $\sim 250$~deg$^2$ and
therefore a factor of $250/41200\approx 0.006$ should be multiplied to
obtain the actual expected observed rate in CHIME. While this
suggests that $\mathcal{O}(1)$ strong lensing events per a few years
are expected from CHIME, we caution that this estimate can easily
change by an order of magnitude or more given the quite large
uncertainties of their event rate and redshift distribution. In
addition, we note that CHIME changes observing regions on the sky
rapidly as the rotation of the Earth, which is not ideal for
identifying multiple images.

\begin{figure}
\begin{center}
\includegraphics[width=8.0cm]{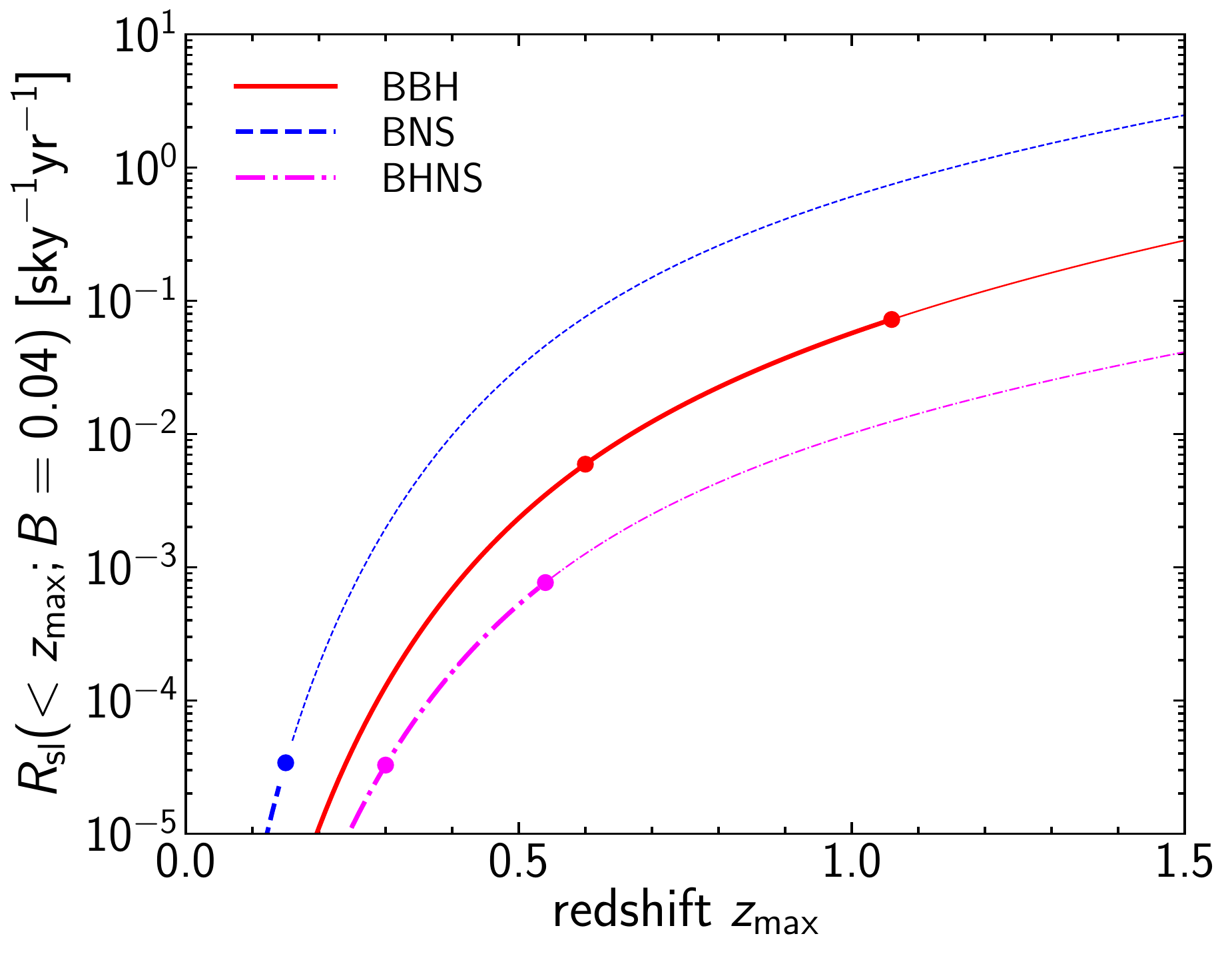}
\caption{Same as Figure~\ref{fig:rates_sn}, but for strong lensing of
  gravitational waves and rough estimates of $z_{\rm lim}$ for
  the Advanced LIGO design sensitivity \cite{2015CQGra..32g4001L}.}
\label{fig:rates_gw}
\end{center}
\end{figure}

Figure~\ref{fig:rates_gw} shows expected observed rates of strong
lensing of gravitational waves along with rough estimates of $z_{\rm
  max}$ for the Advanced LIGO design sensitivity
\cite{2015CQGra..32g4001L}. This result indicates that, albeit the
probability is not very high, it may be possible to detect strongly
lensed gravitational waves of binary black hole mergers in Advanced
LIGO, which is broadly consistent with more detailed calculations
(e.g., \cite{2018MNRAS.476.2220L,2018MNRAS.480.3842O}).

Finally, we discuss typical magnifications of strongly lensed
explosive transients detected in surveys. Table~\ref{tab:lensed_sne}
indicates that magnifications of those observed strongly lensed
supernovae are high in general, $\mu_{\rm tot}\gtrsim 30$, which may
appear odd given the steep magnification probability distribution
of $dP/d\mu \propto \mu^{-3}$. The probability distribution suggests
that such highly magnified events are much rarer than strong lensing
events with modest magnifications, $\mu_{\rm tot}<10$ or so. However, 
this apparent discrepancy can easily be resolved if we take account of
the steep dependence of the expected observed event rate on the
redshift. From equation~(\ref{eq:r_sl_norm}), at sufficiently low
redshifts we have $R_{\rm sl}(<z_{\rm max})\propto z_{\rm max}^\eta$
with $\eta\approx 6$. Therefore the differential distribution of 
$R_{\rm sl}$ at $z=z_{\rm lim}$ is given by
\begin{equation}
\frac{dR_{\rm sl}}{dz_{\rm lim}}\propto z_{\rm lim}^{\eta-1}.
\label{eq:drdz1}
\end{equation}
Also from equation~(\ref{eq:def_zmax}), again at sufficiently low
redshift, we can detect strong lensing events at $z>z_{\rm lim}$ if the
magnification factors satisfy
\begin{equation}
\mu>\left(\frac{z}{z_{\rm lim}}\right)^2.
\end{equation}
Since the cumulative probability distribution of the magnification is
$P(>\mu)\propto \mu^{-2}$, the differential distribution of 
$R_{\rm sl}$ at $z>z_{\rm lim}$ is approximately given by
\begin{equation}
\frac{dR_{\rm sl}}{dz}\propto z^{\eta-1}\left(\frac{z_{\rm
  lim}}{z}\right)^4=z^{\eta-5}z_{\rm lim}^4.
\label{eq:drdz2}
\end{equation}
By taking the ratio of equations~(\ref{eq:drdz1}) and
(\ref{eq:drdz2}), we have 
\begin{equation}
\frac{dR_{\rm sl}/dz}{dR_{\rm sl}/dz_{\rm lim}} \propto
\left(\frac{z}{z_{\rm lim}}\right)^{\eta-5}>1\;\;\;\;(\eta>5).
\label{eq:drdz3}
\end{equation}
Since equation~(\ref{eq:drdz3}) is an increasing function of $z$, we
preferentially observe strong lensing events with $z\gg z_{\rm lim}$
i.e., $\mu\gg1$, which qualitatively explains the high magnification
factors of PS1-10afx and iPTF16geu. In sufficiently deep surveys, on
the other hand, Figure~\ref{fig:slrate} implies that the slope of the
strong lensing rate becomes shallower, $\eta<5$, for which this
argument no longer holds so that strongly lensing events with modest
magnifications are preferentially observed, although the detail
depends also on the shape of the luminosity function. We expect that,
at least for sufficiently shallow surveys such as CHIME for fast radio 
bursts and Advanced LIGO for gravitational waves, we typically observe
highly magnified events with redshifts well beyond the redshift limit
of unlensed events. This point has also discussed in e.g., 
\cite{2018MNRAS.480.3842O} in the context of strongly lensed 
gravitational waves. 

\section{Conclusions}
\label{sec:conclusions}

In this article, we have reviewed the science of strong lensing of
explosive 
transients, specifically focusing on supernovae, gamma-ray bursts,
fast radio bursts, and gravitational waves from compact binary
mergers. Although many strongly lensed quasars and galaxies have
already been identified, strong lensing of these explosive transients
is complementary to traditional strong lensing and enables new
applications that were not possible before. In this article we have
discussed possible applications of these new strong lensing events,
summarized the current status of strong lens searches, and presented
expected rates of strong lensing events adopting a simplified
approach. 

Rapidly evolving light curves of these transients indicate that we
expect a lot of progress in applications of time delays between
multiple images. In particular for gamma-ray bursts, fast radio
bursts, and gravitational waves, thanks to their very short time
scales of $\lesssim 1$~sec we can drastically improve the accuracy of
time delay measurements as compared with the current accuracy for
strongly lensed quasars, $\sim 1$~day. The very accurate
measurements of time delays open new avenues, including improved
constraints on cosmological parameters such as the Hubble constant
$H_0$, a probe of small-scale perturbations from dark matter
substructures, tests of fundamental physics from the propagation
speed, and constraints on the abundance of compact dark matter from
the search of pair events with short time delays. 

The compact sizes of these explosive transients imply that wave
optics effects may play an important role. When the frequency is
comparable to the inverse of the typical time delay, wave optics effects
induce the interference pattern as a function of the source position
and frequency, although in order for this effect to be observed the
source size must be smaller than the width of the interference pattern.
On the other hand, the frequency is much lower than  the inverse of
the typical time delay, lensing magnifications are suppressed due to
diffraction. We have presented detailed discussions on whether
wave optics effects are relevant for strong lensing of these explosive
transients, and argued that wave optics effects can become important for
strong lensing of gravitational waves, and probably for strong lensing
of fast radio bursts as well depending on their actual progenitor
sizes. 

Strong lensing may also help better understand these explosive
transients. Thanks to lensing magnifications, we can detect very high
redshift events that are not accessible without lensing
magnifications. Lensing magnifications can also be used as a
microscope to resolve the fine structure of sources to constrain their
progenitor models.

We have presented the current status and future prospect of the strong
lens search. We have described recent discoveries of strongly lensed
supernovae, PS1-10afx \cite{2014Sci...344..396Q}, SN Refsdal
\cite{2015Sci...347.1123K}, and iPTF16geu \cite{2017Sci...356..291G}.
We have computed expected observed rates of strong lensing of various
explosive transients adopting a simple and concise approach. These
calculations suggest that strong lensing of gamma-ray bursts, fast
radio bursts, and gravitational waves, can be observed in near
future. Using this simple model, we have discussed selection effects,
and showed that highly magnified strong lensing events are
preferentially observed in shallow surveys.

In this review article, we have not covered all explosive transients at
cosmological distances. For instance, the tidal disruption event is a
disruption of a star by the tidal force of the back hole (see e.g.,
\cite{2015JHEAp...7..148K}), which can also act as a source of strong
lensing. Furthermore, time-domain surveys may identify new types of
transients that are similar to strong lensing of explosive transients.
One such example is a caustic crossing. When a star in a
gravitationally lensed galaxy passes through a caustic it is
magnified by a factor of several thousands or more and hence can be
detected even at cosmological distances. Recent discoveries of caustic
crossings of individual stars at $z\sim 1-1.5$
\cite{2018NatAs...2..334K,2018NatAs...2..324R,2019ApJ...880...58K,2019ApJ...881....8C}
have attracted a lot of attention. The caustic crossing has the time
scale of the light curve near the peak as short as several days, and
therefore is definitely an interesting target to find in future
time-domain surveys, in addition to strongly lensed explosive
transients. In coming years, these new time-variable strong lensing
events will deepen our view of the Universe in several ways. 

\section*{Acknowledgments}
I thank Kazumi Kashiyama for useful discussions, and Ariel Goobar,
Shaoqi Hou, and Yufeng Li for useful comments. I also thank anonymous
referees for many useful comments and suggestions.
This work was supported in part by World Premier International
Research Center Initiative (WPI Initiative), MEXT, Japan, and JSPS
KAKENHI Grant Numbers JP18H04572, JP15H05892, and JP18K03693. 

\section*{References}

\newcommand{\newblock}{} 
\bibliographystyle{JHEP} 
\bibliography{review}

\end{document}